\documentclass[tightenlines,a4paper,twocolumn,floatfix,aps,showpacs,prd,epsf]{revtex4}
\usepackage{graphicx}
\usepackage{subeqn}
\usepackage{color}
\usepackage{epsfig}
\usepackage{dcolumn}
\usepackage{bm}
\def\dis{\displaystyle}
\def\beq{\begin{equation}}
\def\eeq{\end{equation}}
\def\barr{\begin{array}}
\def\earr{\end{array}}
\def\shat{\hat s}
\def\that{\hat t}

\def\gsim{\buildrel{\scriptscriptstyle >}\over{\scriptscriptstyle\sim}}
\def\gev{\, {\rm GeV}}
\def\tev{\, {\rm TeV}}
\def\fb{\, {\rm fb}}
\def\pb{\, {\rm pb}}

\begin{document}

\title{Quark Excitations Through the Prism of 
Direct Photon Plus Jet at the LHC}
\author{Satyaki Bhattacharya\footnote{Email Address: bhattacharya.satyaki@gmail.com}, Sushil Singh Chauhan\footnote{Email Address: sushil@fnal.gov}, Brajesh Chandra Choudhary\footnote{Email Address: brajesh@fnal.gov} and  Debajyoti Choudhury\footnote{Email Address: debchou@physics.du.ac.in}}
\affiliation{ \indent Department of Physics and Astrophysics, University of Delhi, Delhi 110007, India.}

\begin{abstract}

The quest to know the structure of matter has resulted in various
theoretical speculations wherein additional colored fermions are
postulated. Arising either as Kaluza-Klein excitations of ordinary
quarks, or as excited states in scenarios wherein the quarks
themselves are composites, or even in theories with extended gauge
symmetry, the presence of such fermions ($q^*$) can potentially be
manifested in $\gamma + jet$ final states at the LHC. Using unitarized
amplitudes and the CMS setup, we demonstrate that in the initial phase
of LHC operation (with an integrated luminosity of $200 \pb^{-1}$) one
can discover such states for a mass upto 2.0 TeV.  The discovery of a
$q^*$ with a mass as large as $\sim$5 TeV can be acheived for an
integrated luminosity of $\sim 140 \fb^{-1}$. We also comment on the 
feasibility of mass determination.
\end{abstract}

\pacs{12.60.Rc, 13.40.-f, 13.85.Qk}
\maketitle
\section{Introduction}
    \label{sec:introd}
The replication of fermion families, while being of profound 
significance in our understanding of fundamental issues such 
as $CP$-violation and baryogenesis, nonetheless is puzzling 
in its own right. 
Despite its enormous success in explaining all observed phenomena in
the regime of particle physics, the Standard Model (SM) has been
entirely unable to proffer any insight into this aspect. 
Indeed, though the observed mass hierarchies and fermion
mixings can easily be accommodated, the SM framework, by its very
structure, is unable to even ask such questions of itself.  This has
led to various speculative ideas seeking to explain these
ill-understood issues. Prominent among these are ($i$) models with
extended (family) symmetry, ($ii$) constructions based on higher
dimensional theory (with or without a string theory motivation) and
($iii$) the possibility of quark-lepton compositeness, namely that the
SM fermions are not elementary at all.

Many of the ideas discussed 
in this article would be equally applicable---possibly with minor 
variations---to theories belonging to 
any of these three classes. However,
for the sake of concreteness, we shall 
consider theories of compositeness as the basic template 
for our discussions. Part of the motivation lies in the fact 
of these theories 
having a more straightforward ultraviolet completion and, furthermore, 
suffering from a fewer number of extra channels, thereby reducing 
possible ambiguities.

In such theories, the fundamental constituents of matter, very often termed
{\em preon}s\cite{preon}, are postulated to 
experience an hitherto unknown force on
account of an asymptotically free but confining gauge
interaction\cite{preon_binding}, which becomes very strong at a
characteristic scale $\Lambda$, thereby leading to bound states
(composites) to be identified with quarks and/or leptons.  In
many such models\cite{comp_mod, Additional_comp}, though not all, 
quarks and leptons
share at least some common constituents.

Accepting this hypothesis would naturally lead to 
the existence of excited states of fermions at a mass scale comparable to
the dynamics of the new binding force.  
In the simplest
phenomenological models \cite{hagi_baur_boud}, the excited fermions
are assumed to have both spin and isospin 1/2 and to have both their
left- and right-handed components in weak isodoublets ({\em i.e.}\
they are vector-like). Similar is the case for, say higher-dimensional 
models wherein the known universe is constrained to be on a 4-dimensional 
subspace (a 3-brane) while the SM fields---in particular, the fermions---live
in all the dimensions. The analogues of the excited fermions would be the 
Kaluza-Klein excitations with the mass scale being identified with the 
inverse of the compactification scale.

Given that the ``excited states'' do suffer the SM gauge interactions,
these may be produced at high-energy colliders and, subsequently, 
would decay,
radiatively, into an ordinary fermion and a gauge boson (photon,
$W$, $Z$ or gluon).  At an $e^+ e^-$ collider, 
charged excited fermions could be 
pair-produced via $s$-channel ($\gamma$ and $Z$) exchanges in 
collisions, while for excited neutrinos only $Z$ exchange
contributes. Although $t$-channel diagrams are also possible 
($W$ for $\nu^*_e$ and $\gamma/Z$ for $e^*$), 
such contributions to the overall pair
production cross-section are generally much smaller 
on account of the smallness of the
cross-couplings~\cite{hagi_baur_boud} between 
the excited state, its SM counterpart and a gauge
boson.  This very same
coupling, on the other hand, 
may be used to singly produce such states (through both
$s$- and $t$-channel diagrams).  The four LEP collaborations have used
these (and other) modes to essentially rule out such excitations
almost upto the kinematically allowed range~\cite{LEPresults}.  At the
{\sc hera}, on the other hand, both excited leptons and quarks may be
produced singly through $t$-channel diagrams and these processes have
been looked at without any positive results~\cite{HERA}.


If quarks and leptons are composite particles made up of smaller
constituents, phenomenological effects may be observable at the LHC
and might even show up at Tevatron once sufficiently large luminosities
are accumulated and analysed.  If the 
compositeness/excitation scale
($\Lambda$) is not too high, then excited quarks can be produced
on shell, while at energies sufficiently below $\Lambda$ such
an eventuality would manifest itself as an effective four fermion contact
interaction invariant under SM guage
transformations~\cite{cont_inter,majhi}. Based on phenomenological studies of flavour independent
contact interactions for two photon production, the lower bound has
been estimated to be $\Lambda_{\pm} >$ 2.88 (3.24) TeV at 95$\%$ CL
for an integrated luminosity of 100 (200) $\fb^{-1}$~\cite{diphoton} at
the LHC.

Both the experiments at the Tevatron, the CDF and D$\O$, have searched
for excited quarks. The latter are assumed to couple
to the SM particles primarily through gauge couplings. 
Their most visible signature could be either pair
production or single excited state production via quark-gluon fusion,
provided the $q^* q g$ coupling is sufficiently large. Enhanced dijet
production rate with an invariant mass peak above the SM continuum
is one of the prominent signals, extensively searched for at the
Tevatron and the D$\O$ collaboration has excluded the
mass range of 200-720 GeV\cite{d0_m}. Similarly, the 
CDF collaboration has excluded a mass range of 80-570 GeV
\cite{cdf_m1, cdf_m2} based on searches with various final states. 
In a similar vein, 
the CDF collaboration has put a lower bound of
$\Lambda \geq 2.81$ GeV at 95$\%$ CL using the $q\bar{q}\rightarrow e \nu$
prcoess\cite{cdf_W} whereas the D$\O$ collaboration rules out
 $\Lambda \leq $ 2.0 TeV at 95$\%$ CL from diject mass peak
searches\cite{d0_dijet}.


At the LHC, both the ATLAS and the CMS collaborations have predicted
the sensitivity in the dijet production mode. The ATLAS collaboration
has xlaimed that the use nof dijet angular distributions would allow
contact interactions to be probed upto $\Lambda=$10 TeV with a
integrated luminosity of 700 $pb^{-1}$. The CMS collaboration, on the
other hand, has estimated that $\Lambda=$ 6.2 TeV can be excluded at
95$\%$ CL with a luminosity of 100 $pb^{-1}$ and that 5$\sigma$
sensitivity could be reached for $\Lambda=$8 TeV with just $1
\fb^{-1}$ of data\cite{lhc_study}. Recently, the possibility of top
quark compositeness has been explored through the $pp\rightarrow
t\bar{t}t\bar{t}$ production process and it has been estimated that a
5$\sigma$ excess can be observed for a new state of 2 TeV
\cite{top_compositness}.


As an effective tool for the measurement of gluon density inside the
colliding hadrons and for precision test of pQCD predictions, the
isolated $\gamma+$jet final state has been studied with great detail
at the Tevatron collider\cite{collider_1,collider_2,collider_3} and
fixed target experiments\cite{fixed_target}. Since $\gamma+jet$ final
state will be one of the key backgrounds for the
$H\rightarrow\gamma\gamma$ search at the LHC, extensive isolation
studies addressing all known issues have been performed with this
process both theoretically and with detailed detector simulations. It
is also an important background for many new physics scenario
e.g. Large Extra Dimensions\cite{LED_1,LED_2}, Randall Sundrum
Gravitons\cite{RSG} etc. It will, thus, be very interesting to
look at $\gamma+$jet as a probe of excited quarks in view of the
unprecedented energy scale at LHC and the in-depth knowledge of this
process gleaned from previous experiments and studies.


The rest of the paper has been organized as follows. In section II
we have discussed the effective Lagrangian for the theory and new
physics contribution. In Sections III and IV, we respectively discuss the
backgrounds and event generation. 
Section V describes the photon and jet algorithms
used for the analysis. In section VI we discuss the smearings
due to detector resolution effects. Section VII gives the details of
kinematical variables used to separate the signal from the
background whereas Section VIII deals with isolation study. The significance of signal and discovery are discussed in section IX and X
respectively. In section XI we have presented the result of the
analysis. Systematics is discussed in detail in section XII
followed by our conclusions and outlook.

\section{New Physics contribution to $\gamma + \it Jet$ production}
With our interest lying not in the pair-production of such excited states, 
but rather on their contribution to the photon plus single jet rates at a 
hadronic collider, we confine ourselves to examining  
only the relevant part of the Lagrangian, namely 
the (chromo-)magnetic transition between 
ordinary and excited states. In general, 
these may be parametrized by
\begin{equation}
{\mathcal L}_{int} = \frac{1}{2 \, \Lambda}\bar{q^*_R} \, \sigma^{\mu\nu}
\left[\sum_i g_i \; b_i \; T_i^a \; G^a_{i \, \mu\nu} \right] q_L + h.c.,
   \label{eq:Lagrangian}
\end{equation}
where the index $i$ runs over the three SM gauge groups, viz. $SU(3)$,
$SU(2)$ and $U(1)$ and $g_i$, $G^a_{i \, \mu\nu}$ and $T_i^a$ are the
corresponding gauge couplings, field strength tensors and generators
respectively. The dimensionless constants $b_i$ are, {\em a priori},
unknown and presumably of order unity. With these determining both the
production rates and the branching into various modes, clearly, the
phenomenology would depend considerably on their (relative) sizes. In
this article, we shall make the simplifying assumption that the
excited states do not couple at all to the weak gauge bosons, but do
so with the gluons and the photon. At first glance, this might seem
incompatible with a $SU(2) \otimes U(1)$ invariant structure. 
However, complicated embeddings could be the answer. More than this, 
since the assumption would not change the results 
qualitatively, it,  at least, has the merit of reducing the number 
of possible couplings and hence simplifying the analysis.

A further point needs to be noted here.  With the Lagrangian of
eq.(\ref{eq:Lagrangian}) being a higher dimensional operator, the
cross sections would typically grow with the center of mass energy,
consequently violating unitarity. This is not unexpected in an
effective theory as the term in eq.(\ref{eq:Lagrangian}) is only the
first term and the loss of unitarity, to a given order, is presumably
cured once suitable higher dimensional operators are included.  An
equivalent way to achieve the same goal is to consider the $b_i$ to be
form factors rather than constants. To this end, we shall define the
$q^* q \gamma$ and $q^* q g$ vertices to be given by
\beq
\barr{rcl}
\overline{q^*} \, q \, \gamma_\mu (p)  : & \qquad & \dis
  \frac{e \, e_q \, f_{1}}{\Lambda} \; 
     \left(1 +  \frac{Q^2}{\Lambda^2}\right)^{-n_1} \;
             \sigma_{\mu \nu} \; p^\nu
\\[2ex]
\overline{q^*} \, q \, g_\mu (p) : & \qquad & \dis
\frac{g_{s} f_{3}}{\Lambda} \; \left(1 + \frac{Q^2}{\Lambda^2}\right)^{-n_3} \;
             \sigma_{\mu \nu} \; p^\nu \; T_{\alpha}
\earr
   \label{eq:vertices}
\eeq
where $Q$ denotes the relevant momentum transfer and $f_i \sim 1$ are  
dimensionless constants related to $b_i$ of eqn.(\ref{eq:Lagrangian}).
It can be checked
that, for $Q^2 = s$, unitarity is restored as long as the constants
$n_i \geq 1$. From now on, eqn.(\ref{eq:vertices}) defines our 
theory$^{\footnotemark[1]}$.
For the rest of our analysis, we shall confine ourselves
to a discussion of $n_i = 1$. 
While this might seem to be an optimistic choice, it is not quite so.
In fact, the collider search limits in the literature actually 
correspond to $n_i = 0$ and, thus, our limits would be more 
conservative.

\footnotetext[1]{While a Lagrangian formulation leading to such vertices
would necessitate a seemingly non-local Lagrangian, this is not
unexpected in an effective theory.}

With the aforementioned Lagrangian, the width of the $q*$ is given by
\beq
\barr{rcl}
\Gamma(q*) & = & \dis 
\Gamma_{q + g} + \Gamma_{q + \gamma}
\\[2ex]
\Gamma_{q + g} & = & \dis 
\frac {2 \alpha_{s} f_{3}^{2} } {3} \, \Gamma_0
\\[2ex]
\Gamma_{q + \gamma} & = & \dis
\frac {e_{q}^{2} \alpha f_{1}^{2} } {2 } 
\, \Gamma_0
\\[2ex]
\Gamma_0 & \equiv & \dis
\frac {M_{q^{*}}^{3}}{ \Lambda^{2}} \left(1- \frac{4 m_{q}^{2}}{M_{q^{*}}^{2}} \right)  \left( 1- \frac{m_{q}^2}{M_{q^{*}}^2} \right)^{2}
\earr
\label{eq:gamma_1}
\eeq
and can be very large for a heavy $q^*$ (see Table \ref{Table:widths}). 
As a fat resonance is often 
difficult to observe, this will turn out to have profound consequences.
\begin{table}[!h]
\caption{ $\Gamma(q*)$ as a function of $M_{q*}$(=$\Lambda$) 
for different coupling strengths. Both $\alpha_{s}$ and $\alpha_{em}$
are evaluated at $M_{q^*}$.}  
\begin{ruledtabular} 
\begin{tabular}{ccc} 
$M_{q*}$ &  \multicolumn{2}{c}{$\Gamma$ (GeV)}\\
\cline{2-3}
\\
(TeV) & $f_1 = f_3 = 1.0 $ & $f_1 = f_3 = 0.5 $  \\
\hline
0.5 &  34.4 & 8.61\\

1.0 &  63.6 & 15.9 \\

2.0 & 118 & 29.6 \\

3.0 & 170 & 42.6 \\

4.0 & 221 & 55.2 \\

5.0 & 271 & 67.6 \\

6.0 & 319 & 79.8 \\

7.0 & 367 & 91.8 \\
\end{tabular}
\end{ruledtabular}
   \label{Table:widths}
\end{table}


\begin{figure}[htbp]
\vspace*{-8ex}
\scalebox{0.20}{\includegraphics{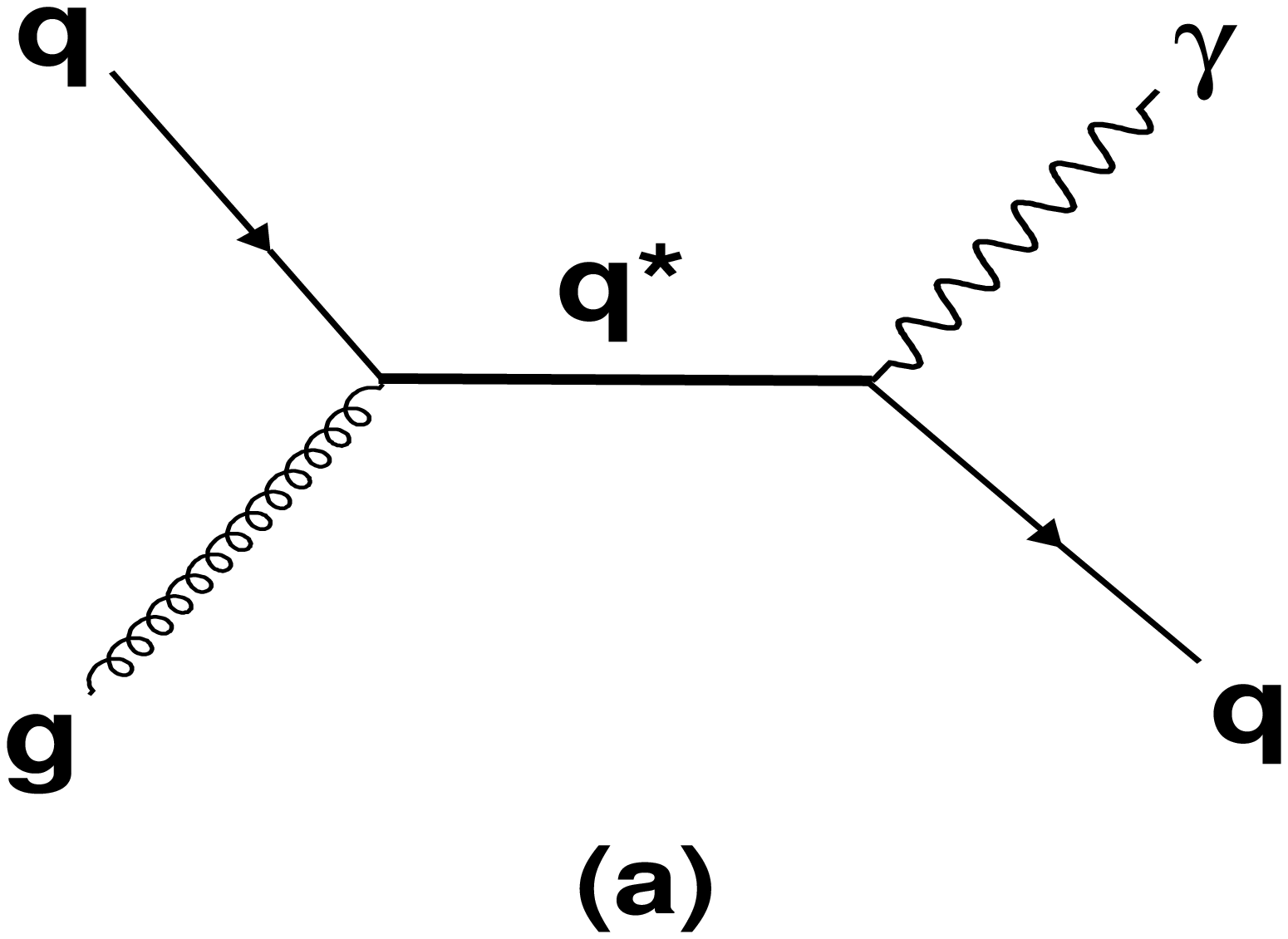}}                      
\scalebox{0.20}{\includegraphics{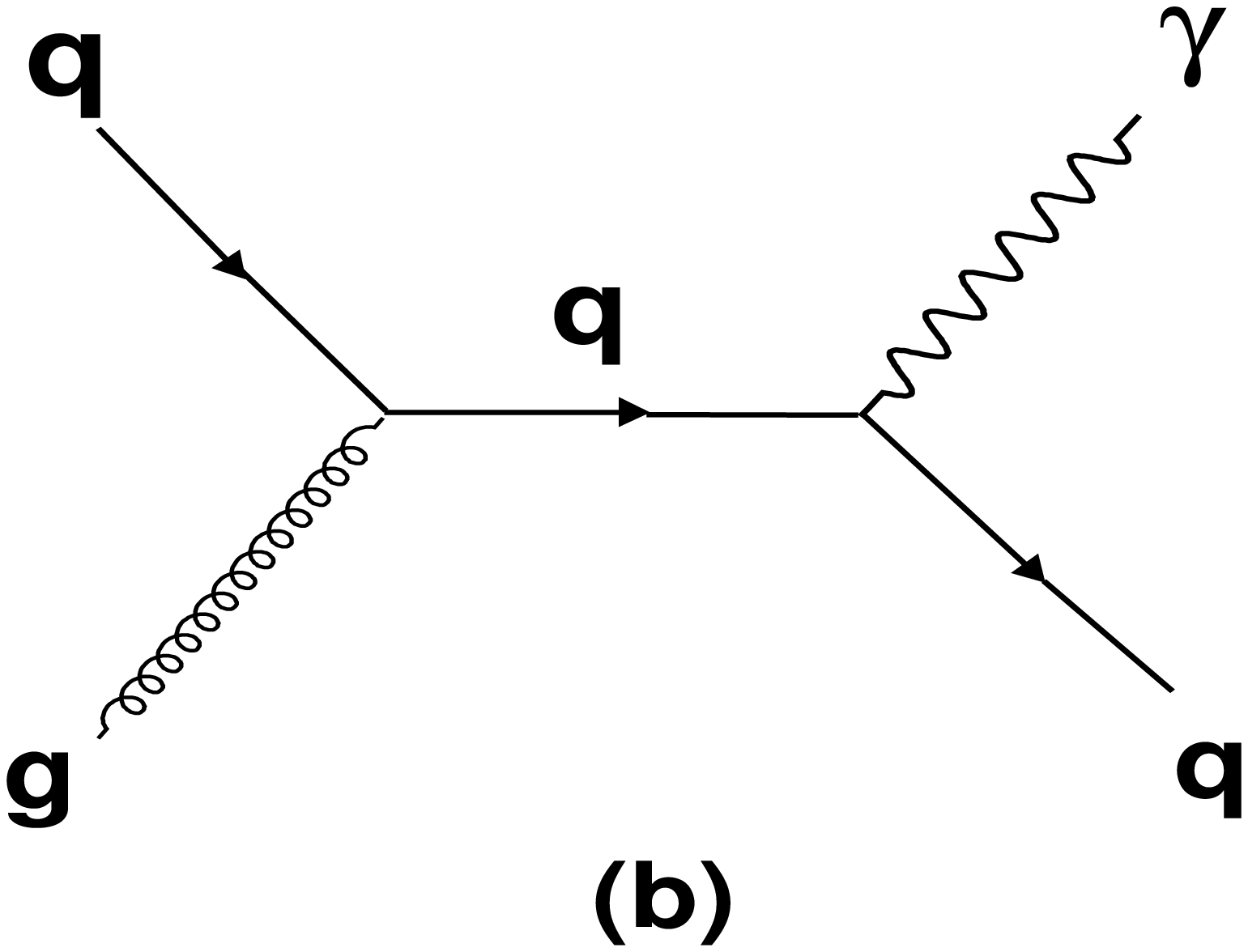}}                                 
\vspace*{-20ex}
\scalebox{0.20}{\includegraphics{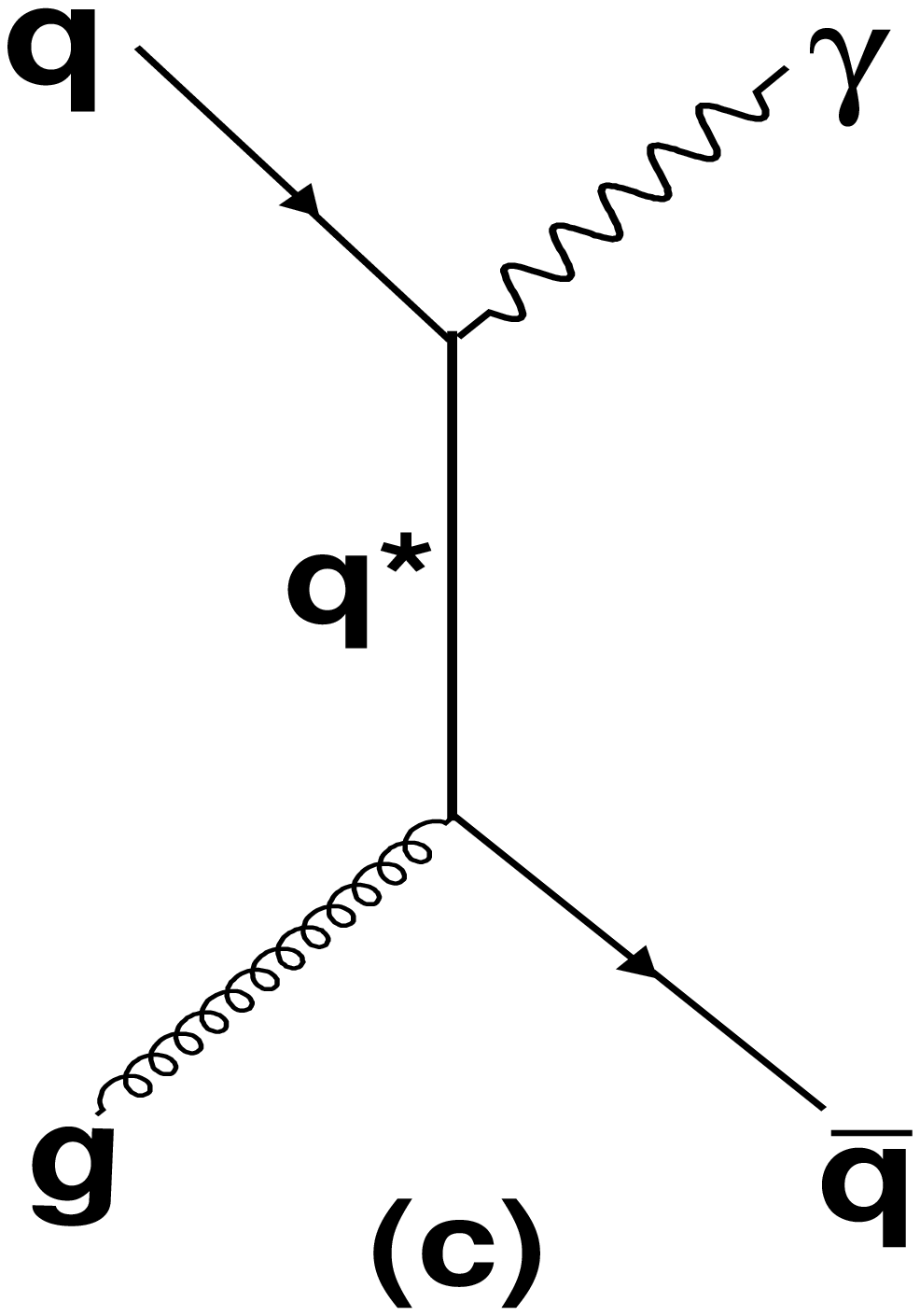}}
\scalebox{0.20}{\includegraphics{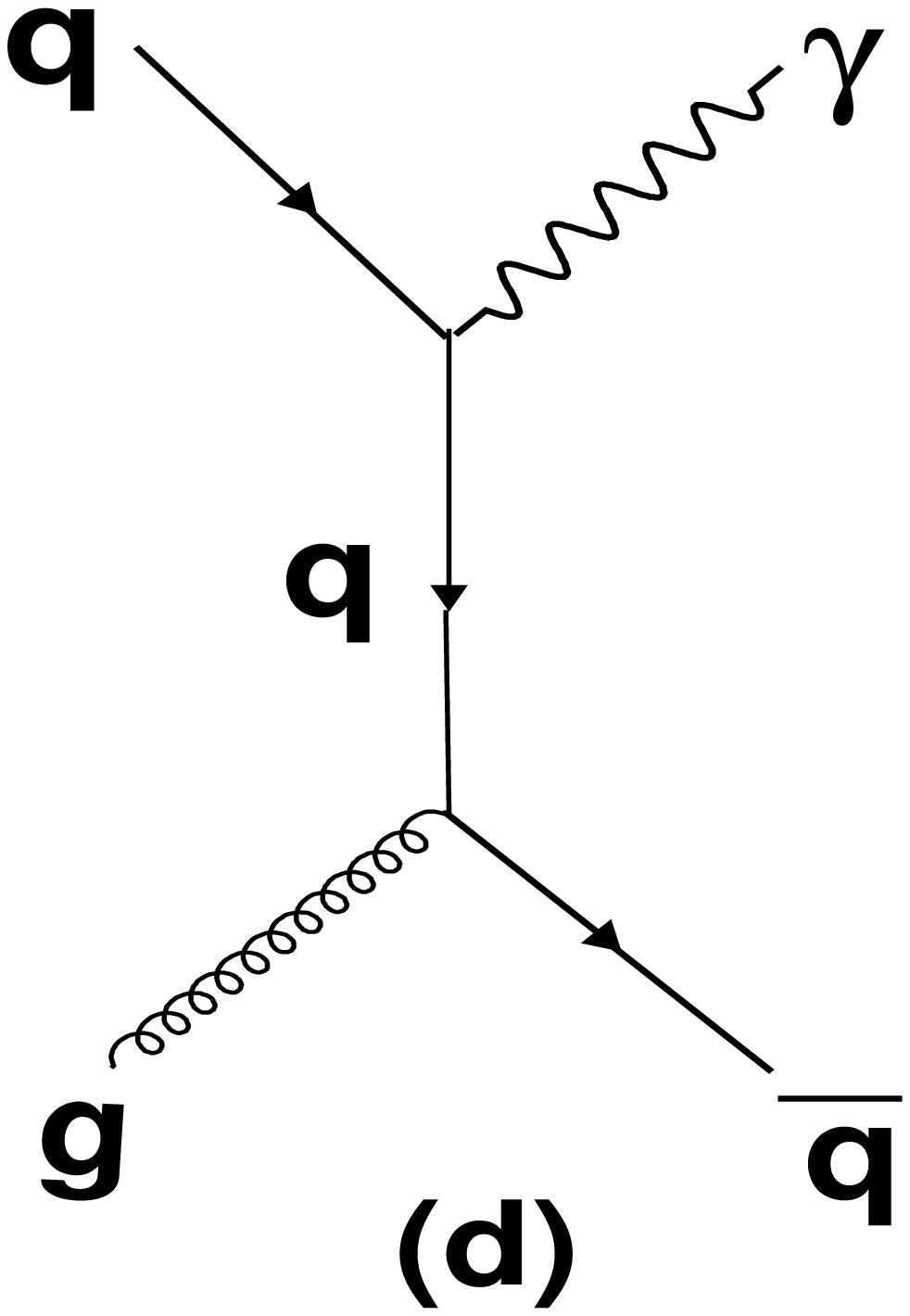}}
 
\vspace*{15ex}
\caption{\label{fig1}Production of $\gamma + Jet$ final state through 
excited quark mediation (a \& c) as well as 
SM processes (b \& d).}
\label{fig:qg_gamq}
\end{figure}

With the introduction of these  (flavour-diagonal) 
vertex as in eq.(\ref{eq:vertices}), the 
subprocess $q g \to q \gamma$ acquires a new contribution as 
portrayed in Fig.\ref{fig:qg_gamq}(a). Adding this contribution to the 
pure QCD one, the ensuing differential cross section reads
\begin{equation}
\barr{rcl}
\dis \left. \frac{d \sigma}{d \that} \right|_{q g \to q \gamma}
& = & \dis
\frac{-\pi \, \alpha \, \alpha_{s} \, e_{q}^{2}} {3 \, \shat^2} \, 
        \left[ C_{sm}+ 2 \frac {f_{1} f_{3}}{\Lambda^{2}} C_{I} 
           + \frac {f_{1}^{2} f_{3}^{2}} {\Lambda^{4}} C_{Q} \right]
\\[3ex]

C_{sm} & \equiv & \dis \frac{\hat{u}}{\hat{s}} + \frac {\hat {s}}{\hat {u}} 
   
\\[3ex]

C_{I} & \equiv & \dis 
\frac {\hat{s}^{2} \, (\hat{s}-M_{q^{*}}^{2}) \, {\cal F}_s }
      {(\hat{s}-M_{q^{*}}^{2})^{2} + \Gamma^{2} M_{q^{*}}^{2}} 
+ \frac{\hat{u}^{2} \,{\cal F}_u }{\hat{u}- M_{q*}^{2}} 
\\[3ex]
C_{Q} & \equiv & \dis  \left ( \hat s \hat u + M_{q^{*}}^{2} t  \right) 
\\[1ex]
&  & \dis 
\Bigg[ 
\frac {\hat s^{2} \; {\cal F}_s^2 } 
      {(\hat{s}- M_{q^{*}}^{2} )^{2} + \Gamma^{2} M_{q^{*}}^{2} } 
 +
\frac { \hat{u}^{2} \; {\cal F}_u^2 } 
       {(\hat{u} - M_{q^{*}}^{2})^{2}} 
    \, 
\Bigg]
\\[3ex]
& + & \dis 
2 M_{q^{*}}^{2} \frac{\hat{s} \hat{t} \hat{u}}{\hat{u}-M_{q^{*}}^{2}} \;
\frac {(\hat{s}- M_{q^{*}}^{2}) \, {\cal F}_s \, {\cal F}_u}
      { (\hat{s}- M_{q^{*}}^{2})^2 +\Gamma^{2} M_{q^{*}}^{2}} 
\\[3ex]
{\cal F}_s & \equiv & \dis 
     \left(1 + \hat s / \Lambda^{2} \right)^{-(n_{1}+n_{3})}
\\[2ex]
{\cal F}_t & \equiv & \dis 
     \left(1 - \hat t / \Lambda^{2} \right)^{-(n_{1}+n_{3})}
\\[2ex]
{\cal F}_u & \equiv & \dis 
     \left(1 - \hat u / \Lambda^{2} \right)^{-(n_{1}+n_{3})} 
\earr
\label{eq:matrix_1}
\end{equation}
with the SM result being recovered in the limit $\Lambda \to \infty$.
The new physics contribution to the differential cross section thus
depends on four parameters, namely $f_1, f_3, \Lambda$ and the mass of
the excited state $M_{q^*}$.  For simplicity, we assume these to be
flavour-independent (within a generation, it obviously has to be
so). For eq.(\ref{eq:Lagrangian}) to make sense as an effective
Lagrangian, the masses have to be less than $\Lambda$
(Ref.\cite{Hasenfratz:1987tk} requires that $M_{q^*} < \Lambda /
\sqrt{2}$). Note that as long as $\Lambda \gg \hat s$, one of $f_{1,
3}$ can always be absorbed in $\Lambda$. In our analyses, we would be
considering only moderate values for these parameters.

For $q \bar q \to g \gamma$, the Feynman diagrams are as in 
Fig.\ref{fig:qg_gamq}(c-d); the differential 
cross-sections are related to those in eqn.(\ref{eq:matrix_1}) by crossing 
symmetry and are given by 
\begin{equation}
\barr{rcl}
\dis \left. \frac{d \sigma}{d \that}\right|_{q \bar q \to g \gamma}
& = & \dis
\frac{ 8 \pi \, \alpha \, \alpha_{s} \, e_{q}^{2}} {9 \, \shat^2}  
\left[ B_{sm} - 2 \frac {f_{1} f_{3}}{\Lambda^{2}} B_{I} 
       + \frac {f_{1}^{2} f_{3}^{2}} {\Lambda^{4}} B_{Q} \right]
\\[3ex]
B_{sm} & \equiv & \dis \frac {\hat u } {\hat t } + \frac{\hat t }{\hat u}
\\[3ex]
B_{I}  & \equiv & \dis  
      \frac {\hat t^{2} \, {\cal F}_t }{\hat t - M_{q^*}^{2}} 
    + \frac {\hat u^{2} \, {\cal F}_u }{\hat u - M_{q^*}^{2}} 
\\[3ex]

B_{Q} & \equiv & \dis \hat t \hat u  \,
      \left[ \frac {\hat t^{2} \, {\cal F}_t^2 }{(\hat t - M_{q^*}^{2})^{2}} 
            + \frac {\hat u^{2} {\cal F}_u^2 }{(\hat u - M_{q^*}^{2})^{2}}  
      \right]
\\[3ex]
& + & \dis \hspace*{0em}
 M_{q^*}^{2} \, \hat s  \, 
   \left[ \frac {\hat t \, {\cal F}_t}{\hat t-M_{q^*}^{2}} 
        + \frac {\hat u \, {\cal F}_u}{\hat u-M_{q^*}^{2}} 
   \right]^{2}
\earr
\label{eq:matrix_2}
\end{equation}

\section{Backgrounds}
     \label{sec:bkgd}

\begin{figure}[htbp]
\scalebox{0.20}{\includegraphics{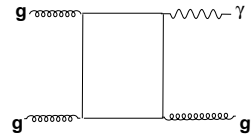}}
\vspace*{-8ex}
\caption{\label{fig3} Type-I Background: Additional contribution from 
gluon fusion. Lowest order background emanates from (b) and (d) in 
Fig.\ref{fig:qg_gamq}.}
\label{fig:bkg_1}
\end{figure}

\begin{figure}[htbp]
\scalebox{0.20}{\includegraphics{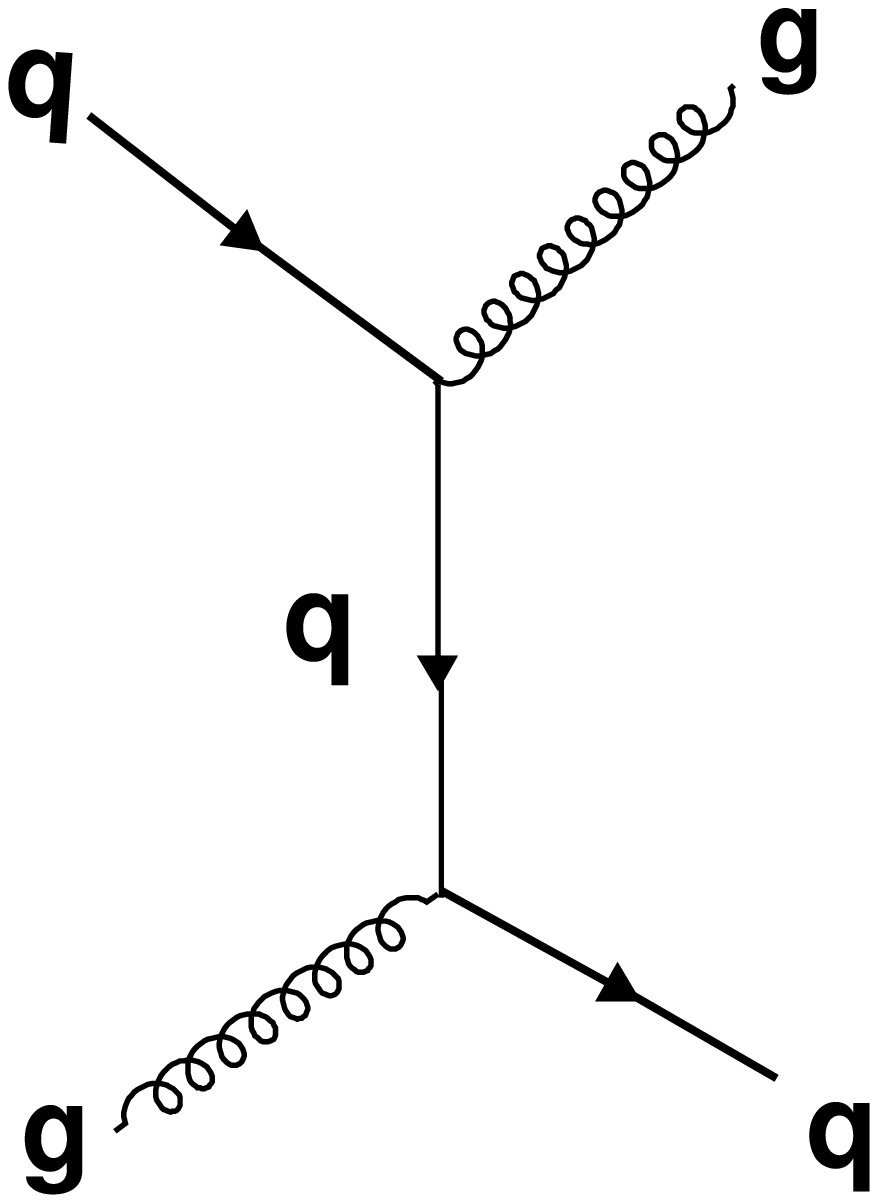}}
\scalebox{0.20}{\includegraphics{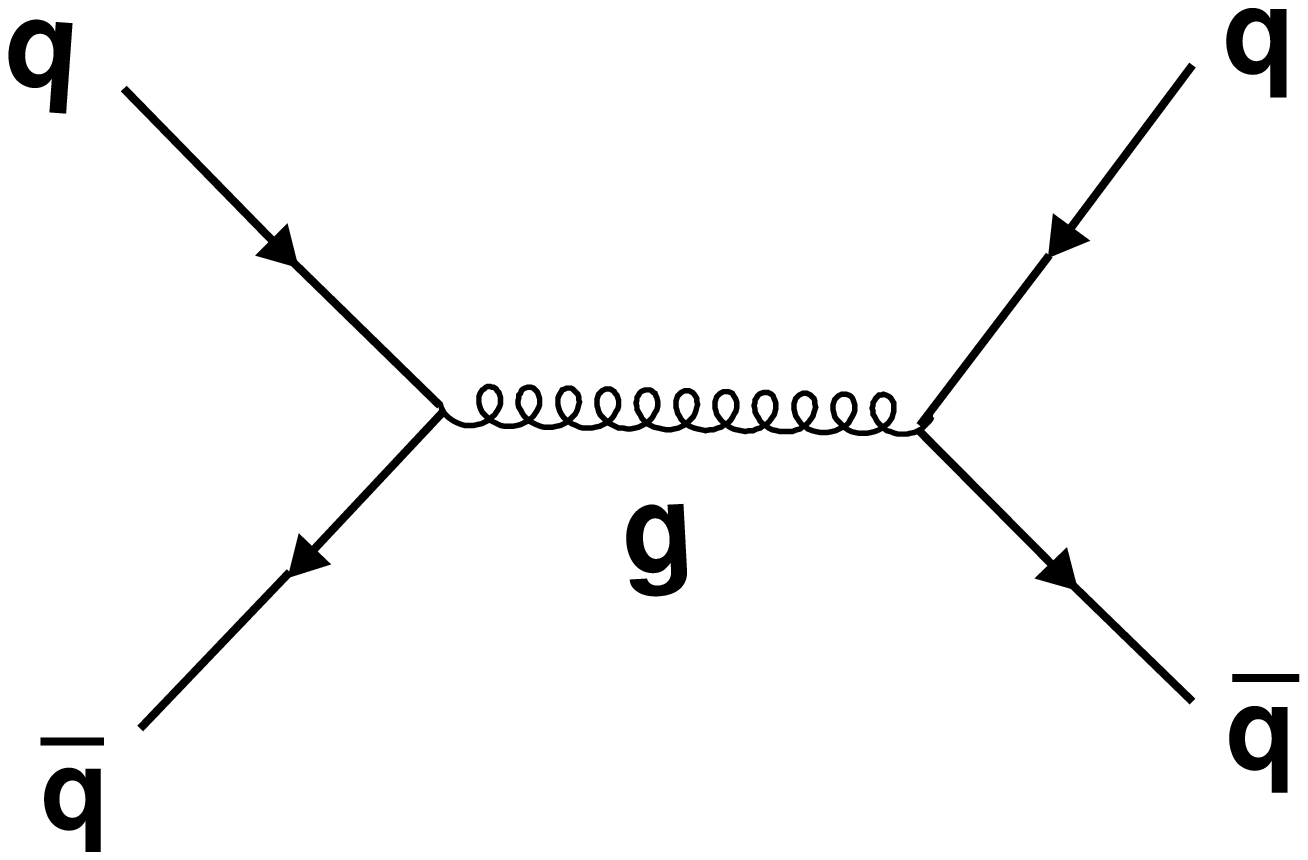}}
\vspace*{-6ex}
\scalebox{0.20}{\includegraphics{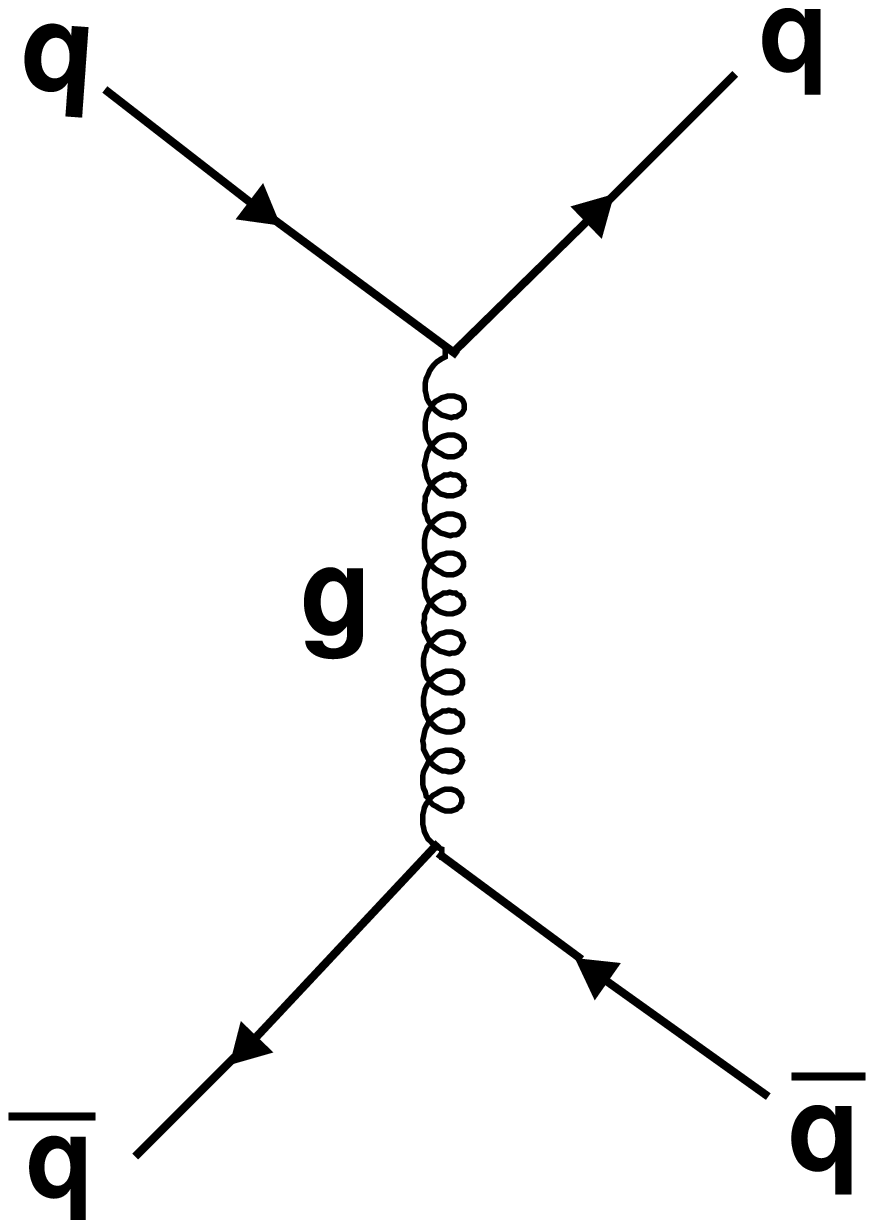}}
\scalebox{0.20}{\includegraphics{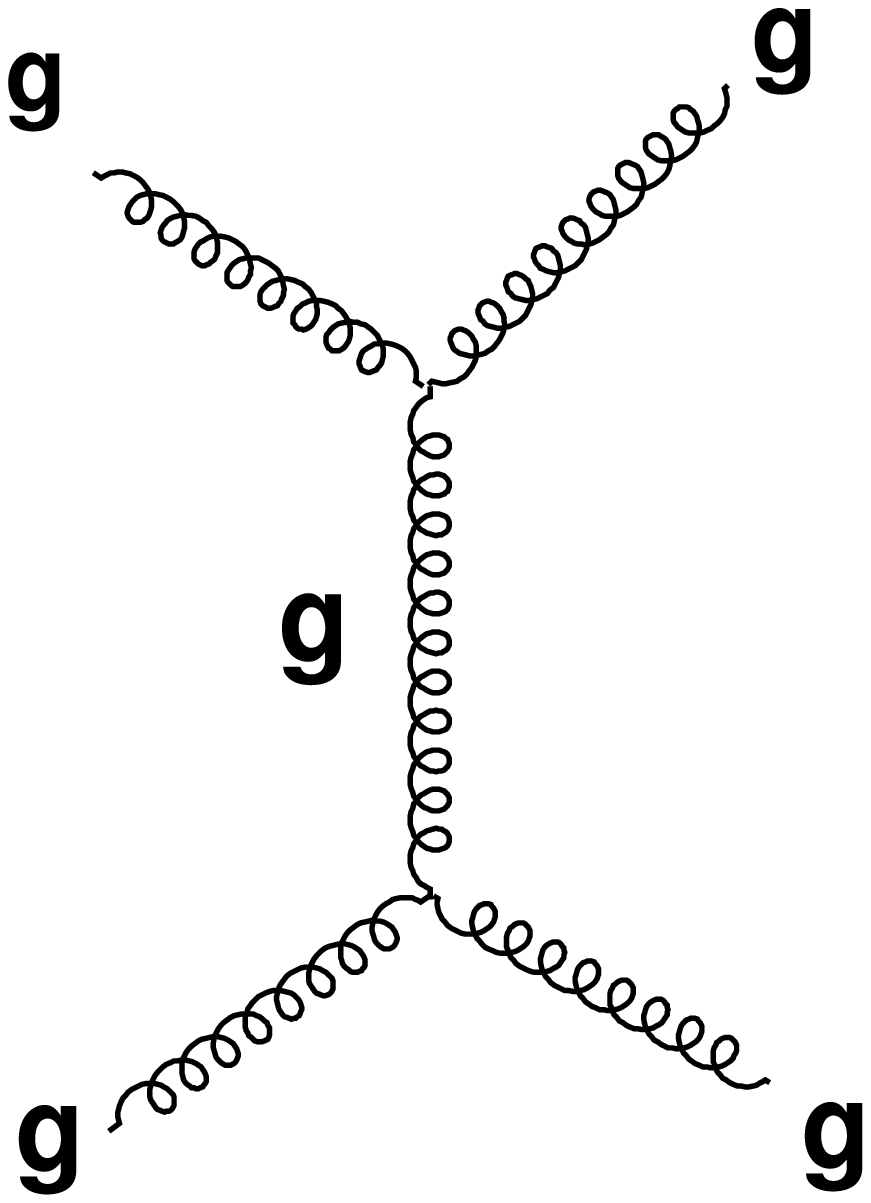}} 
\caption{\label{fig4} Type-II Background: QCD jet production where a jet fakes 
a photon giving a $\gamma + Jet$ final state.}
\label{fig:bkg_2}
\end{figure}

\begin{figure}[htbp]
\vspace*{-8ex}
\scalebox{0.20}{\includegraphics{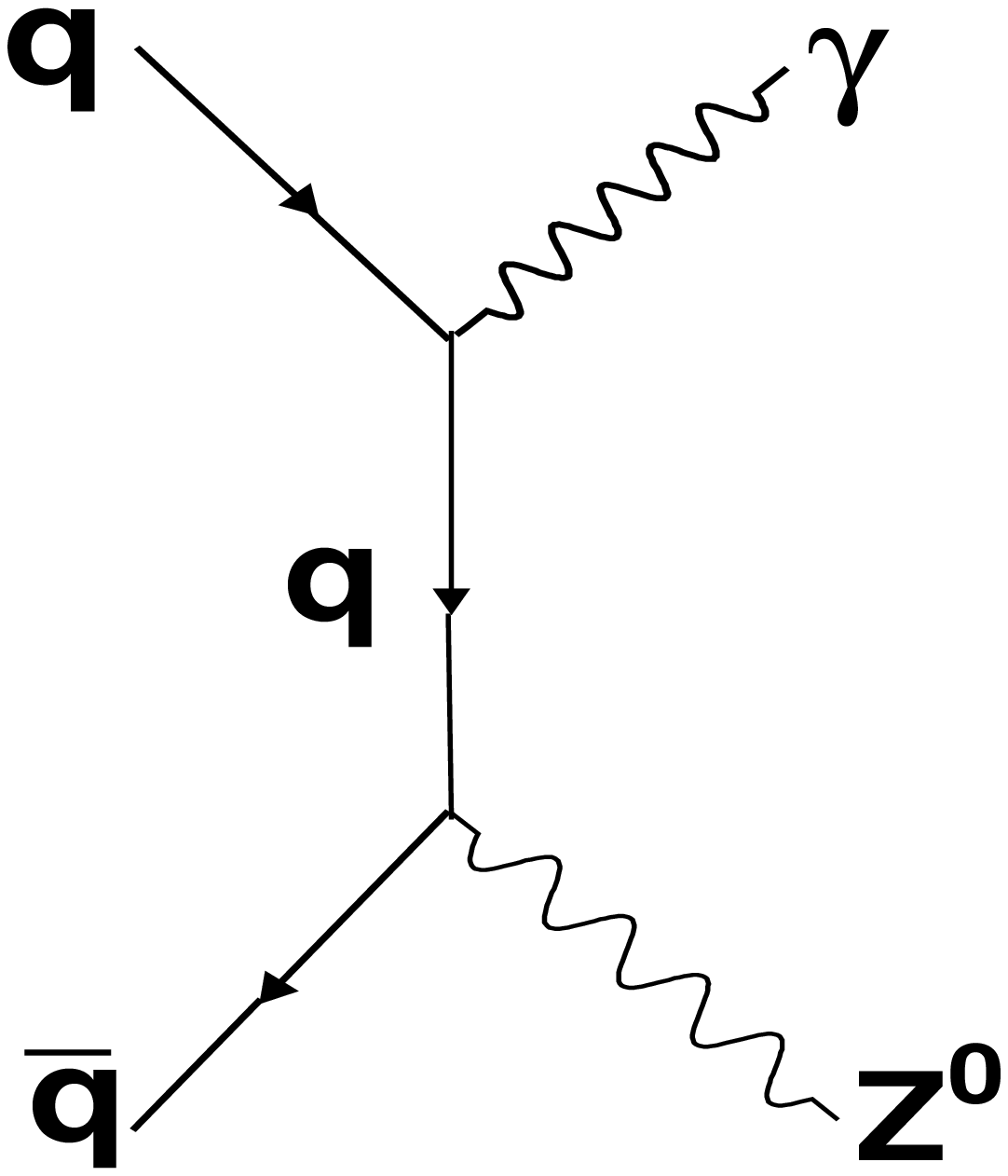}}
\scalebox{0.20}{\includegraphics{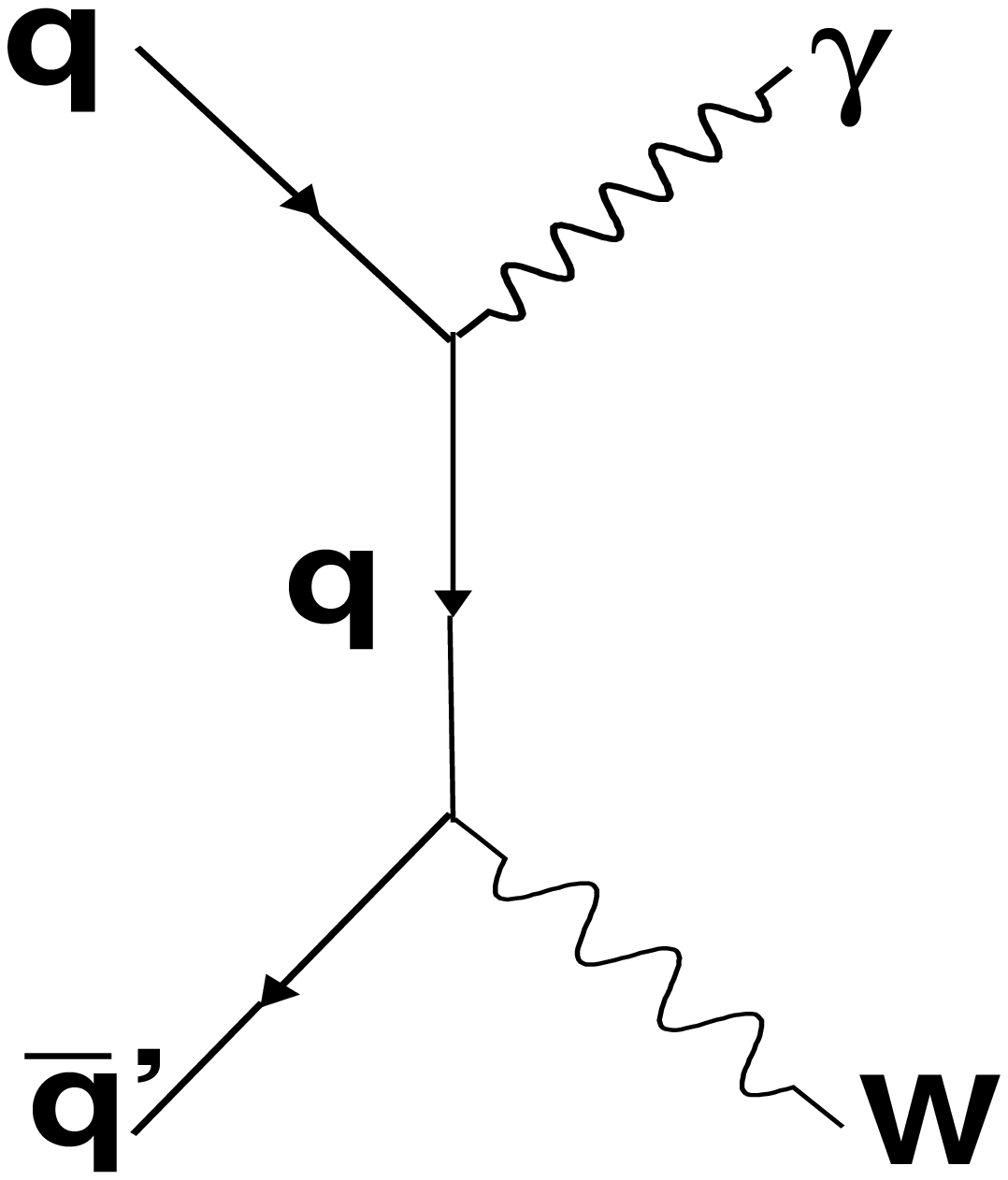}}
\vspace*{-3ex}
\caption{\label{fig5} Type-III Background: Here $W/Z^{0}$ decays to two jets 
and only one of the jets passes the trigger threshold.}
\label{fig:bkg_3}
\end{figure}

 The $\gamma+$jet final state can be mimicked by many known processes
of the SM. We consider all the leading contributions (both the physics
backgrounds as well as the detector ones) and broadly categorize these
into three classes viz., \begin{itemize}

\item Type-I: where a photon and a hard jet is produced in the hard
scattering.

\item Type-II: QCD dijet, where one of the jets fragments into a high
$E_{T}$ $\pi^{0}$ which then decays into a pair of overlapping photons
and, hence, is registered as a single photon. Moreover, in some cases
the electromagnetic fraction of a jet can mimic a photon in the detector.

\item Type-III: Photon + dijet production, where one of the jets is
either lost or mismeasured. This could proceed either from hard 
processes such as $q \bar q \to q \bar q \gamma, g g \gamma$ (with 
all possible interchanges of initial and final state partons) 
or result from $W/Z + \gamma$ production with the heavy bosons 
decaying into a pair of jets. 

\end{itemize}
While the leading contributions to the Type-I background emanates from the SM amplitudes 
of Fig. \ref{fig:qg_gamq}, a further contribution is displayed in 
Fig.\ref{fig:bkg_1}. 
In Figs.\ref{fig:bkg_2} \& \ref{fig:bkg_3},  
we show the major contributing Feynman diagrams
for the Types II \& III backgrounds respectively.


\begin{table*} 
\caption{ Production cross section  in different $\hat P_{T}$ bins for various Standard Model backgrounds with $\gamma + Jet$ final state.}  
\begin{ruledtabular} 
\begin{tabular}{|cccccccc|} 
Subprocess & 50-100 GeV & 100-200 GeV & 200-400 GeV 
     & 400-600 GeV & 600-1000 GeV &1000-1500  GeV& 1500 GeV and above\\
  & (pb)  & (pb) & (pb) & (pb) & (pb) & (pb) & (pb)\\                                            
\hline
& & & & & & & \\
$qg \rightarrow \gamma q$ & $7.22 \times 10^3$  & 569   & 36.3   & 1.53  & 
     $ 2.22  \times 10^{-1}$  & $ 1.19 \times 10^{-2}$  & 
     $ 7.6 \times 10^{-4}$\\ 

$q\bar{q} \rightarrow \gamma g$ &  652 & 65.3  & 5.56  & $ 3.18 \times 10^{-1}$  & $ 5.67 \times 10^{-2}$  & $ 3.76 \times 10^{-3}$  &  $ 2.8 \times 10 ^{-4}$
\\ 

$gg \rightarrow \gamma g$ & 1.79  & $ 8.6 \times 10^{-2}$ & 
   $ 3.1 \times 10^{-3}$ & $ 7.04 \times 10^{-5}$ & $ 6.32 \times 10^{-6}$ & 
   $ 1.75 \times 10^{-7}$ & $ 5.8 \times 10^{-9}$\\ 

\hline
& & & & & & & \\
QCD Jet   & $ 1.71 \times 10^{7}$  & $ 9.70  \times 10^{5}$  
          & $ 4.44  \times 10^{4}$ & $1.39 \times 10^3$  & 171 
          & 8.19  & 5.34 $ \times 10^{-1}$\\ 

\hline
& & & & & & & \\
\footnotemark[1]$Z(jj) + \gamma$  & 5.08   & $8.49  \times 10^{-1}$ & 
    $9.50 \times 10^{-2}$ & $6.23 \times 10^{-3}$ & $ 1.16 \times 10^{-3}$ 
    & $ 8.48 \times 10^{-5}$ &  $ 6.46 \times 10^{-6}$\\ 

\footnotemark[1]$W(jj) + \gamma$  & 4.80   & $ 6.93 \times 10^{-1}$ & 
   $ 6.19 \times 10^{-2}$ & $ 4.16 \times 10^{-3}$ & $ 7.39 \times 10^{-4}$ & 
   $ 4.67 \times 10^{-5}$ & $ 2.99 \times 10^{-6}$\\ 
\end{tabular}
\footnotetext[1]{Here the branching fraction is taken into account.}
\end{ruledtabular}
\end{table*}


At the LHC,  the Type-I background is dominated by 
the Compton process ($qg\rightarrow\gamma q$), while the other two
subprocesses, namely 
anihilation($q\bar{q}\rightarrow\gamma g$) and gluon fusion
($gg\rightarrow\gamma g$) contribute only a small fraction in the low
transverse momentum ($P_{T}$) ranges. For higher $P_T$, 
the annihilation subprocess can contribute upto about $\sim 23\%$ 
of the total Type-I background.
 
 The Type-II background accrues mainly from $qg\rightarrow qg$,
$q\bar{q}\rightarrow q\bar{q}$ and $gg\rightarrow gg$. For $\hat P_{T}
\geq $ 200 GeV (here $\hat{P_{T}}$ is the $P_{T}$ of the outgoing
partons in center of momentum frame in a $2 \rightarrow 2$ hard
scattering process), the total production cross section is $\sim
10^{4}$ times larger than the Type-I background. However, the fact of
the probability of a jet faking a photon in the detector being $\sim
10^{-3} -10^{-4}$ reduces the Type-II background to the same order as
the Type-I.  Moreover, for high transverse momenta the QCD dijet falls
very rapidly $(\sim P_{T}^{-4})$, thereby suggesting a simple
mechanism of suppressing this background

 Although the total Type-III background is very small compared to the others, 
for the $P_{T}$ range under consideration in this analysis, 
it turns out to be of the same order as $gg\rightarrow \gamma g$. And while 
nonresonant subprocesses (such as the ${\cal O}(\alpha_s^2 \alpha)$ 
contributions to $q \bar q \to q \bar q \gamma$ or $g g \to q \bar q \gamma$) 
can, in principle, be substantial, note that these, in some sense 
are related to the much larger Type-I and Type-II backgrounds. Consequently, 
the former are typically suppressed when 
appropriate phase space cuts  are imposed to reduce the latter.

 In Table-II we show the production cross section for all
backgrounds in different $\hat P_{T}$ ranges.

\section{Event Generation \& Cuts}
     \label{sec:monte} 

The event generation for signal and different background processes was
done with {\sc pythia-}v6.325~\cite{pythia}. For signal event
generation the matrix elements of Eq. (\ref{eq:matrix_1}) and
(\ref{eq:matrix_2}) were implemented inside the {\sc pythia}
framwork. We used the following common parameters and {\sc pythia}
switches:

\begin{itemize} 
\item 
 Parton Distribution Function(PDFs): CTEQ 5L~\cite{cteq_1};
\item 
 $Q^{2}=$ $\hat {s}$;
\item
MultiParton Interaction(MPI): ``ON";
\item
Initial State Radiation(ISR) and Final State Radiation (FSR): ``ON".

\end{itemize}

To get enough statistics for both the signal and the backgrounds, we
divided the whole analysis into three phase space regions determined 
by the value of the
$P_{T}$ of the final state $\gamma$ and the jet. For this purpose,
the following $\hat {P_{T}}$ ({\sc ckin}(3) parameter of {\sc pythia})
criteria were used for different mass points of signal:

\begin{itemize}
\item
 $\hat {P_{T}} \geq 180 \gev$: 1.0--3.0 TeV, 
\item
 $\hat {P_{T}} \geq 450 \gev$: 3.5--4.5 TeV, 
\item
 $\hat {P_{T}} \geq 950 \gev$: 5.0--6.0 TeV. 
\end{itemize}

A total of 16 mass points, 11 for coupling strength $f_1 = f_3 = 1.0$
(with a step size of 0.5 TeV) and 5 for $f_1 = f_3 = 0.5$ were
generated. The different backgrounds were also generated in various
$\hat {P_{T}}$ range. No pseudorapidity restriction was applied while
generating the events as the large $\hat P_{T}$ cut requirement
naturally restricts the events to well within $|\eta|< 5.0$. We must
also mention here that in the final selection of $\gamma$, we have
used the fiducial volume of the electromagnetic calorimeter of the
CMS detector i.e. $|\eta| \leq 2.5$ with 1.444 $\leq |\eta| \leq$ 1.566
excluded on account of the insensitive region between the barrel and
the endcaps\cite{BE_gap}. The jets were selected up to $|\eta| \leq 3.0$
only, because of poor resolution in the forward calorimeter.

\begin{figure}[!h]
\vspace*{-7ex}
\scalebox{0.41}{\includegraphics{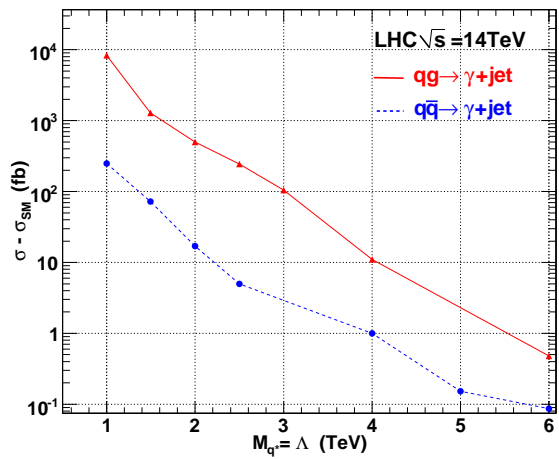}}
\caption{Deviation of cross section from SM with $M_{q^*} (=\Lambda)$ 
at $\sqrt{s}=14$ TeV.} 
\label{fig:cs}
\end{figure}

 Fig.~\ref{fig:cs} shows the deviation in the total cross-section of
$qg\rightarrow \gamma+jet$ as function of $M_{q^*} (=\Lambda)$.
Clearly, the variation is well-approximated by a $\Lambda^{-2}$
contribution superimposed upon a constant (the SM value). This is
reflective of the fact that, for large $\Lambda$, the new physics
contribution is dominated by the interference term in
Eqs.(\ref{eq:matrix_1},\ref{eq:matrix_2}) rather than the pure
$\Lambda^{-4}$ term . Only if we had imposed harder cuts on the
photons, would the latter term have dominated (albeit at the cost of
reducing event numbers and hence the sensitivity).

\section{Photon and Jet Candidates at the Generator Level}
    \label{sec:photon}

Although a mass peak in the signal will appear as an excess of events
over the continuum SM background, the significance for such an
observation will depend on the size of this continnum. Hence, to
enhance the signal peak, it is necceassary to reduce the background as
much as possible. For the signal under investigation, QCD dijet is the
largest background, as it mimics $\gamma+$jet final state when one of
the jet fakes a photon. To estimate this background reasonably at the
generator level, it is a must to have a proper understanding of the
reconstruction algorithm for a specific design of the detector rather
than limit ourselves to only partonic level photon and jets from the
final state in an event.

 Taking this into consideration, we have used a clustering algorithm
to account for fake photons arising from jets~\cite{cluster_algo}. The
electromganetic calorimeter (ECAL) of the CMS detector is made up of $
PbWO_{4}$ crystals and each crystal covers $0.0175 \times 0.0175$
(equivalently, $1^{\circ}$) in the $\Delta \eta -\Delta \phi$ space
($\phi$ being the azimuthal angle). For photon reconstruction, we have
used the ``hybrid'' algorithm\cite{ecal_1} where we consider only
those final state electromagnetic particles (i.e., $\gamma,e^{+}$ and
$e^{-}$) in the $\eta-\phi$ space such that neither of the distances
$\Delta\eta$ and $\Delta\phi$ from the seed object exceeds 0.09. This
extension is equivalent to a $10 \times 10$ crystal size in the CMS
detector. The seed for such a clustering must have a minimum $P_{T}$
of 1 GeV. A photon candidate is either a direct photon or other
electromagnetic obejcts such as $\pi^{0}\rightarrow \gamma\gamma,
\rho^{0}\rightarrow \gamma\gamma$ etc. The main contribution of fake
photons comes from $\pi^{0}\rightarrow \gamma\gamma (\sim 81 \%)$,
$\eta\rightarrow\gamma\gamma (\sim 12 \%)$ while other sources give
only a small contribution. A detailed discussion of this
reconstruction algorithm at the generator level can be found in a
previous work ~\cite{qstar_diphoton}.


 For jet reconstruction, various algorithms have been used by
different collider experiments. These include the
MidpointCone \cite{MidPoint_1}, IterativeCone
\cite{IterCone_1,IterCone_2}, and the $K_{t}$ algorithms
\cite{KtCone_1,KtCone_2,KtCone_3}. The $K_{t}$ and MidPoint algorithms
are used mostly for offline analysis. Since we have used the CMS setup
in our analysis, we use the IterativeCone algorithm to reconstruct
jets at the generator level. Being much faster, this is commonly used
for software based triggers.  While the first algorithms for the jets
at the hadron colliders started with simple cones in the $\Delta
\eta-\Delta \phi$ space ~\cite{cone_1}, clustering techniques have
greatly improved in sophistication over the last two decades
\cite{MidPoint_1,KtCone_1}.

  For a real detector, the first step in the reconstruction, before
invoking the jet algorithm, is to apply noise and pile-up suppression
with a set of cuts on $E_{T}$. To make ``perfect detector jets", we
used a seed $P_{T}$ cut on the $P_{T}$-ordered final state particles
and selected only those which had a transverse momentum above the
required minimum$^{\footnotemark[2]}$ of $P_{T seed} \geq $1.0 GeV. Once the
seed is selected, we search around for all the particles in a cone of
$\Delta R \leq 0.5$.  The objects inside the cone are used to
calculate a {\it proto-jet} direction and energy using the
E-scheme($\sum P_{i}$). The computed direction is then used to seed a
new proto-jet. The procedure is repeated until both the energy and the
direction of the putative jet is stable between iterations. We
quantify this by requiring that the energy should change by less than
$1\%$ and the direction by less than $\Delta R = 0.01$.  When a stable
proto-jet is found, all objects in the proto-jet are removed from the
list of input objects and the stable proto-jet is added to the list of
jets. The whole procedure is repeated until the list is bereft of 
objects with an $E_{T}$ above the seed threshold. The cone size and
the seed threshold are the parameters of the algorithm.
\footnotetext[2]{The seed threshold can vary from 0.5 to ~ 2.0 GeV
depending on the energy of reconstructed jet.}

\section{Smearing Effects}
    \label{sec:smearing}

While a detailed and full-scale detector simulation is beyond the
scope of this work, realistic detector effects can easily be
approximated. To this end, we smear the generator level information
with ECAL and HCAL resolutions of the CMS detector.

 For the ECAL resolution function, we use the form 
\begin{equation}
\nonumber
\frac{\delta E}{E} = \frac{a} {\sqrt{E}}  \oplus  \frac{a_{n}}{E} \oplus C  
  \label{eq:ecal_reso}
\end{equation}
where $a$ denotes the stochastic term, $a_{n}$ is the white noise term and 
$C$ is the constant term, with the three contributions being 
added in quadrature. For each of these terms, we 
use values identical to those for the electromagnetic calorimeter
of the CMS \cite{ecal_1}, namely
\[
\barr{lcl}
C & = & 0.55 \%  \\
a_{n} & = & \left\{ \begin{array}{lcl} 
                   2.1 \times 10^{-3} \gev & \quad \qquad & |\eta| < 1.5 \\[1ex]
                   2.45 \times 10^{-3} \gev & \quad \qquad & 1.5 \leq |\eta| \leq 2.5
                    \end{array}
                  \right.
\\[2.5ex]
  a & = & \left\{ \begin{array}{lcl} 
              2.7 \times 10^{-2} \gev^{1/2} & \qquad  & |\eta| < 1.5 \\[1ex]
              5.7 \times 10^{-2} \gev^{1/2} & \qquad & 1.5 \leq |\eta| \leq 2.5
 \ .
                    \end{array}
                  \right.
\earr
\]
The resolutions for $\Delta \eta$ and $\Delta\phi$ 
were taken to be 0.02 and 0.001 respectively
for both the barrel and endcap.

For the hadronic calorimeter, the resolutions were once again 
assumed to be the same as those for the CMS HCAL~\cite{hcal_1,AN_1}, namely,
\begin{itemize}
\item Barrel:
\[
\nonumber
\frac{\delta E}{E}= \frac{65 \%} {\sqrt{E/ \gev}} \oplus 5 \%, \Delta \eta = 0.04, \Delta \phi = 0.02
\]

\item Endcaps:
\[
\nonumber
\frac{\delta E}{E}= \frac{83 \%} {\sqrt{E/ \gev}} \oplus 5 \%, \Delta \eta =0.03, \Delta \phi = 0.02
\]

\item Forward region:
\[
\frac{\delta E}{E}= \frac{100 \%} {\sqrt{E / \gev}} \oplus 5 \%, \Delta \eta = 0.04, \Delta \phi =0.04 \ .
\]
\end{itemize}

\begin{figure}[htbp]
\vspace*{-2ex}
\scalebox{0.45}{\includegraphics{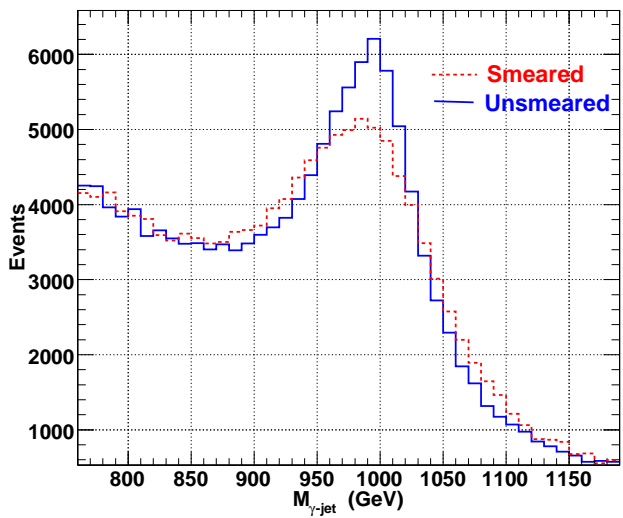}}
\caption{Effect of smearing on the mass peak for an excited quark of 1 TeV.}
\label{fig:smeared_unsmeared}
\end{figure}

The four momenta of the photon and jet were recalculated after
applying these resolution effects using an appropriate Gaussian smeared
function. In Fig. \ref{fig:smeared_unsmeared}, we show the effect of
resolution on the mass peak for a $M_{q^*}$ of 1 TeV. 


It should be noted that the  ATLAS detector at the LHC has a
better jet energy resolution with the constant term being $\sim$2
$\%$~\cite{atlas_tdr} compare to $\sim$5 $\%$ in the CMS detector. On
the other hand, the CMS ECAL has a better resolution 
than the ATLAS one owing to a smaller constant term.
However, with the resolving power being dominated by the jet energy 
resolution, ATLAS should do somewhat better. In other words, our results 
correspond to a {\em conservative} choice.

\section{Kinematical Variables}
    \label{sec:kinematics}

\begin{figure*} 
\vspace*{-15ex} 
\scalebox{0.43}{\includegraphics{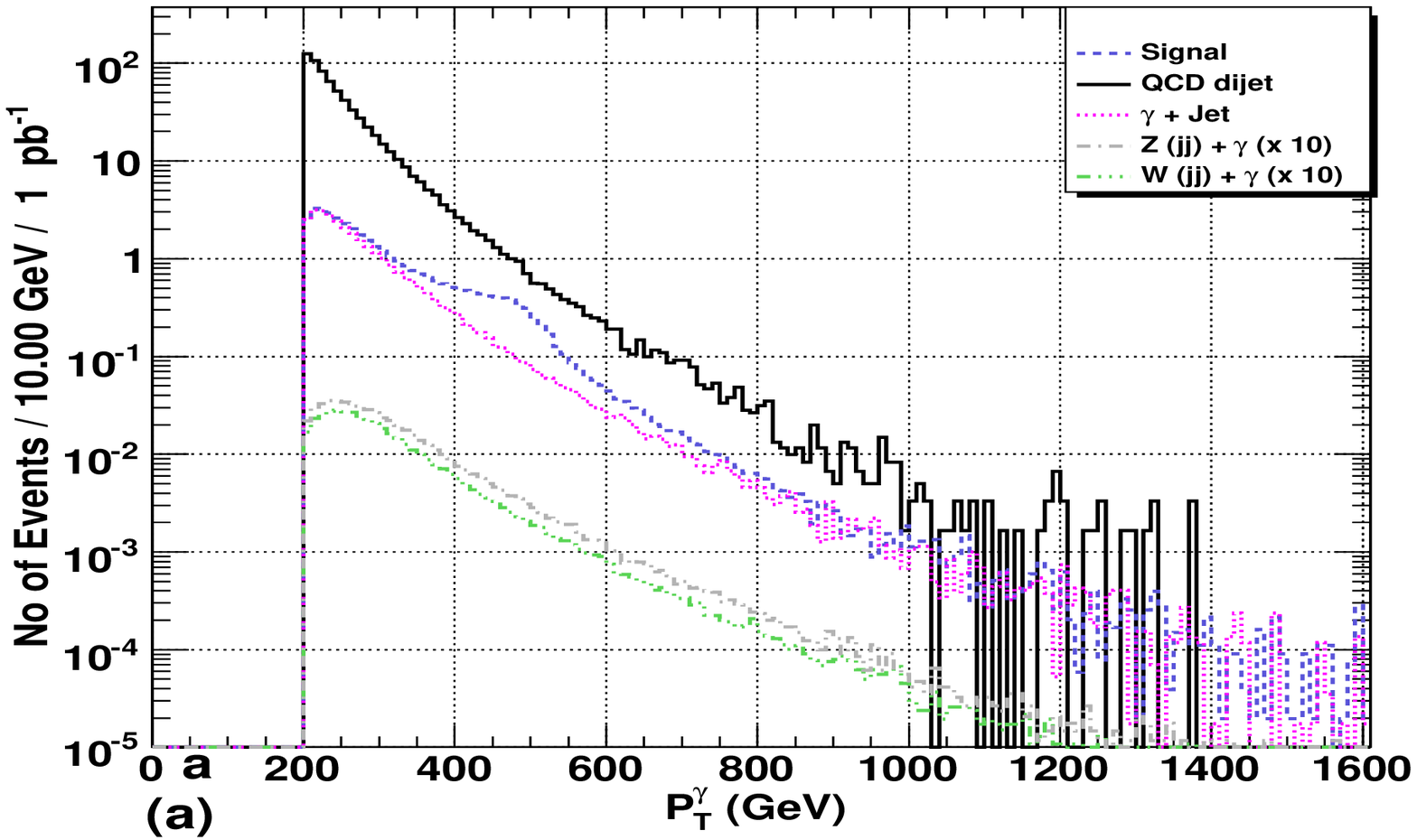}} 
\scalebox{0.43}{\includegraphics{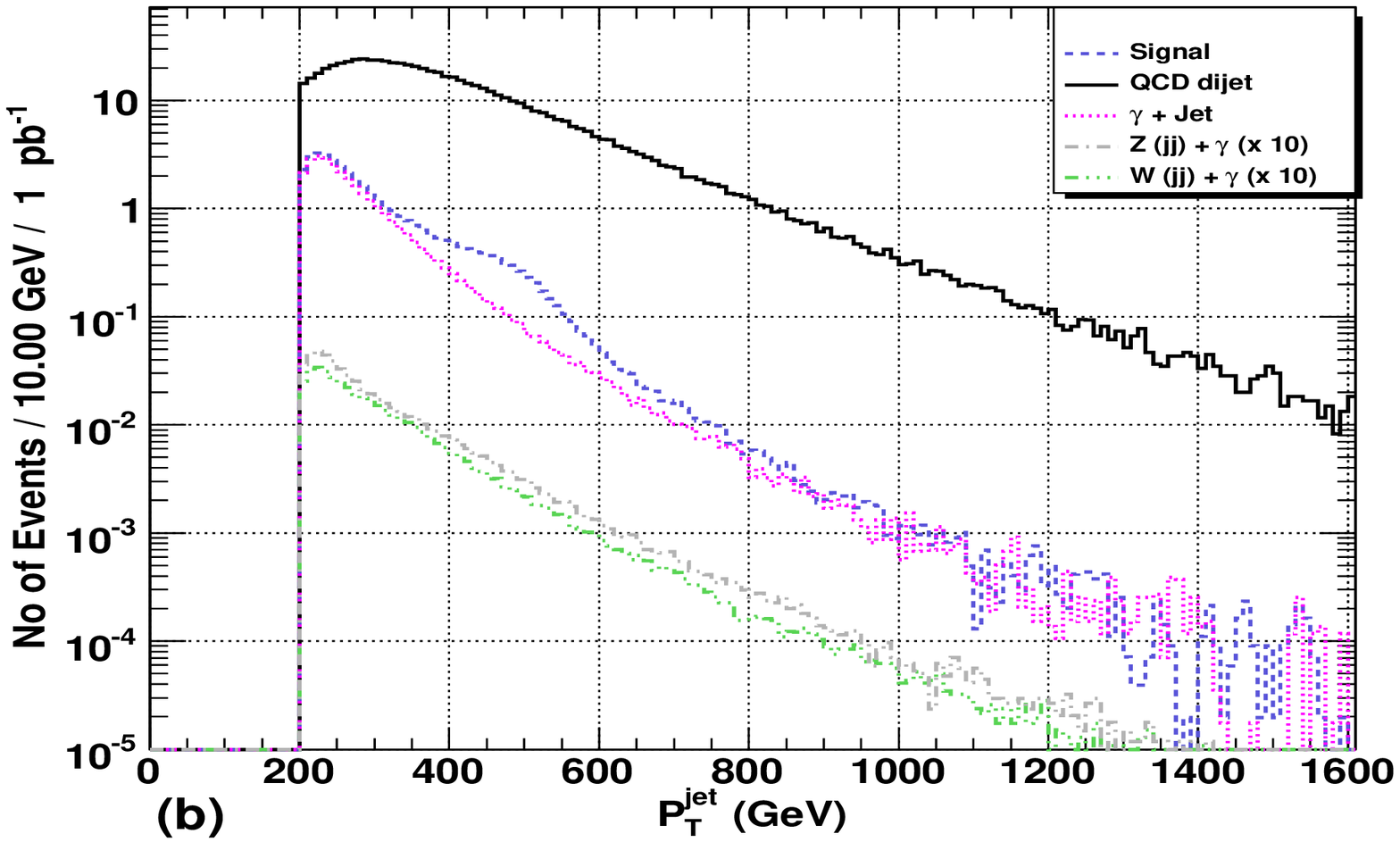}} 
\vspace*{-19ex} 
\scalebox{0.43}{\includegraphics{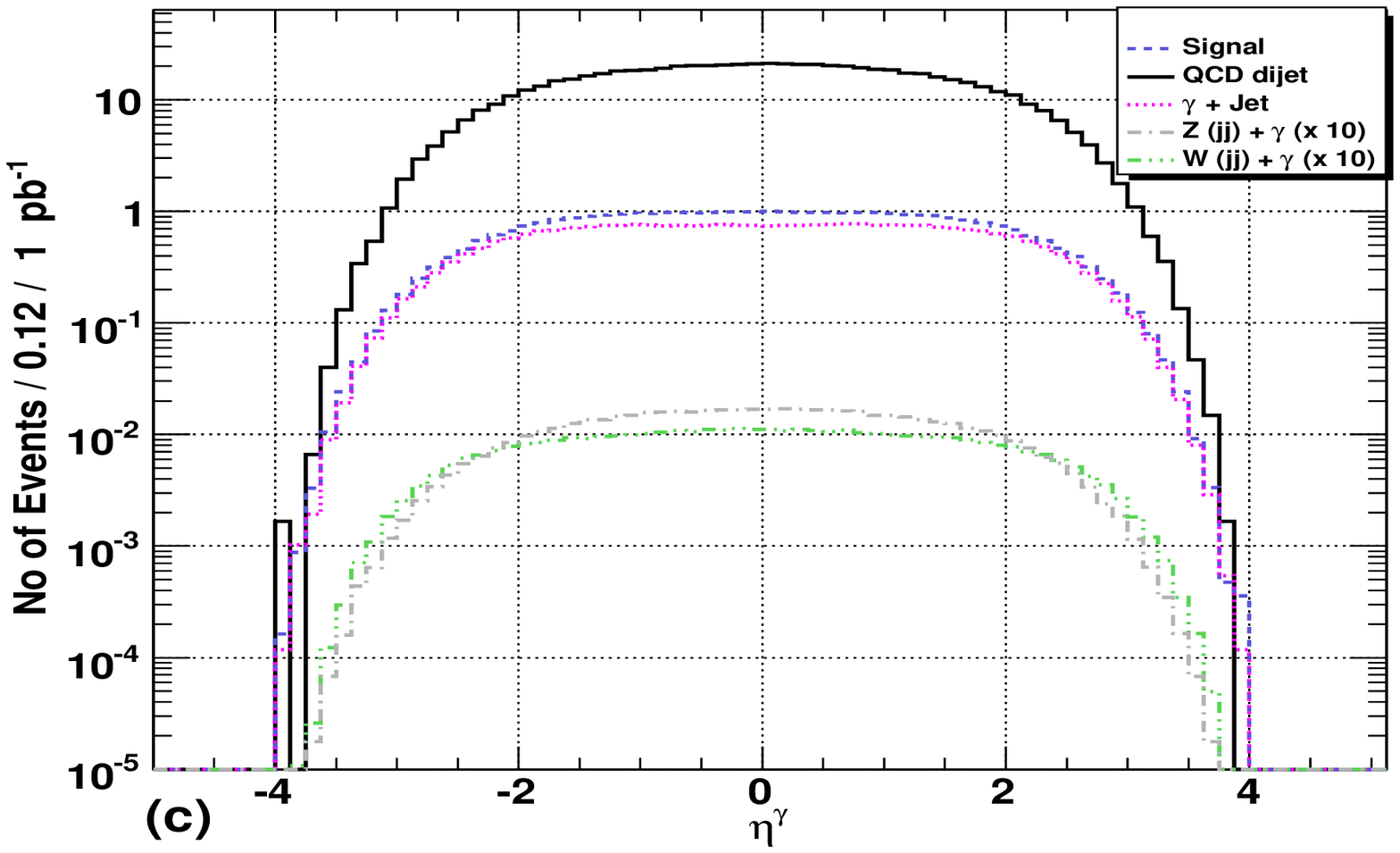}} 
\scalebox{0.43}{\includegraphics{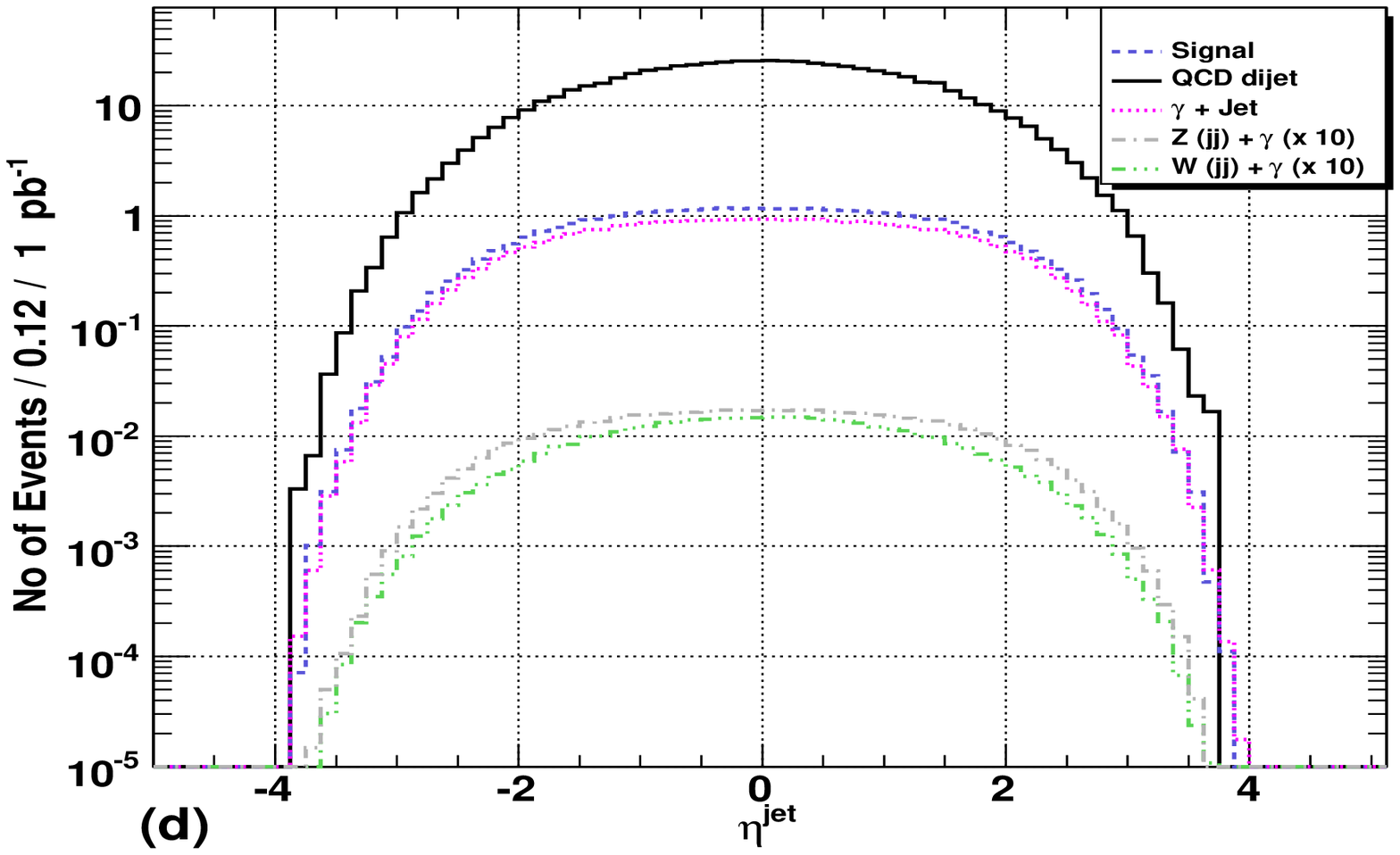}} 
\vspace*{-19ex} 
\scalebox{0.43}{\includegraphics{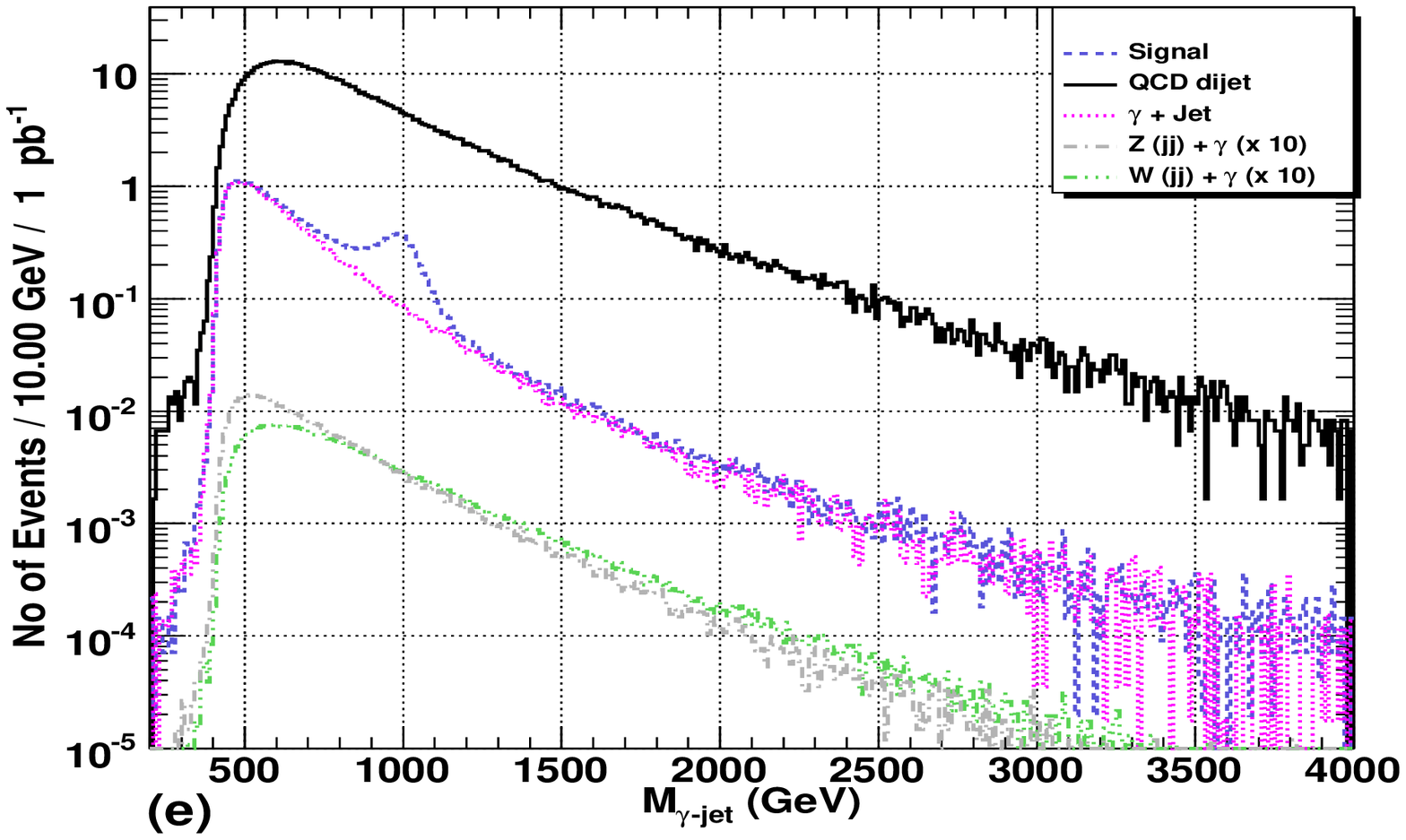}} 
\scalebox{0.43}{\includegraphics{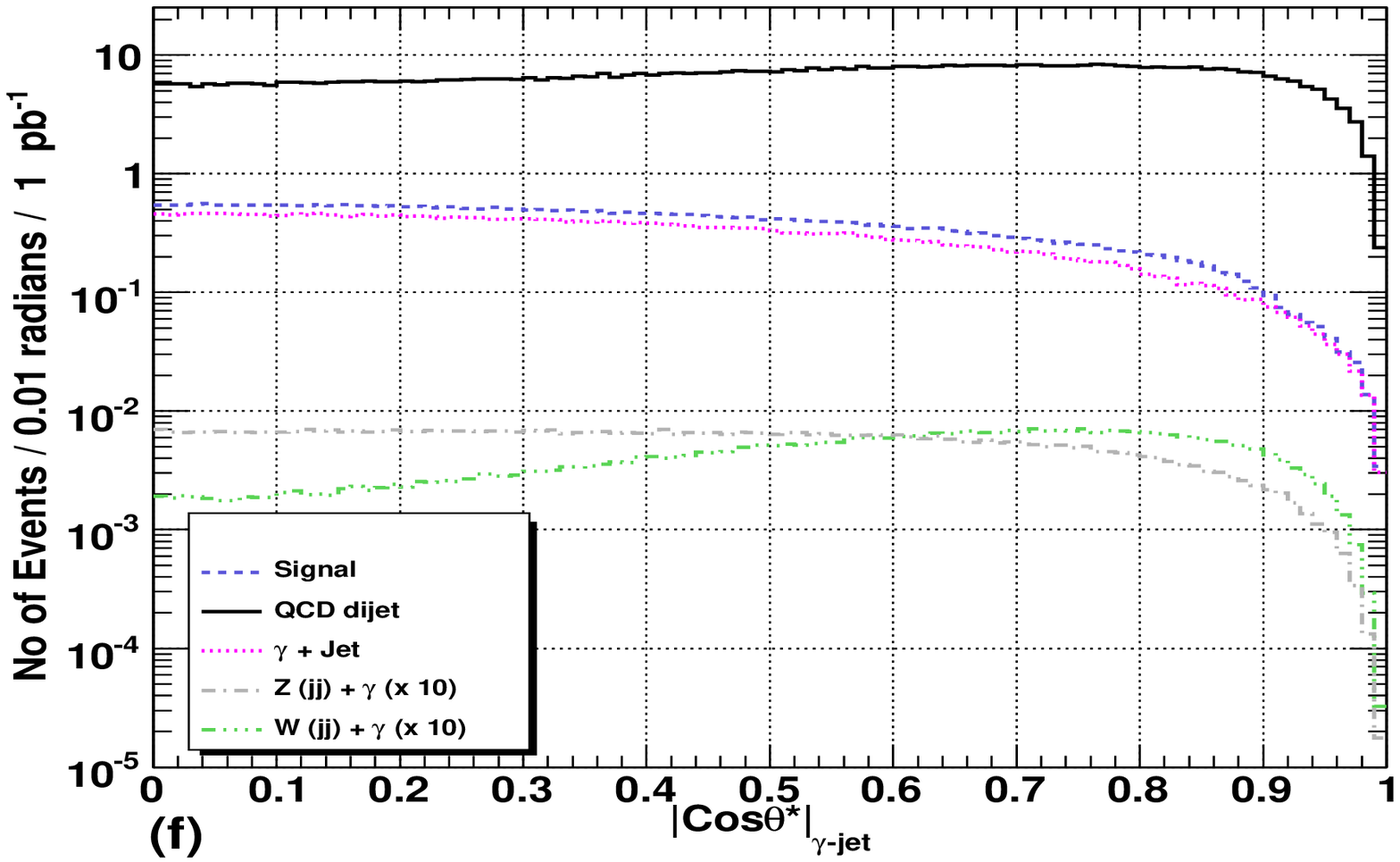}} 
\vspace*{18ex} 
\caption{Kinematic variable distributions after 
200 GeV pre-selection cut on $P_{T}$ (a) 
$P_{T}^{\gamma}$ distribution (b) $P_{T}^{jet}$ 
distribution (c) $\eta^{\gamma}$ distribution (d) 
$\eta^{jet}$ distribution (e) $M_{\gamma-jet}$ 
distribution and (f) 
$|\cos\theta^*|_{\gamma-jet}$. The signal corresponds to $M_{q^*} = 1 \tev$.} 
\label{fig:kine200} 
\end{figure*}

In Fig.~\ref{fig:kine200}, we show the kinematical 
distributions for the leading photon and the leading jet 
for $P_{T}^{\gamma,jet}\geq$ 200 GeV dataset for 
the background and the signal for $M_{q^*} = $1 TeV. 
For purpose of visual clarity, the distributions 
for $Z+\gamma$ and $W+\gamma$ backgrounds have been scaled up 
by a factor of 10. 
The bump in the 
transverse momentum distributions are primarily 
driven by the on-shell production of the $q^*$ and, therefore,
are centred slightly below $M_{q^*} / 2$. 
As is evident from Fig.~\ref{fig:kine200}(e), an excess in the 
invariant mass spectrum would be 
quite prominent for even $\int$\textit{$ 
{L}.dt =$}1$pb^{-1}$. 
The $t$-channel contribution has been included 
and manifests itself in the elongation of the side bands. 
It may be noted that the QCD dijet background is 
more than 10 times as large as the signal, but falls very rapidly 
with $P_{T}^{\gamma/{\rm jet}}$ (the mistagging probability has already 
been included). 
Fig.~\ref{fig:kine200}(f) shows  the  distribution in the subprocess 
center of mass scattering angle, with 
$\cos\theta^{*} = 
\tanh[(\eta^{\gamma}-\eta^{jet})/2.0]$. 
Note that major differences between the 
signal and background profile occur only for $p_T$ and 
invariant mass distributions, whereas the other 
phase space variables are not very sensitive discriminants.

\begin{table} \caption{Preselection-efficiency and geometrical acceptence for various SM backgrounds and few signal points. }
\begin{ruledtabular} \begin{tabular}{|c|ccccc|} Selection Cut & Signal & $\gamma+Jet$ &QCD & $Z+\gamma$ & $W+\gamma$\\
                             &  $\%$ & $\%$ & $\%$ & $\%$ & $\%$\\
\hline

 $P_{T}^{\gamma, jet}\geq 200 GeV$ & 48.7    & 44.2 & 0.90  & 38.4 &  37.1\\
                                  & [1 TeV]  &       &        &       &       \\
                                  &          &       &        &       &       \\

 $P_{T}^{\gamma, jet}\geq 500 GeV$ & 40.2 &39.8 & 0.42 & 50.4 & 50.6\\
                                  & [4 TeV] &        &         &      &           \\
                                  &         &        &         &      &          \\

 $P_{T}^{\gamma, jet}\geq 1 TeV$   & 47.4  & 46.0  & 0.51 & 58.8  & 59.9\\
                                   & [5 TeV] &       &         &      &         \\
                                   &         &       &         &       &       \\

\hline

$|\eta^{\gamma}| \leq 2.5$, $|\eta^{jet}| \leq 3.0,$ & 42.4  & 38.2  & 0.81  & 32.8 & 33.2 \\
$|\eta^{\gamma}| \not \in [1.4442,1.5666]$&[1 TeV] &        &        &       &       \\
                                          &        &            &          &       &\\
                                           & 38.2   & 37.8    &  0.40   & 47.4   & 48.4 \\
                       & [4 TeV]  &            &            &           &    \\
                       &          &            &            &           &         \\

               & 46.4  & 45.0    &  0.50   & 56.3    & 58.7 \\
               & [5 TeV]  &            &            &           &         \\
               &          &            &            &           &         \\
\end{tabular}
\end{ruledtabular}
\end{table}

\begin{figure*} 
\vspace*{-15ex} 
\scalebox{0.43}{\includegraphics{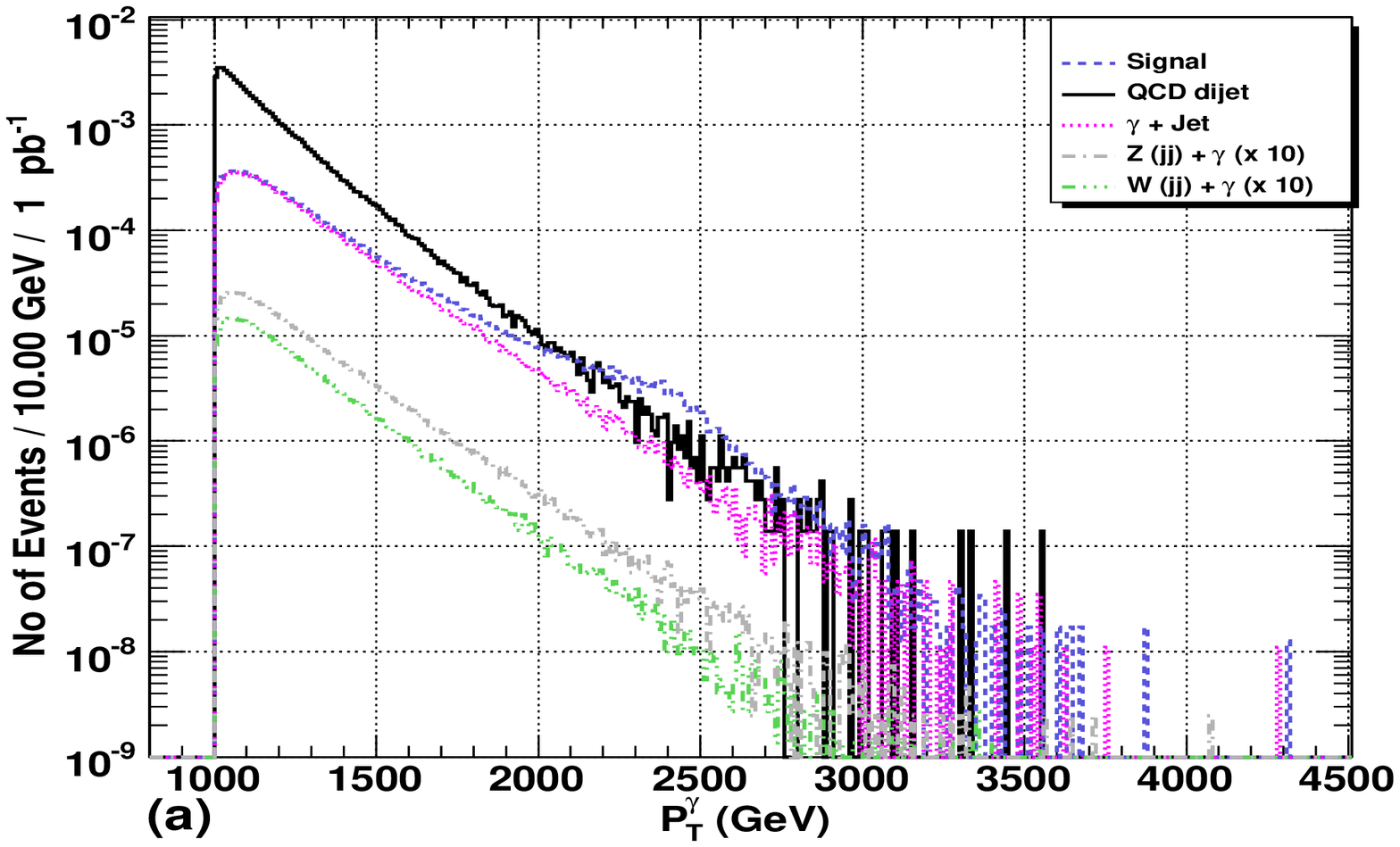}} 
\scalebox{0.43}{\includegraphics{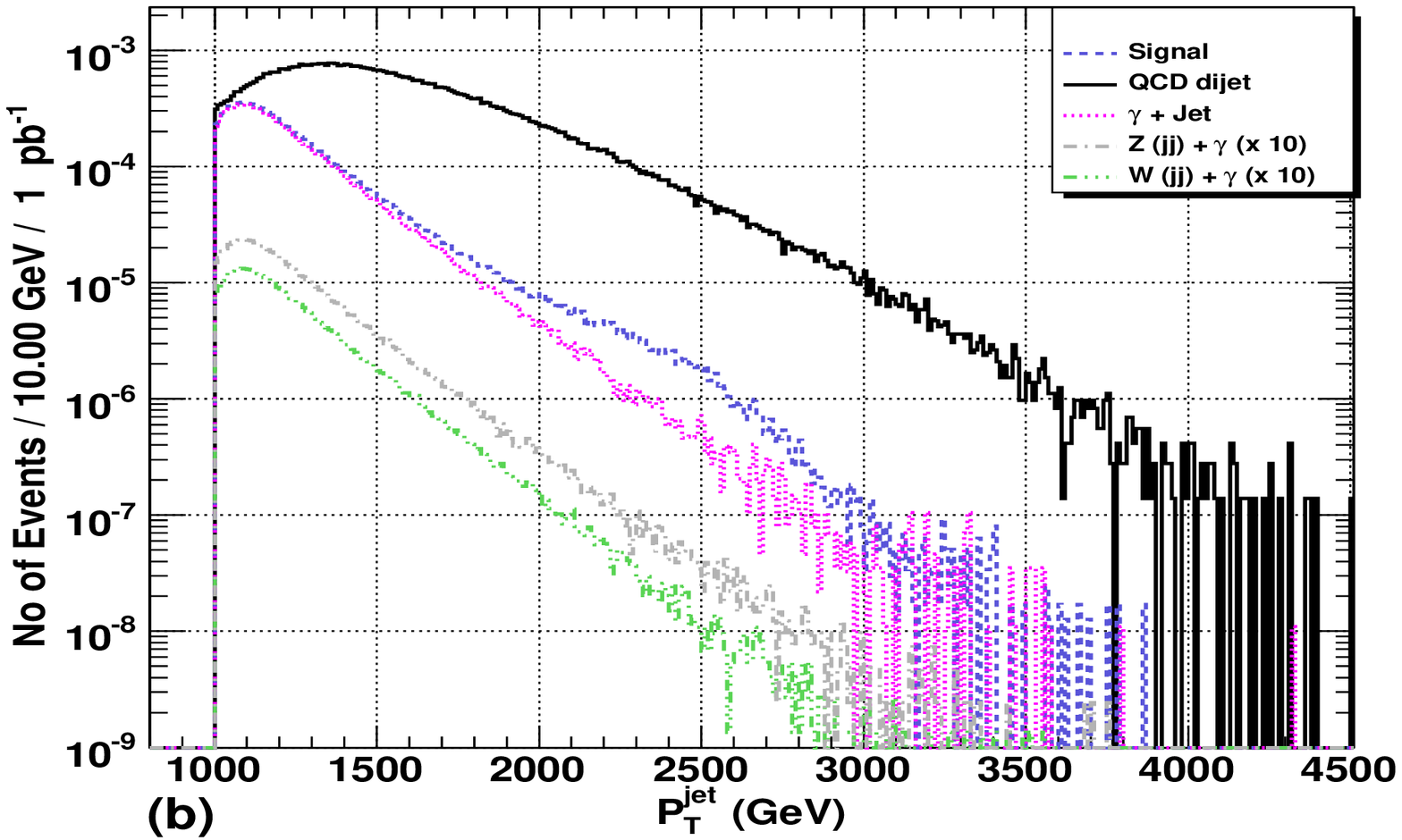}} 
\vspace*{-19ex} 
\scalebox{0.43}{\includegraphics{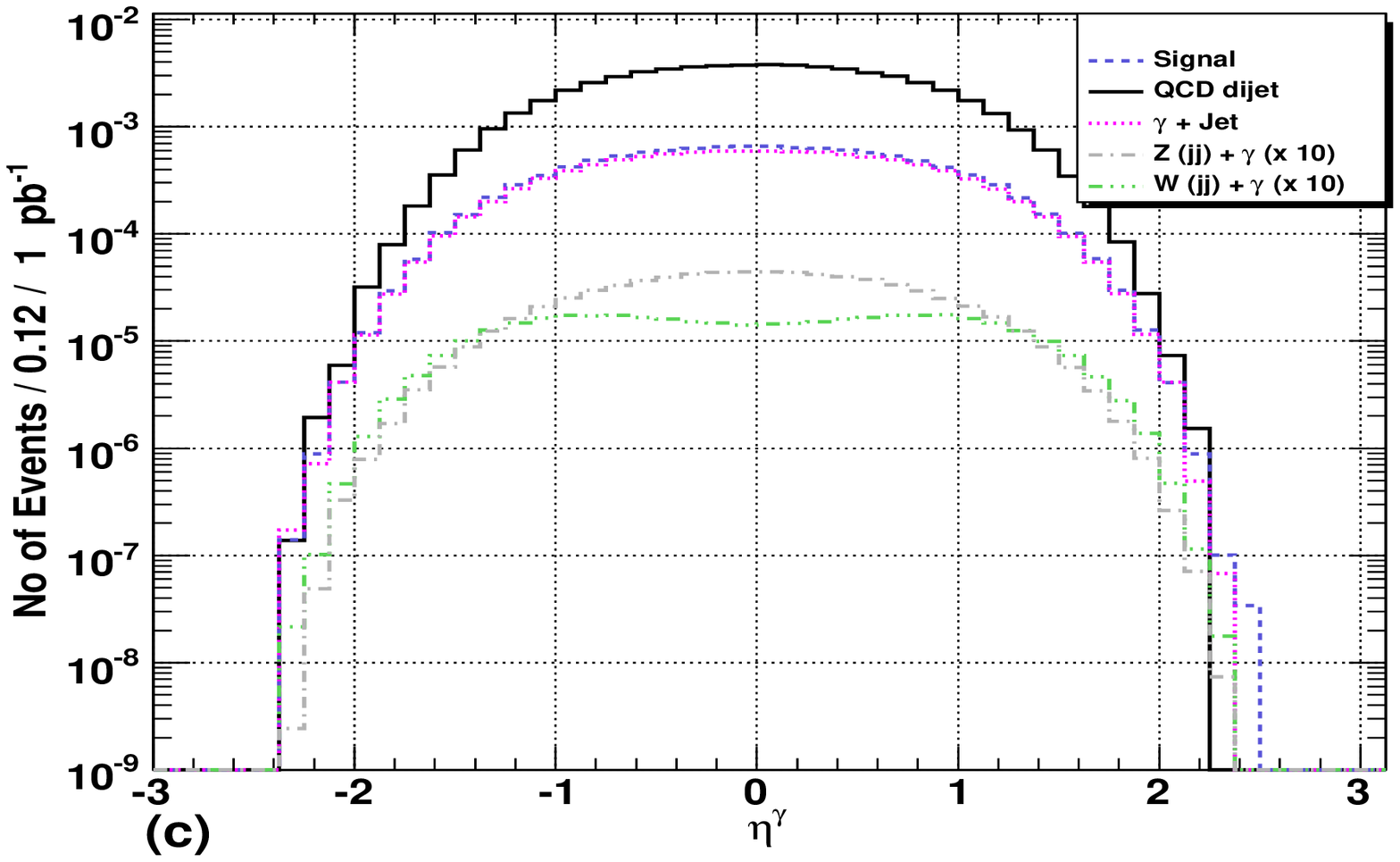}} 
\scalebox{0.43}{\includegraphics{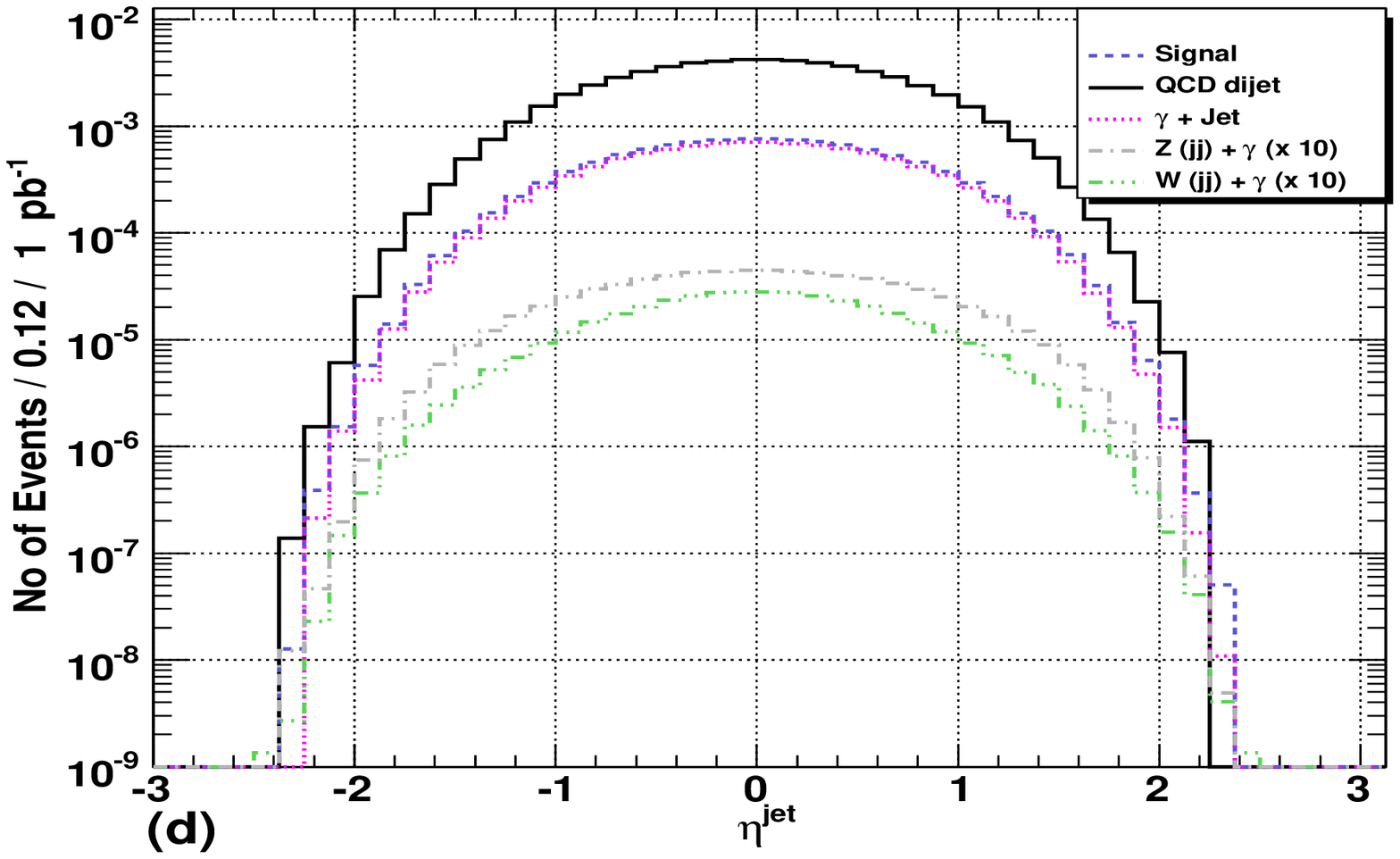}} 
\vspace*{-19ex} 
\scalebox{0.43}{\includegraphics{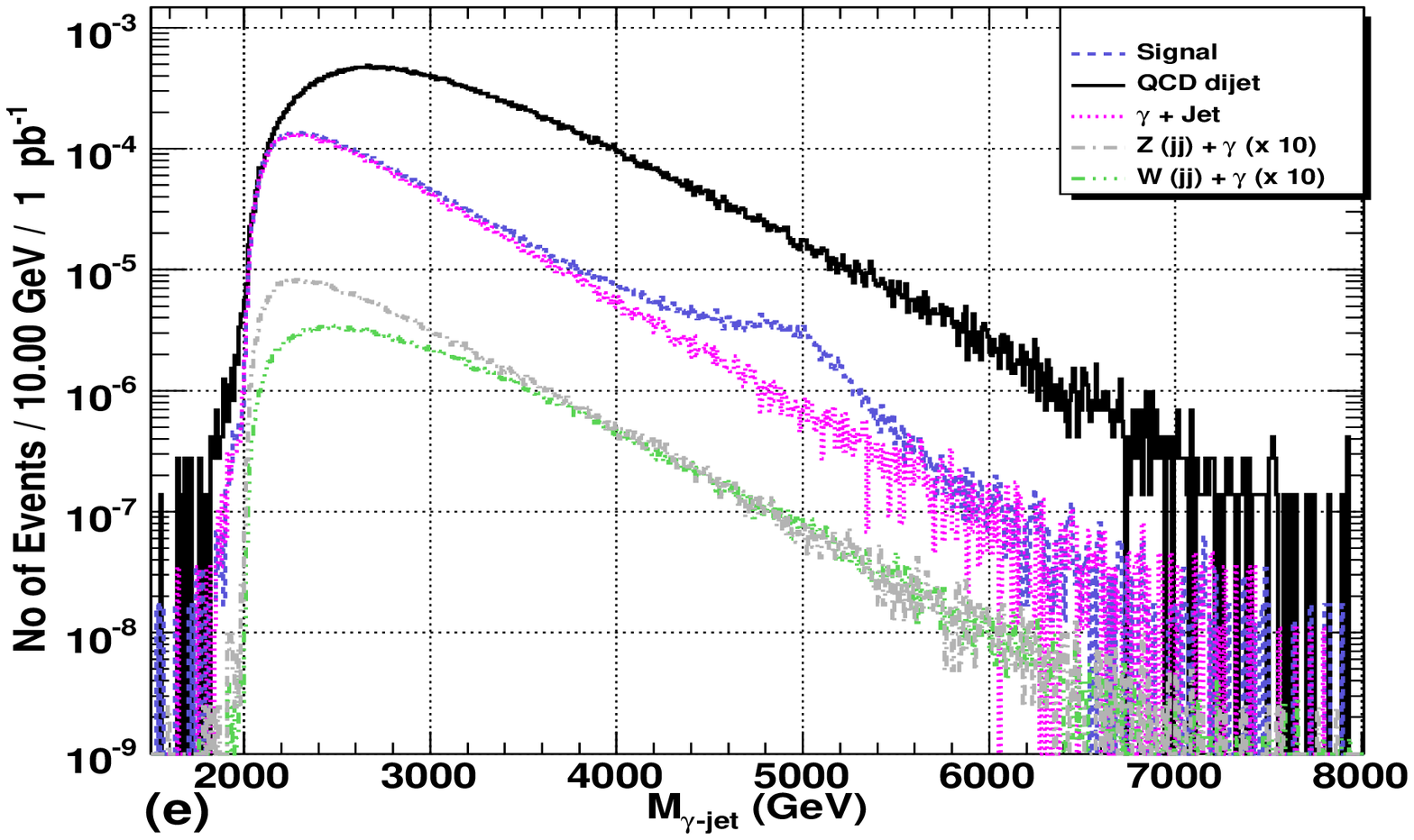}} 
\scalebox{0.43}{\includegraphics{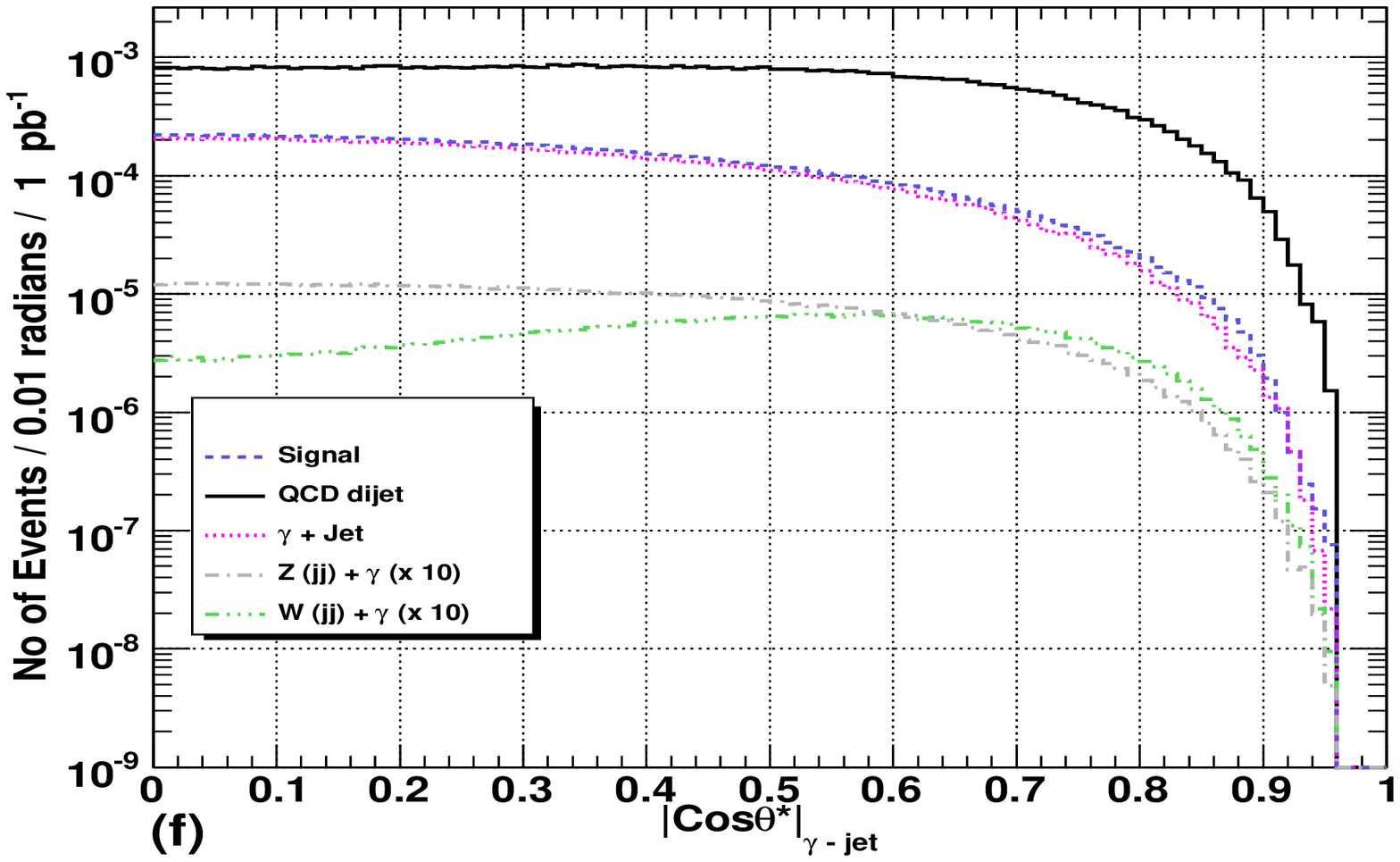}} 
\vspace*{18ex} 
\caption{As in Fig.\ref{fig:kine200} but with a 
1 TeV pre-selection cut on $P_{T}$ and a signal corresponding
to $M_{q^*} = 5 \tev$ instead.} 
\label{fig:kine1000} 
\end{figure*}

Fig.~\ref{fig:kine1000} shows similar distributions as in
Fig.~\ref{fig:kine200} but for the $M_{q^*} =$ 5 TeV point
instead. For these distributions we have used a $P^{\gamma,jet}_{T}$
cut of 1 TeV at the pre-selection level. With the $P_{T}$ spectrum for
the photon from QCD background falling very rapidly, the signal
dominates over the background above 2 TeV even without isolation
cuts. As for the corresponding invariant mass ($M_{\gamma-jet}$)
distribution---see Fig.~\ref{fig:kine1000}(e)--- a combination of the
large natural width and smearing effects results in a broad bump
rather than a sharp one.  Once again, the other distributions do not
discriminate between the signal and background in any forceful manner.
While the the slight dip in the central $\eta^{\gamma}$ region for the
$W+\gamma$ process might seem intriguing (especially in the absence of
any such dip in the $Z+\gamma$ distribution), it is but a
straightforward reflection of the well-known Radiation-Amplitude-Zero
(RAZ) present in the former~\cite{Mikaelian:1979nr, Wg_dip}.  That
the RAZ in the angular distributuion is apparent only for the high
$P_{T}^{\gamma,jet}$ cutoff case can be understood by realising that
the rapidity of the photon as measured in the laboratory can be
related to the rapidity (scattering angle) in the partonic subprocess
center of mass frame through
\[
\eta(\gamma)=\frac{1}{2} ln 
\left(\frac{x_{1}}{x_{2}}\right)+\eta^{*}(\gamma) 
\]
where $x_{i}$ are the momentum fractions of the incoming partons. For
small $\hat{s}$ values (hence lower {\sc ckin}(3) cuts) the parton
densities are maximized when the (anti-)quark acquire small(large)
momentum fractions respectively.  This leads to a considerably large
contribution to $\eta^\gamma$ from the boost, thereby smearing the
original double peaked $\eta^{\gamma}$ distribution into centrally
peaked one.  On the contrary, for typically high $\hat{s}$ ({\sc
ckin}(3)$ \geq$ 1 TeV) values the $x_{i}$ tend to be not too different
thereby reducing the smearing on this account.

In Table III we show the pre-selection efficiencies and geometrical
acceptences for the CMS detector for various backgrounds and signal of
$M_{q^*}=$1, 4 and 5 TeV against the total generated events.

\section{Isolation Variables}
   \label{sec:isolation}

In a detector, a photon candidate is reconstructed by summing the
electromagnetic energy deposition in ECAL towers in a limited region
 of space, with the sum being required to be above a certain $E_{T}$
threshold. For the sake of simplicity, this limited region can be
visualized as a cone in $\Delta \eta-\Delta\phi$ space given by
$\Delta R\equiv \sqrt{\Delta\phi^{2} +\Delta\eta^{2}}$, and containing
most of the energy of the electromegnetic object.

A jet fragmenting into neutral and charge hadrons with a $\pi^{0}\rightarrow \gamma\gamma$ (with
overlapping photons) carrying maximum momentum can also lead to fake single-photon candidates. To
remove such events, photons are required not to have associated charged
tracks within a cone of size $R_{\rm iso}$. This is implemented by
requiring that the scalar/vector sum of energy/transverse momentum within
$R_{\rm iso}$ should be below a certain threshold. For example, the D$\O$
and the CDF experiments demand that the $E_{T}$ due to charged tracks
within a cone of $\Delta R= 0.4$ around the photon should be less than
a certain value. In this anlaysis, we closely follow the CMS detector
simulation studies ~\cite{pg} and consider the following isolation
variables:

\begin{itemize}
\item
the number of tracks ($N_{trk}$) above a certain threshold 
inside a cone around the photon candidate.
\item
the scalar sum of transverse energy ($E_{TSUM}$) inside a cone
around the photon. Although, in a full detector simulation the $E_{TSUM}$ is measured separately for ECAL and HCAL, but working at the
generator level, we combine them into a single variable with all kinds
of electromagnetic and hadronic objects around the photon being taken
into account.  
\end{itemize}



\subsection{Track Isolation}
\begin{figure}[!h]
\vspace*{-2ex}
\scalebox{0.40}{\includegraphics{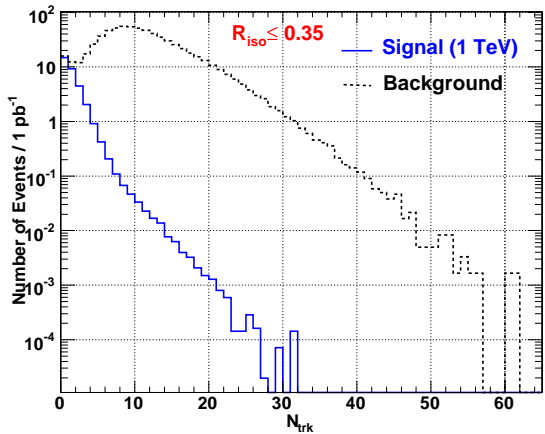}}                             
\caption{Number of tracks($N_{trk}$) for the signal ($M_{q^*}=$1 TeV) and the background events around the photon.}
\label{fig:Trk_N_Lead}
\end{figure}

 For the purpose of track isolation, only `stable' charged particles
e.g. $\pi^{\pm}$, $K^{\pm}$, $e^{\pm}$ and $P^{\pm}$ were
considered. The other particles were found to have only a negligible
contribution. Indeed, $\pi^{\pm}$ alone contribute more than 80$\%$ of
the charged tracks. Fig ~\ref{fig:Trk_N_Lead} shows the distribution
of number of charged tracks ($N_{trk}$) around the leading photon
within a cone of size $\Delta R\leq 0.35$ for a $M_{q^*}=$ 1 TeV
signal as well as for the total background. Since the leading photon
is the {\it true} photon for signal events, most of them are
associated with zero tracks ($N_{trk}=$0) and the distribution falls
off very rapidly for larger $N_{trk}$ values. For background events
though, the distribution peaks at $N_{trk} \sim$7-8 and then falls
slowly. The small rise at $N_{trk}=0$ is due to the fact that
$\gamma+{\it jet}$(SM) and $W/Z+\gamma$ backgrounds have true photons
as the leading photon in the event and have no tracks around them,
while the rising part along with the extended tail is mainly
contributed by the QCD dijet events where the fake photon typically
has a large number of tracks around itself. In this study, we accept a
photon to be an isolated one if there is no track with minimum
transverse momentum ($P_{Tmin}^{trk}$) within a given cone around
it. It should be noted that comparative distributions of signal and
total background, as shown in Fig~\ref{fig:Trk_N_Lead}, is not overly
sensitive to moderate changes in the $P_{Tmin}^{trk}$ value (the exact
values are discussed at a later stage).

 In $pp$ collisions at the LHC, a large number of soft tracks (in the
range of a few MeVs to a few GeVs) will be produced in each event. The
main sources of such soft tracks are ISR, FSR, minimum bias and
underlying events. For a direct photon emerging from the hard
interaction, such soft tracks could actually be in the near vicinity
of the photon. Labelling such photons as non-isolated ones could
potentially reduce the signal efficiency, and many interesting events,
such as those in this study, could be lost. To prevent such loss,
tracks are usually required to pass certain minimum selection
criteria, with a minumum threshold on the transverse momentum being a
common requirement~\cite{cluster_algo,pg,trkpt2}. Adopting this strategy, we
investigate the dependence of the signal efficiency and the
signal/background (S/B) ratio on the chosen $P_{T}$ threshold ($P_{T
min}^{trk}$), varying the latter between 1-3 GeV.

\begin{figure}[!t]
\vspace*{-14ex}
\scalebox{0.45}{\includegraphics{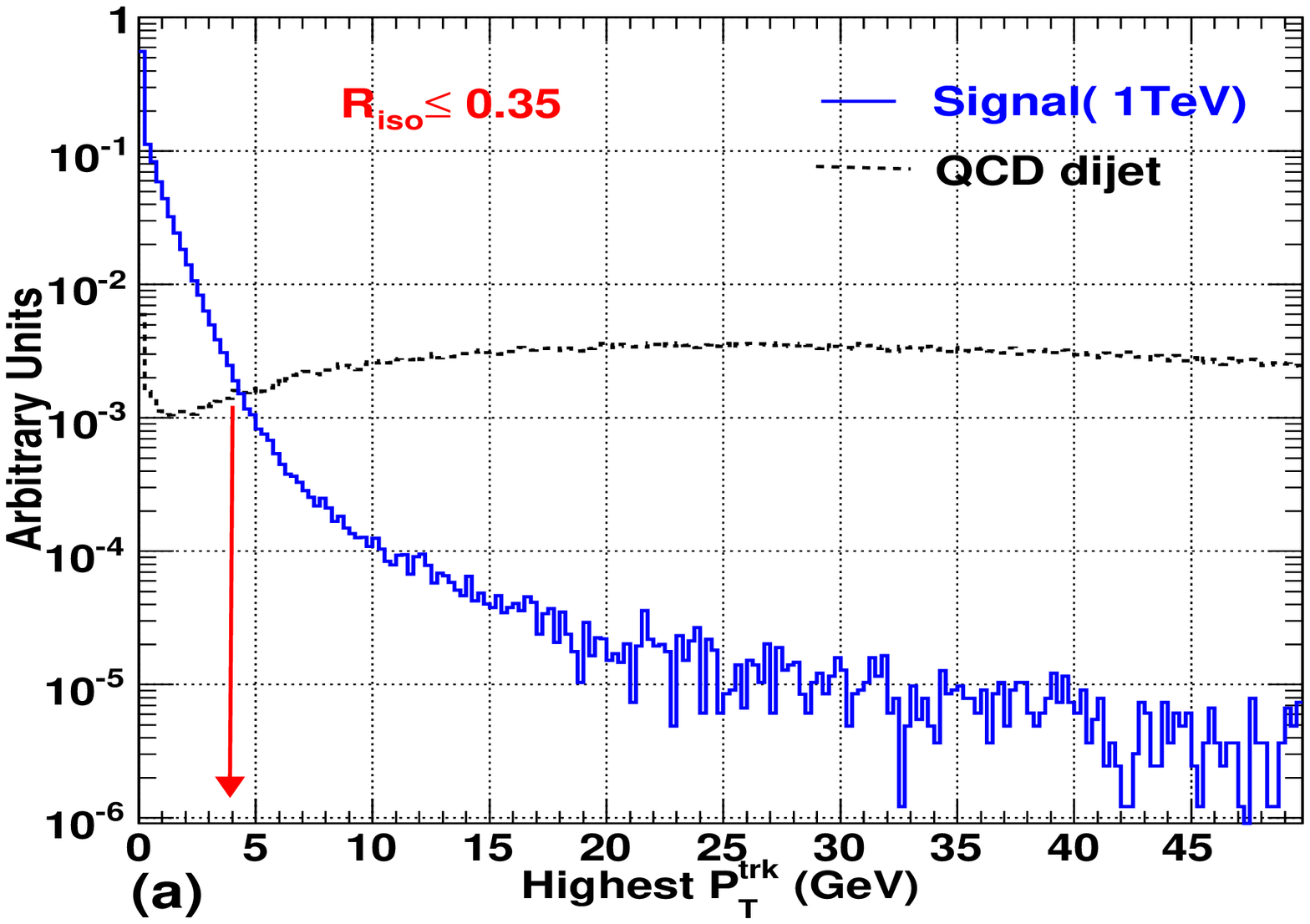}}                                                                              
\vspace*{-10ex}
\scalebox{0.45}{\includegraphics{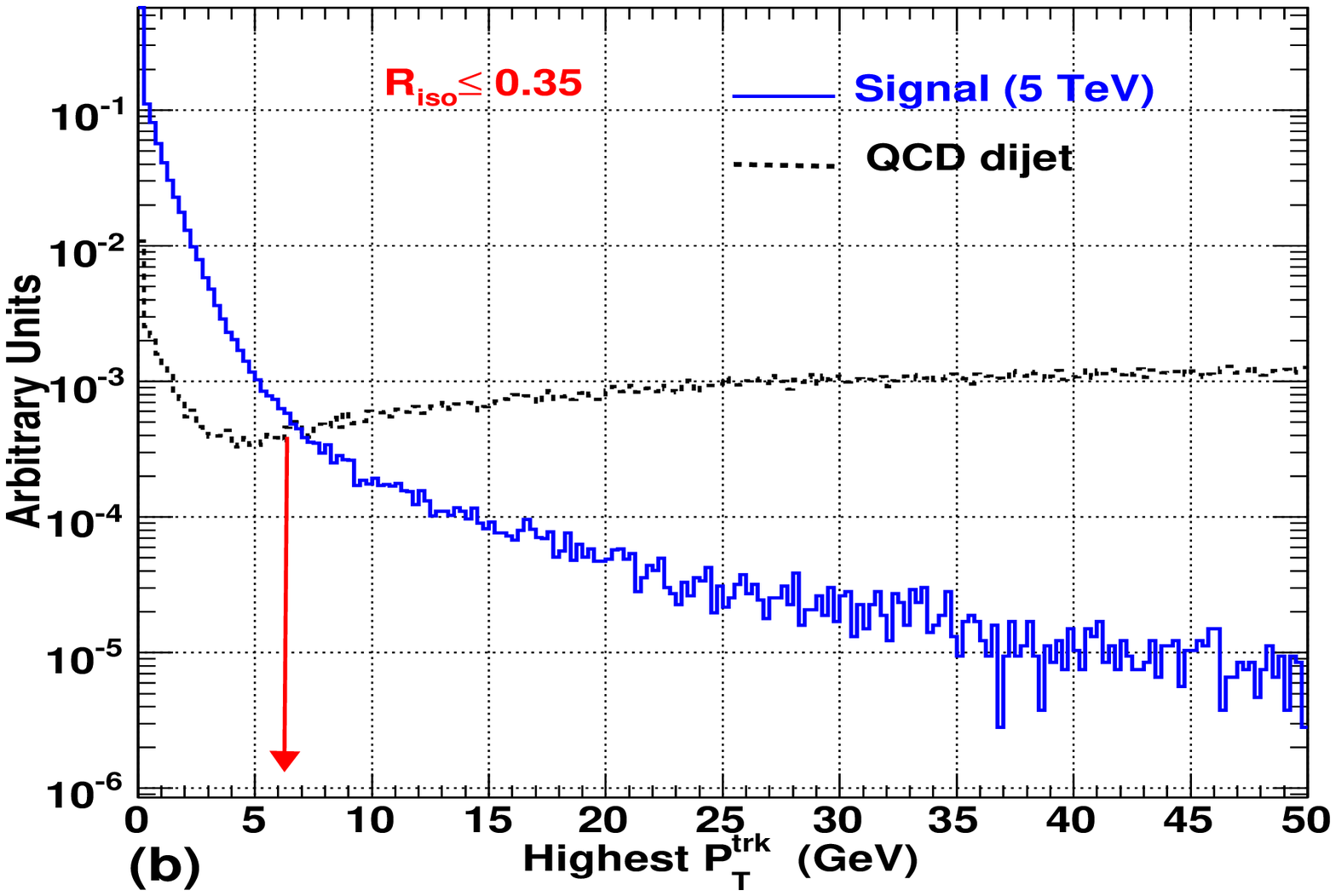}}

\vspace*{-2ex}
\caption{Highest $P_{T}$ track around leading photon for the signal and
the QCD background a) $M_{q^*}=$1 TeV b) $M_{q^*}=$5 TeV. An isolation cone of
size 0.35 has been used. Both distributions are normalized to unity for
the sake of comparison. 
Note that the background differs between the
two panels on account of the differing requirements on $\hat P_T$ 
(vide Sect.\ref{sec:monte}).}
\label{fig:trk_highest} 
\end{figure}
\begin{figure}[!b]
\vspace*{-16ex}
\scalebox{0.45}{\includegraphics{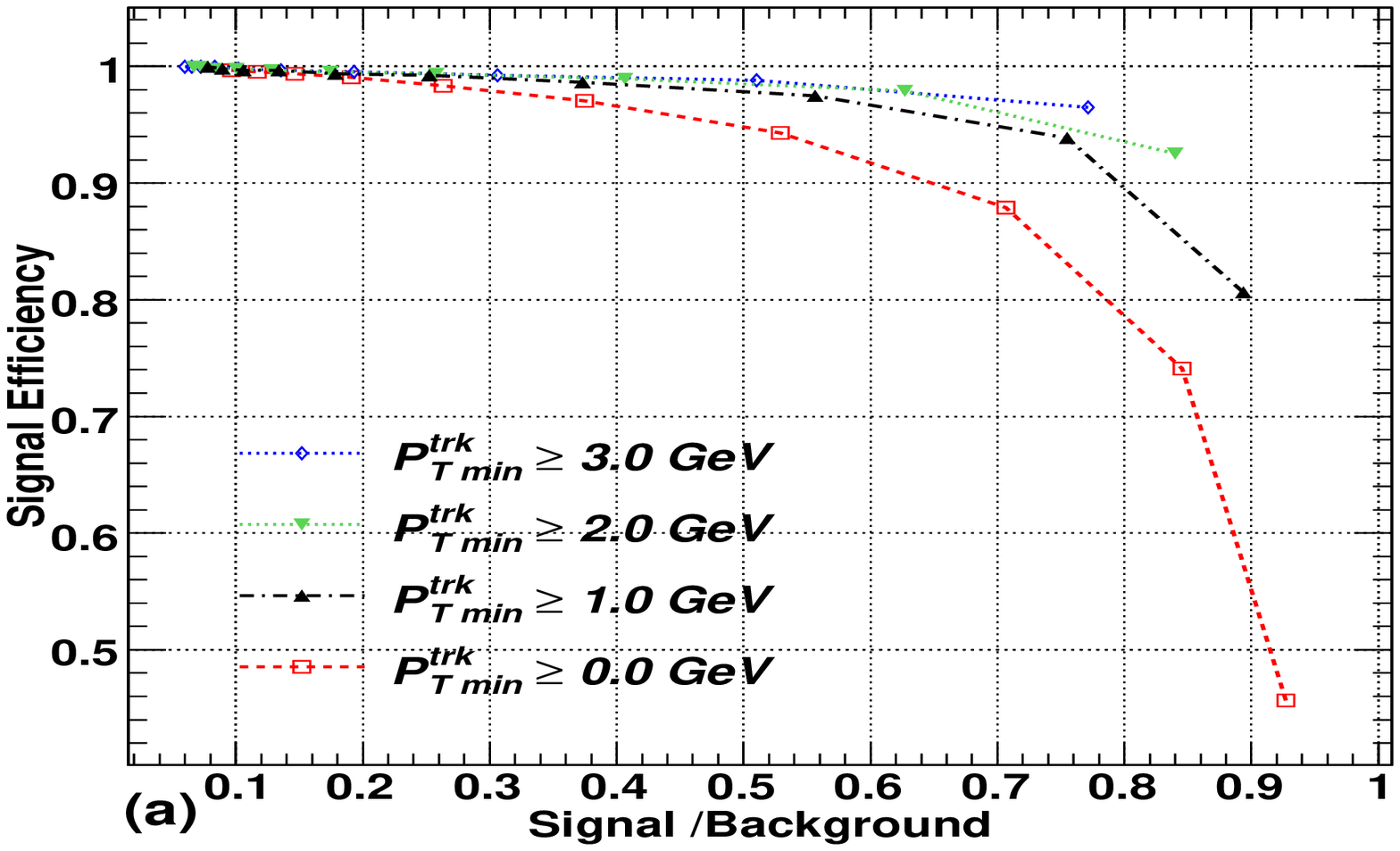}}                                                                     
\vspace*{-19ex}
\scalebox{0.45}{\includegraphics{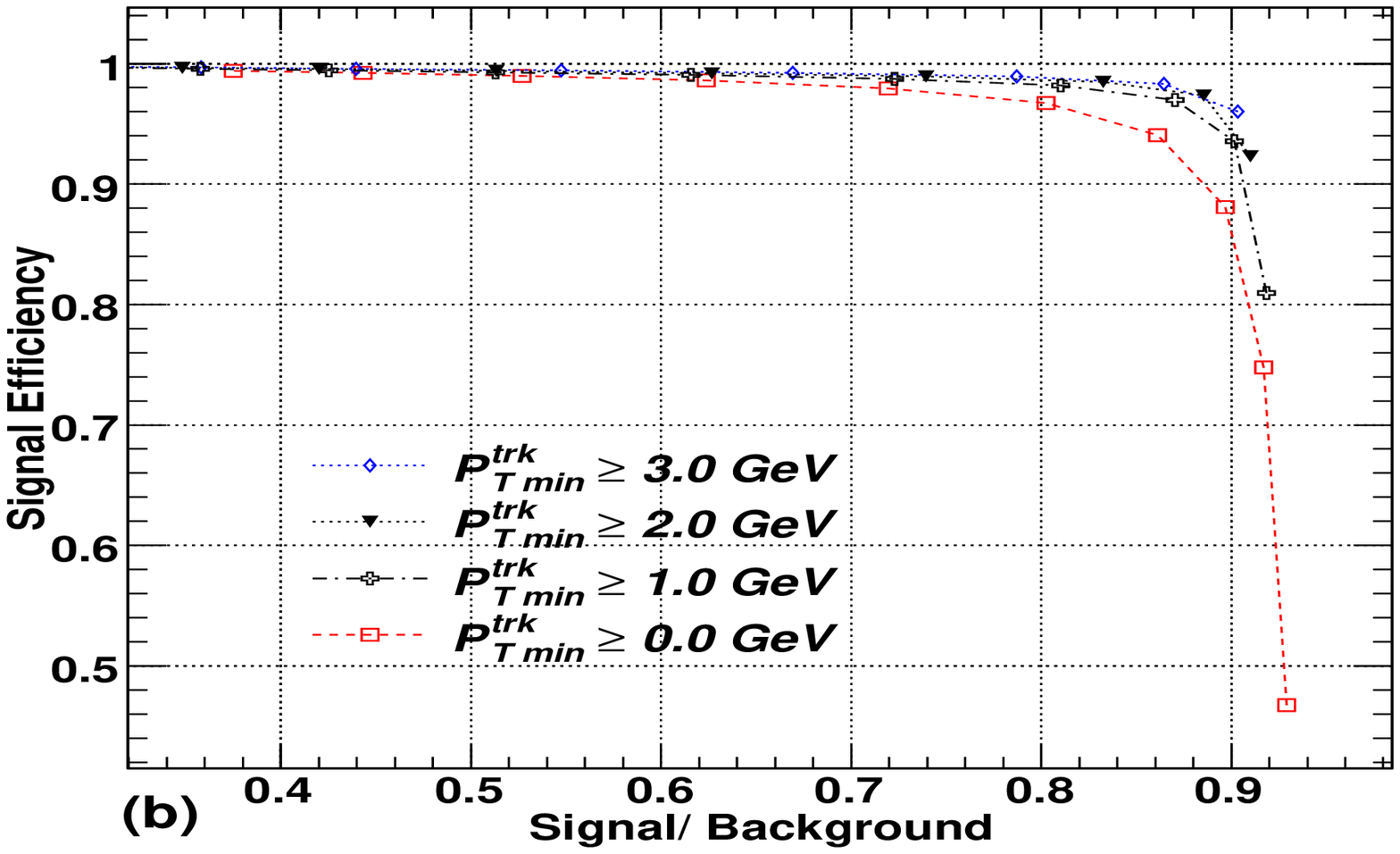}}
\vspace*{1ex}
\caption{ Effect of $P_{T min}^{trk}$ choice 
on signal efficiency vs  $S/B$ for the photons 
from  a) 1 TeV signal b) 5 TeV signal. 
For a given threshold ($P_{T min}^{trk}$), the individual points
correspond to differing values of the number of tracks, $N_{trk}$,
allowed in a cone starting with 0 tracks (for the rightmost point) 
and increasing in steps of one. }
\label{fig:trk_min}
\end{figure}

To optimize the value of $P_{T min}^{trk}$, 
it is useful to examine both the signal and the QCD
dijet background in terms of the 
distribution for the highest-$P_{T}$ track. In Fig ~\ref{fig:trk_highest}(a) and (b) respectively, 
we display this distribution for the signal for $M_{q^*}=$ 1
TeV and 5 TeV. Accompanying these, in each case, are the 
corresponding QCD dijet background.
For ease of comparision both the
distributions are nomralized to unity. As is evident, any tracks accompanying
the photon in a signal event tend to have a low $P_T$, whereas for the 
background events, the distribution is a very wide one. An indicative 
value for the optimal $P_{T min}^{trk}$ is given by the point of intersection 
of the two normalised distributions (signal and background). The optimal
choice does depend on the signal profile (determined, in a large measure 
by the typical momentum transfers), as is evident from the crossover points 
being $\sim$4 (6) GeV for signals corresponding to $M_{q^*}$ of 1 (5) TeV.
Thus, characterizing only those tracks, around a photon, with a 
$P_T \gsim 4 \gev$ as true tracks (or, in other words, accepting
photons with accompanying tracks satisfying 
$p_T \leq P_{T min}^{trk} \sim 4  \gev$ as true photons) would mean 
that a very large fraction of the signal is retained while a significant 
fraction of the background is rejected. 
In
Fig.~\ref{fig:trk_min}, we display the consequent interplay between 
signal efficiency and the signal to background ratio ($S/B$), for two different
signal points ($M_{q^*} =$1 TeV and 5 TeV). 
It is evident from the distributions that adopting a higher
threshold would remarkably increase the signal efficiency  with only 
a small loss in the S/B ratio. 
More importantly, the track isolation requirement reduces the 
fake photon events with 
the major effect  showing up in the QCD dijet background. As is obvious, 
the strict requirement of 
$N_{trk}=0$ in a given cone around the photon reduces only a small
fraction of the signal whereas the S/B ratio
is improved considerably. 

To keep the analysis simple, we then dispense with 
a $M_{q^*}$-dependent choice of the threshold, 
and instead demand 
$P_{Tmin}^{trk}= 3$ GeV and $N_{trk}=0$ irrespective of the 
mass of the $q^*$ being looked for.
Although a choice of $P_{Tmin}^{trk}= 4$ GeV would have led 
to better results (see Fig~\ref{fig:Trk_N_Lead}), we make a more 
conservative choice to account for the fact that, 
in a real detector,  the tracking efficiency is usually less than 100 $\%$.


\subsection{$E_{T}$ Sum Isolation}

\begin{figure}[!b]
\vspace*{-16ex}
\scalebox{0.45}{\includegraphics{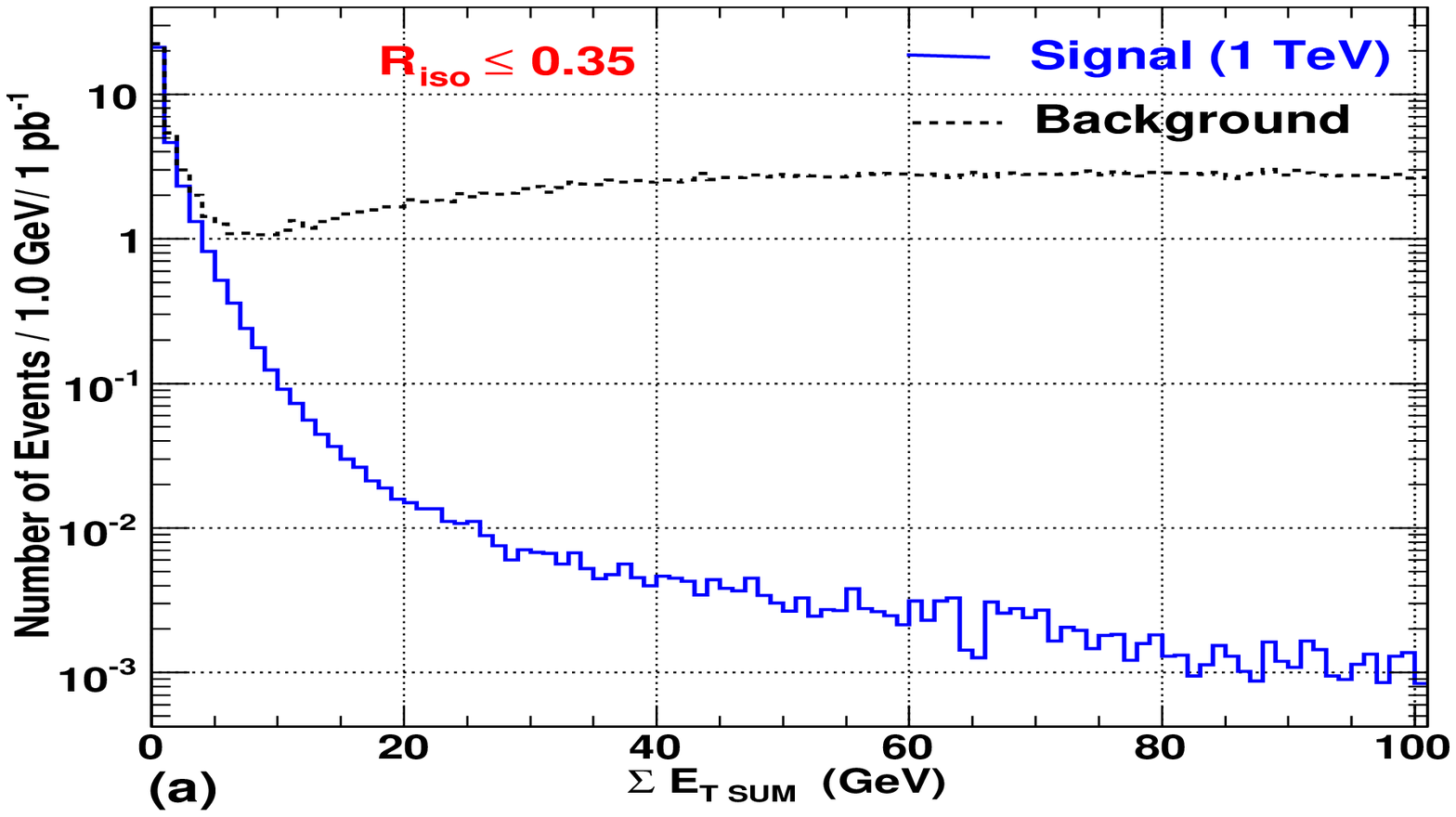}} 
                                                                    
\vspace*{-19ex}
\scalebox{0.45}{\includegraphics{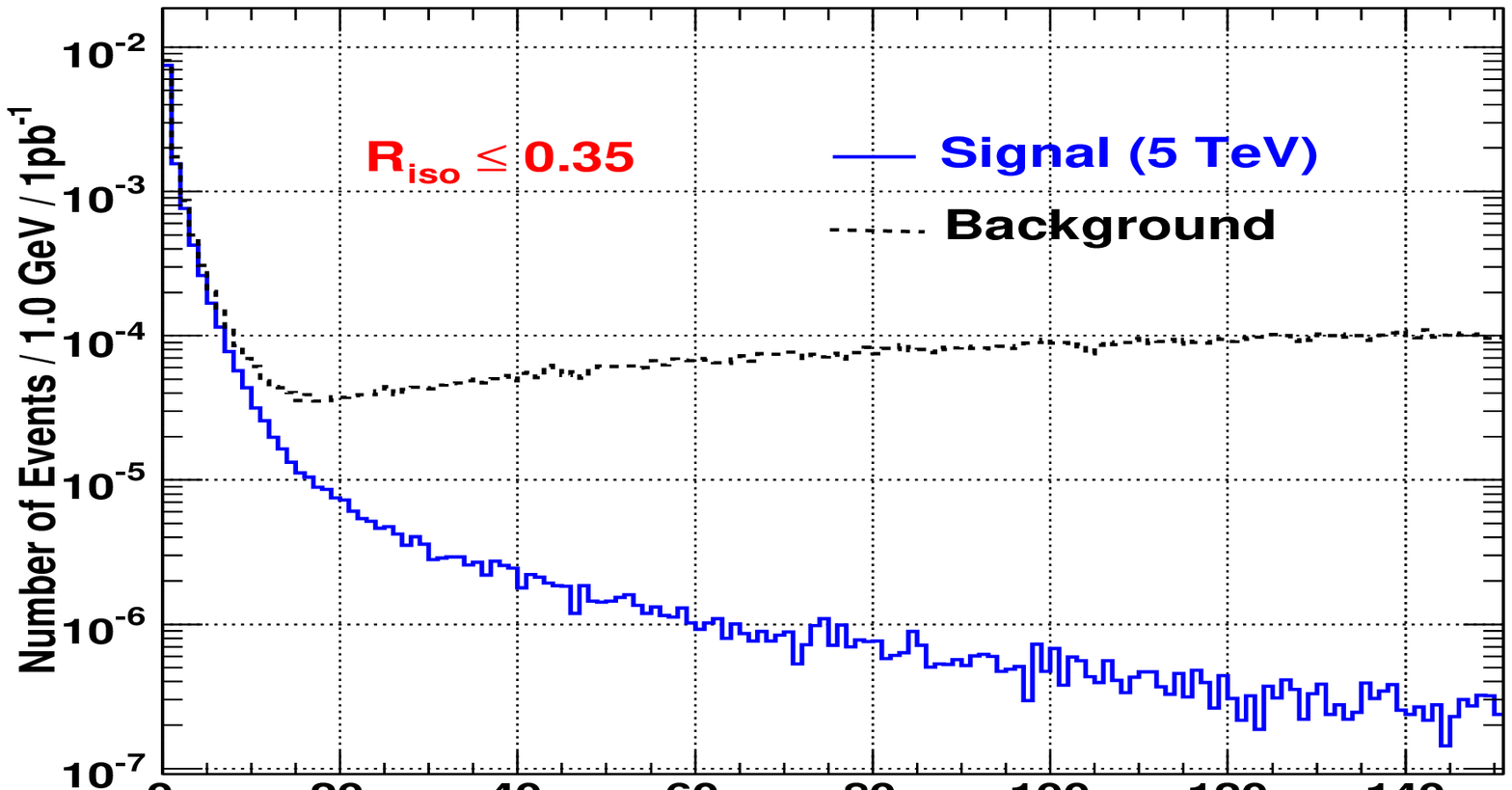}}                            
\vspace*{1ex}
\caption{$E_{T SUM}$ for the background and the signal events around photons for a) $M_{q^*}=$1 TeV b) $M_{q^*}=$5 TeV Signal. 
Distributions are normalized for $\int{L dt}=1 \pb^{-1}$.}
\label{fig:etsum}
\end{figure}

We now discuss the next isolation variable, namely the scalar sum of
transverse energy inside a cone around the photon candidate.  
Figs.\ref{fig:etsum}a(b) show respectively the $E_{T SUM}$
distribution for the leading photon for $M_{q^*}= 1 \, (5) \tev$
for a cone of size $\Delta R=0.35$. Both distributions are
normalized for an integrated luminosity of $1 \pb^{-1}$. It is evident
that, in either case, a large fraction of signal events have $E_{T
SUM}\leq 5.0 \gev$ whereas the background events generically have $E_{T
SUM}\geq 5 \gev$. For  $M_{q^*} =5 \tev$, the
discriminating point is even slightly higher.

\begin{figure*}[!t]
\vspace*{-15ex}
\scalebox{0.42}{\includegraphics{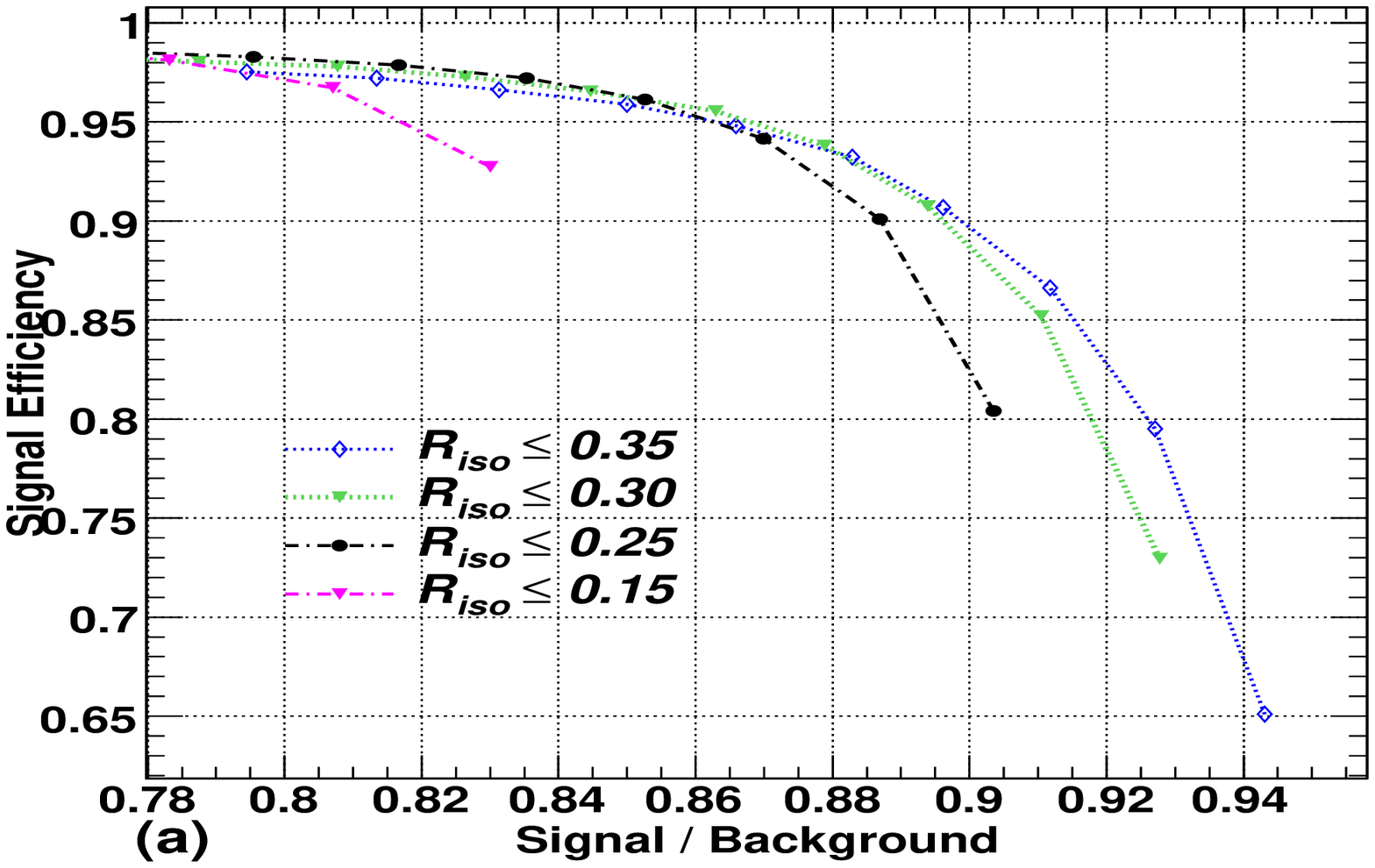}} 
\scalebox{0.42}{\includegraphics{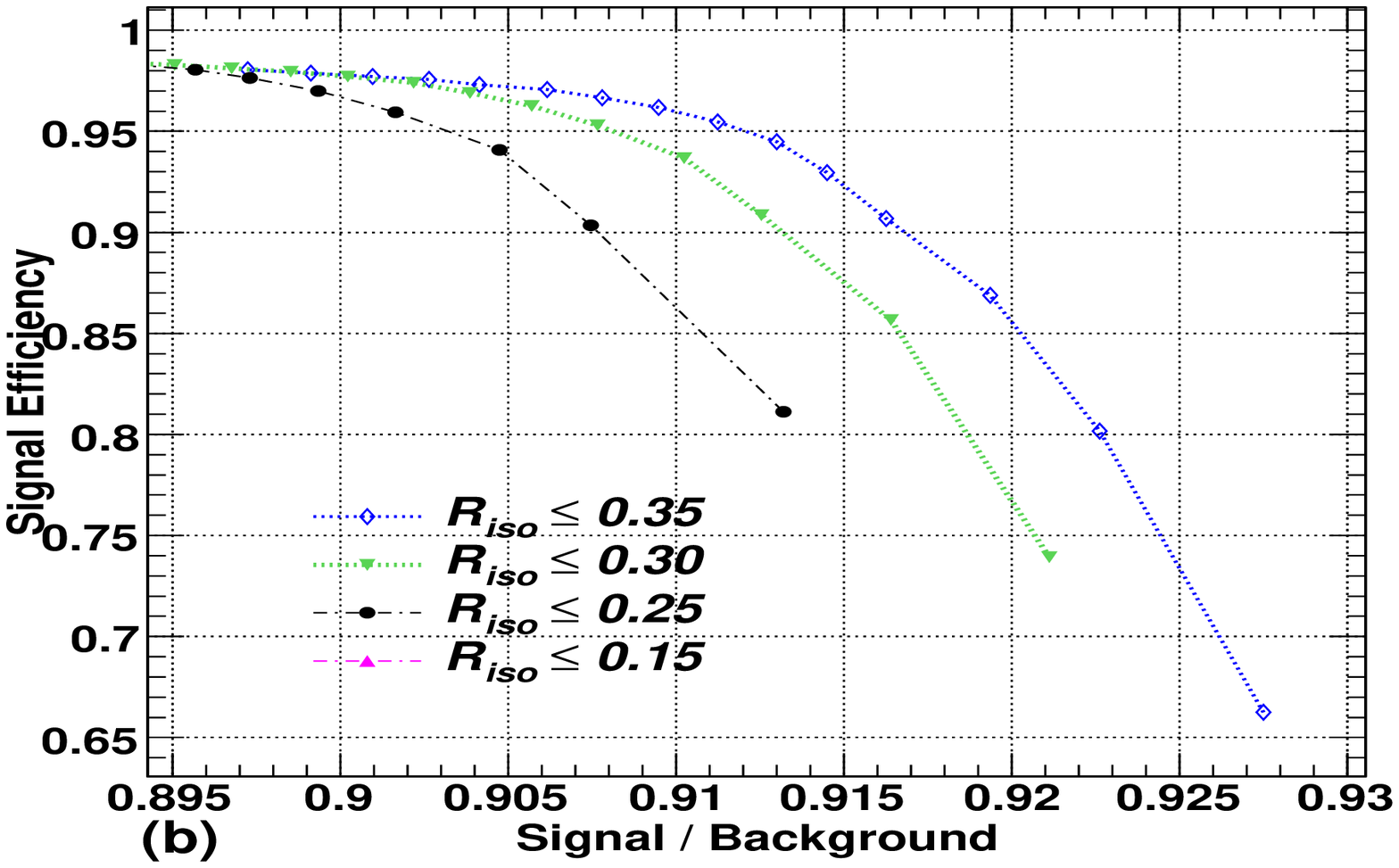}}
\caption{Signal efficiency vs. S/B ratio for different cone sizes for
different choices of the $E_{T SUM}$ threshold around the
leading photon for (a) $M_{q^*}=$1 TeV and (b) $M_{q^*}=$ 5 TeV. For each choice of the cone size, 
individual points correspond to a particular choice for the $E_{T SUM}$ threshold in that cone, starting with 
1 GeV at the rightmost point and going up in steps of 1 GeV.}
\label{fig:etsum2}
\end{figure*}

As in the previous subsection, we next study the dependence of signal
efficiency and S/B ratio on the choice of the $E_{T SUM}$ threshold.
Fig ~\ref{fig:etsum2}(a) and (b) respectively show the
signal efficiency vs. S/B ratio for 1 TeV and 5 TeV signal points. 
It is evident that, for a given signal
efficiency, a higher S/B ratio can be attained for larger cone sizes. 
For example, demanding $\Delta R \leq$0.35 leads to a 
large signal efficiency
($\sim 92 \%$) and $S/B > 0.88$ for either choices of $M_{q^*}$.
On the other hand, any relaxation beyond $E_{T SUM}> 5.0 \gev$ 
reduces S/B 
considerably with only a very small gain in signal efficiency. Several 
$E_{TSUM}$ thresholds for different cone sizes were analyzed along with track
isolation requirements to optimize signal efficiency along
with the S/B ratio.

\begin{table*} \caption{Fraction  of events surviving for the signal and various backgrounds for different isolation cuts (after $P_{T}$ Cut).}
\begin{ruledtabular} \begin{tabular}{|cccccccccc|} $R_{iso}$ & $N_{trk}$ &  $P^{trk}_{T min}$ & $E_{TSUM}^{max}$ & S &  QCD & $\gamma + Jet$ & $Z+\gamma$ & $W+\gamma$ &  (S/B) \\
 &  & (GeV) & (GeV) &  (\%)   &  (\%)   &  (\%)    &  (\%)  & (\%)  & \\
\hline
              &&&&&&&&&\\
&&&& $M_{q^*}= 1.0$ TeV &&&&&\\
\hline

0.30&  0&  1.0&  5.0&     73.6   &  0.93    &  72.9    &  75.0     &  71.9     & 0.970\\  

    &   &     &  6.0&     73.7   &  0.94    &  73.1    &  75.1     &  72.0     & 0.967\\
    
    &   &  2.0&  5.0&     81.2   &  1.12    &  80.6    &  82.9     &  79.4     & 0.951\\

    &   &     &  6.0&     81.7   &  1.15    &  81.0    &  83.3     &   79.9    & 0.946 \\

    &   &  3.0&  5.0&     83.0   &  1.19     &  82.3    &  84.8    &   81.2    & 0.941\\

    &   &     &  6.0&     83.7   & 1.25      &  83.1    &  85.6    &   81.9    & 0.930\\

\hline

0.35&  0&  1.0&  5.0&     69.8   & 0.82     & 69.3     & 71.1      &  68.2     & 0.984\\ 
   
    &   &     &  6.0&     70.1   & 0.83     & 69.5     & 71.4      &  68.5     & 0.982\\

    &   &  2.0&  5.0&     79.0   & 1.01     & 78.4     & 80.5      &  77.2     & 0.967\\

    &   &     &  6.0&     79.8   & 1.05     & 79.1     & 81.4      &  78.0     & 0.960\\

    &   &  3.0&  5.0&     81.0   & 1.08      & 80.4     & 82.6      &  79.2     & 0.957\\     
   
    &   &     &  6.0&     82.2   & 1.14      & 81.6      & 83.9      &  80.4     & 0.947\\                                                                  
\hline
              &&&&&&&&&\\
&&&& $M_{q^*} = 5.0$ TeV &&&&&\\
\hline
0.30&  0&  1.0&  5.0&    82.9   &  1.82   & 83.1     &  83.3     &  81.6      & 0.955\\

    &   &     &  6.0&    83.1   &  1.83   & 83.2     &  83.5     &  81.8      & 0.954 \\

    &   &  2.0&  5.0&    91.1   & 2.11    & 91.1      & 91.5     &  89.5      & 0.950 \\

    &   &     &  6.0&    91.5   & 2.14    & 91.6      & 91.9     &  89.9      & 0.950\\

    &   &  3.0&  5.0&    92.9   &  2.17   & 93.0     &  93.4     &  91.3      & 0.949 \\

    &   &     &  6.0&    93.7   &  2.22   & 93.7     & 94.2      &  92.1      & 0.947\\

\hline

0.35&  0&  1.0&  5.0&    78.8  &  1.63    &  79.0    & 79.3      & 77.7     & 0.960 \\

    &   &     &  6.0&    79.0  &  1.64    &  79.2    & 79.5      & 77.9     & 0.960 \\

    &   &  2.0&  5.0&    88.6  &  1.94    &  88.7    & 89.0      & 87.2     & 0.956 \\

    &   &     &  6.0&    89.3  &  1.97    &  89.4    & 89.8      & 87.9     & 0.956 \\

    &   &  3.0&  5.0&    90.6  &  1.99    &  90.7    & 91.1      & 89.2     & 0.955 \\

    &   &     &  6.0&    91.9  &  2.04    & 91.9     & 92.4      & 90.5     & 0.954 \\

\end{tabular}
\end{ruledtabular}
\end{table*}

\subsection{Final Selection Cuts}

In Table IV, we show the efficiencies for signal and background for all
the isolation variables with differing thresholds. Since we aim to
observe any excess as a mass peak over the SM continuum and, 
in the early phase of the LHC
operation, would be able to identify a signal only for low masses, 
it is rather important to have a large signal efficiency
and S/B ratio for smaller $M_{q^*}$. Hence we have used the 
isolation criteria befitting a 1 TeV signal point (note that this also works
reasonably for higher $M_{q^*}$s), and performed this analysis
for all the different signal points
considered in this study. Note that, while it is indeed possible 
to have yet other criteria to select different threshold based 
on real detector simulation, the qualitative differences in the results
are small.

Based on these studies, the final selection cuts applied are
as follows (the $P_T^{\gamma,jet}$ requirements being determined by the range 
of $M_{q^*}$ being investigated, vide Sect.\ref{sec:monte}):

\begin{itemize}
\item  $P_{T}^{\gamma }, \, P_{T}^{jet} \geq 200$ GeV(500 GeV, 1 TeV);

\item  $|\eta^{\gamma}| \leq 2.5 \quad$ \& 
     $\quad|\eta^{\gamma}| \not \in [1.4442,1.5666]$;
\item  $|\eta^{jet}| \leq 3.0 $;

\item  $N_{trk}=0$ for $P_{T}^{trk} \geq 3.0$ GeV within $R_{iso}\leq 0.35$;

\item $E_{T SUM}< 5.0$ GeV  within $R_{iso}\leq 0.35$.
\end{itemize}


\begin{widetext}
In Table V, we show the expected number of events for
$M_{q^*}=\Lambda= 1 \tev$ for an integrated luminosity of 100 $\pb^{-1}$ for
various combinations of isolation variables discussed above.

\begin{table*}
   \caption{Number of events surviving for the $M_{q^*} = \Lambda =$1 TeV 
            signal and the backgrounds for $\int{L dt}= 100 pb^{-1}$ 
            for different isolation cuts.}
\begin{ruledtabular} 
\begin{tabular}{|ccccccccccc|} 
  $R_{iso}$ & $N_{trk}$ &  $P^{trk}_{T min}$ & $E_{TSUM}^{max}$ & S 
&  QCD & $\gamma + Jet$ & $Z+\gamma$ & $W+\gamma$ & Tot.Background & (S+B) 
\\
 &  & (GeV) & (GeV) & & & &  & & &  \\
\hline
0.30&  0&  1.0&  5.0& 2734  &  626.2  &  2185  & 4.10  & 2.97 & 2818 & 3368\\  

    &   &     &  6.0& 2740  &  634.4  &  2190  & 4.11   & 2.98 & 2831   & 3381\\

    &   &  3.0&  5.0& 3085  &  803.0   &  2467  & 4.64  & 3.36 & 3278 & 3896\\

    &   &     &  6.0& 3112  &  845.9  &  2490  & 4.68  & 3.39 & 3344 & 3966\\

\hline

0.35&  0&  1.0&  5.0& 2596  & 554.0  &  2076   & 3.89   & 2.82   & 2637 & 3157\\ 
   
    &   &     &  6.0& 2604  & 560.8  &  2083   & 3.91    & 2.83  & 2650  & 3172\\

    &   &  3.0&  5.0& 3011  & 727.4  &  2409   & 4.52   & 3.28  & 3144 & 3747\\     
   
    &   &     &  6.0& 3054  &  772.1 &  2444   & 4.59   & 3.32   & 3224 & 3834\\

\end{tabular}
\end{ruledtabular}
\end{table*}

In Fig.~\ref{fig:finalcuts_mass} we have shown the invariant mass
distribution for both signal+background(S+B) and background(B)
after the all selection cuts.

\begin{figure*}[!h]
\vspace*{-10ex}
\scalebox{0.45}{\includegraphics{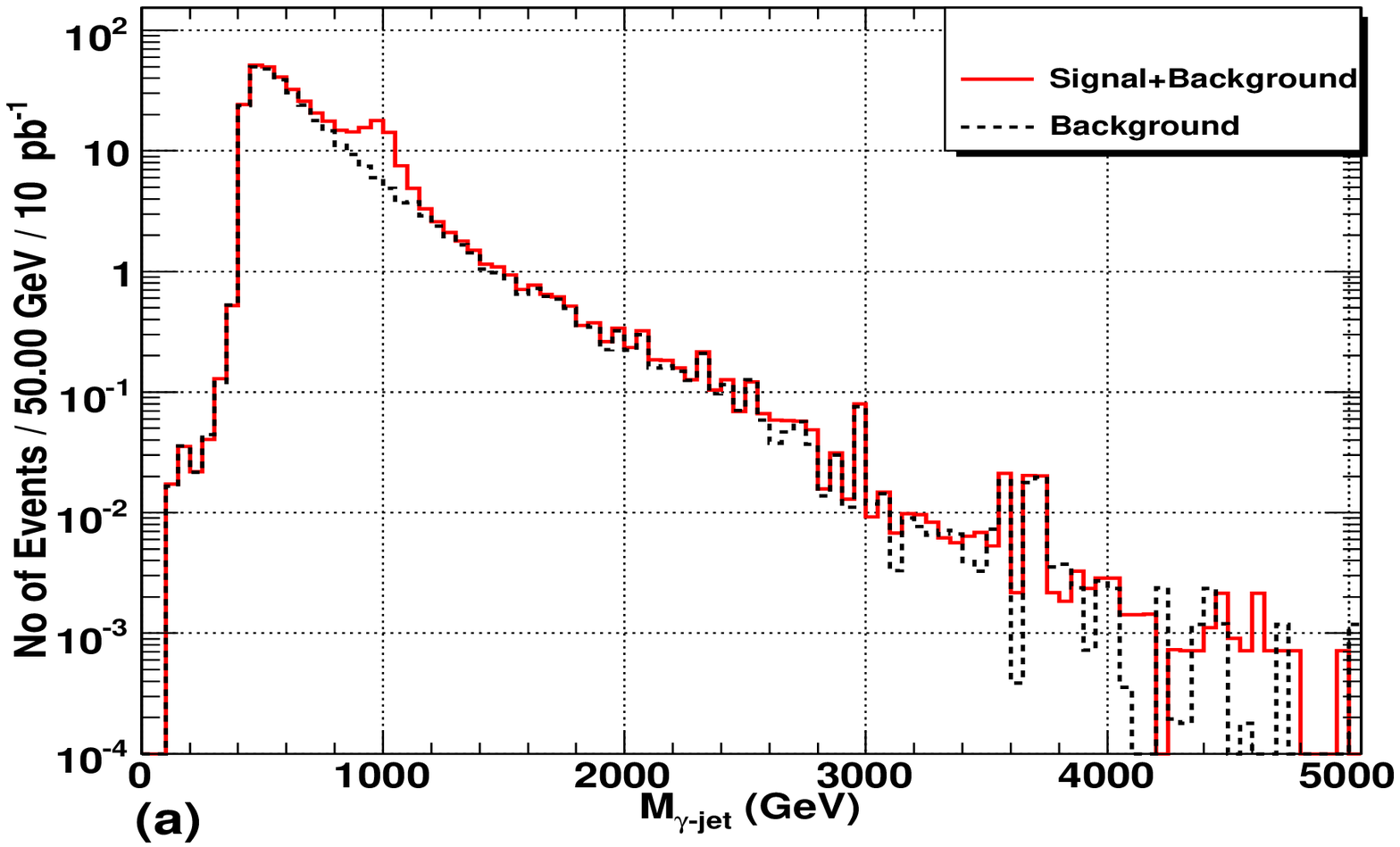}
\includegraphics{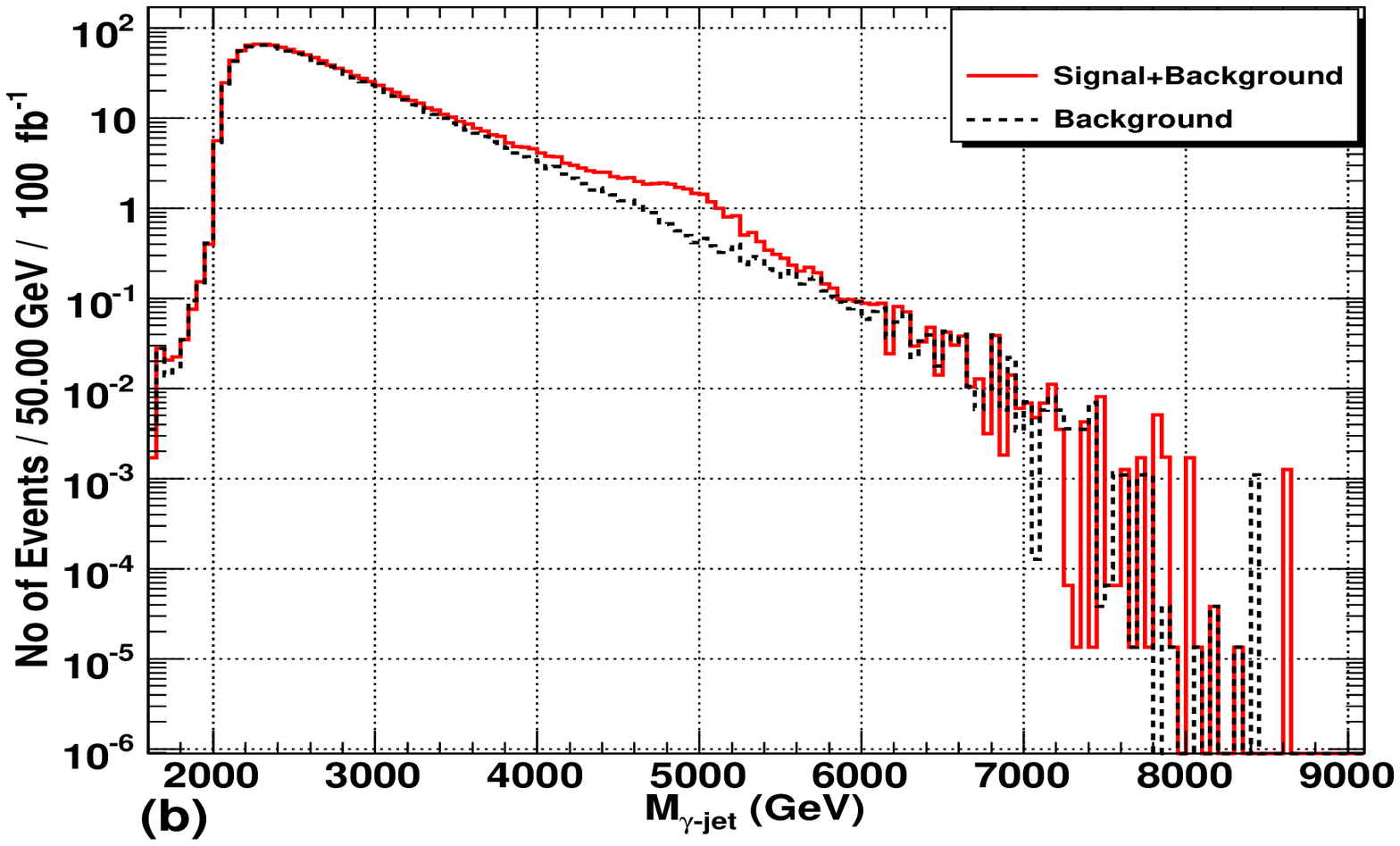}}
\caption{ Invariant mass of $\gamma-$jet system for Signal+Bacgkround and Background after all the isolation and kinematical cuts. (a) $M_{q^*}=$ 1TeV  
(b) $M_{q^*}=$ 5TeV.}
\label{fig:finalcuts_mass}
\end{figure*}
\end{widetext}

\section{Significance Level}


For reporting a discovery significance, we adopt a frequentist
Monte-Carlo technique based on a method of hypothesis testing
originally due to Neyman and Pearson~\cite{cluster_algo,newman,LHWG}. 
 The aim is to
determine which one of two competing hypotheses, the so called null
hypothesis($H_{0}$) and the alternative hypothesis($H_{1}$) is
favoured by the data. In the present context, the SM only case
(background) constitutes the null hypothesis ($H_0$) and the presence
of new physics (i.e. excited quark contribution to the final state)
alongwith the SM is the alternative hypothesis ($H_1$). Heretofore,
$H_0$ and $H_1$ will also be referred to as background only (B) and
signal plus background (S+B) hypotheses.

 In the Neyman-Pearson method one aims to design a test which
minimizes the probability $\beta$ of erroneously rejecting an
alternative hypothesis when it is actually true. Understandably,
$1-\beta$ is defined as the
power of a test and the most powerful (MP) test is the one which maximizes 
the power for a given value of the probability $\alpha$ of rejecting the
null hypothesis as false, when it is true instead. According to the 
Neyman Pearson
lemma~\cite{newman}, the condition for the MP test is obtained as a
condition on the log likelihood ratio (LLR) of a given dataset coming
from the null or the alternative hypothesis.  Even when a MP test does
not exist, the  LLR statistic can be used for testing between two hypothesis
due to its statistically desirable properties~\cite{kandall}. One
accepts or rejects $H_0$ based on the value of the LLR computed from the 
data. If the value of the LLR falls
within a range of values (the critical region) which is unlikely to come from $H_{0}$
 then $H_0$ is rejected. Now, $\alpha$ as defined above is clearly
the probability of the LLR value falling in the critical region when
$H_0$ is true. Hence, $1-\alpha$ is reported as the significance level of
rejecting $H_0$ (i.e., the SM in our case), or, in other words,
 this is the discovery confidence level.

In general, a LLR test can be constructed out of one or
many discriminating quantities (e.g. $P_{T}$, angular separation 
etc.). In this analysis, 
the LLR has been constructed out of a single discriminating variable,
namely $M_{\gamma-jet}$, the invariant mass of the leading $\gamma$ and jet.  
While this is obviously the most sensitive discriminant in the case 
of on-shell production, it plays an important 
role even for virtual exchanges\cite{qstar_diphoton}.

 The likelihood ratio is defined as the ratio of Poisson probabilities:

\begin{equation} 
\barr{rcl}
Q & = & \dis \frac{P_{poiss} (data | S+B)}{P_{poiss} (data|B)}
\\[3ex]
 P_{poiss}(data|S+B) & \equiv & \dis
\frac{(s_{i}+b_{i})^{n_{i}} e^{-(s_{i}+ b_{i})}} 
                             {{n_{i}!}}
\\[3ex]
 P_{poiss}(data|B) & \equiv & \dis
 \frac{(b_{i})^{n_{i}} e^{-(b_{i})}} { {n_{i}!}}
\earr
\end{equation}
where $s_{i}+b_{i}$ are the number of events expected in the $i^{\rm th}$
bin of the $M_{\gamma-jet}$ histogram according to the S+B
hypothesis whereas $b_{i}$ correspond  to the B
hypothesis. $n_{i}$ here denote the number of events in the 
$i^{\rm th}$ bin of $M_{\gamma-jet}$ histogram from ``{\it data}". 
All efficiencies are to be 
folded in $s_{i}$, $b_{i}$ and $n_{i}$.

  The LLR statistic is then given by the expression,
\begin{equation}
\nonumber\\
 -2\ln Q = 2 \sum_{i=1}^{n_{bins}} s_{i} - 2 \sum_{i=1}^{n_{bins}} n_{i} \ln \Big( 1+ \frac{s_{i}}{ b_{i}}\Big)
\end{equation}
The ``{\it data}" $n_{i}$ in the $i^{th}$ bin of the test variable is
generated as a random Poisson fluctuation around the mean value of the
$i^{th}$ bin of the theoretical $M_{\gamma-jet}$ histogram. 

The significance level
is defined as,
\begin{equation}
              \alpha = 1-CL_{B}=P(Q\leq Q_{obs}|B),
\end{equation}
the fraction of experiments in a large ensemble of background only
experiments which would produce results more signal-like than the
observed data. 
By definition, a S+B hypothesis is ``confirmed" at the
5$\sigma$ (3$\sigma$) level if $\alpha \textless 2.8 \times 10^{-7}(
1.35 \times 10^{-3})$ ~\cite{CL_1}.

\begin{figure}[!h]
\vspace*{-7ex}
\scalebox{0.41}{\includegraphics{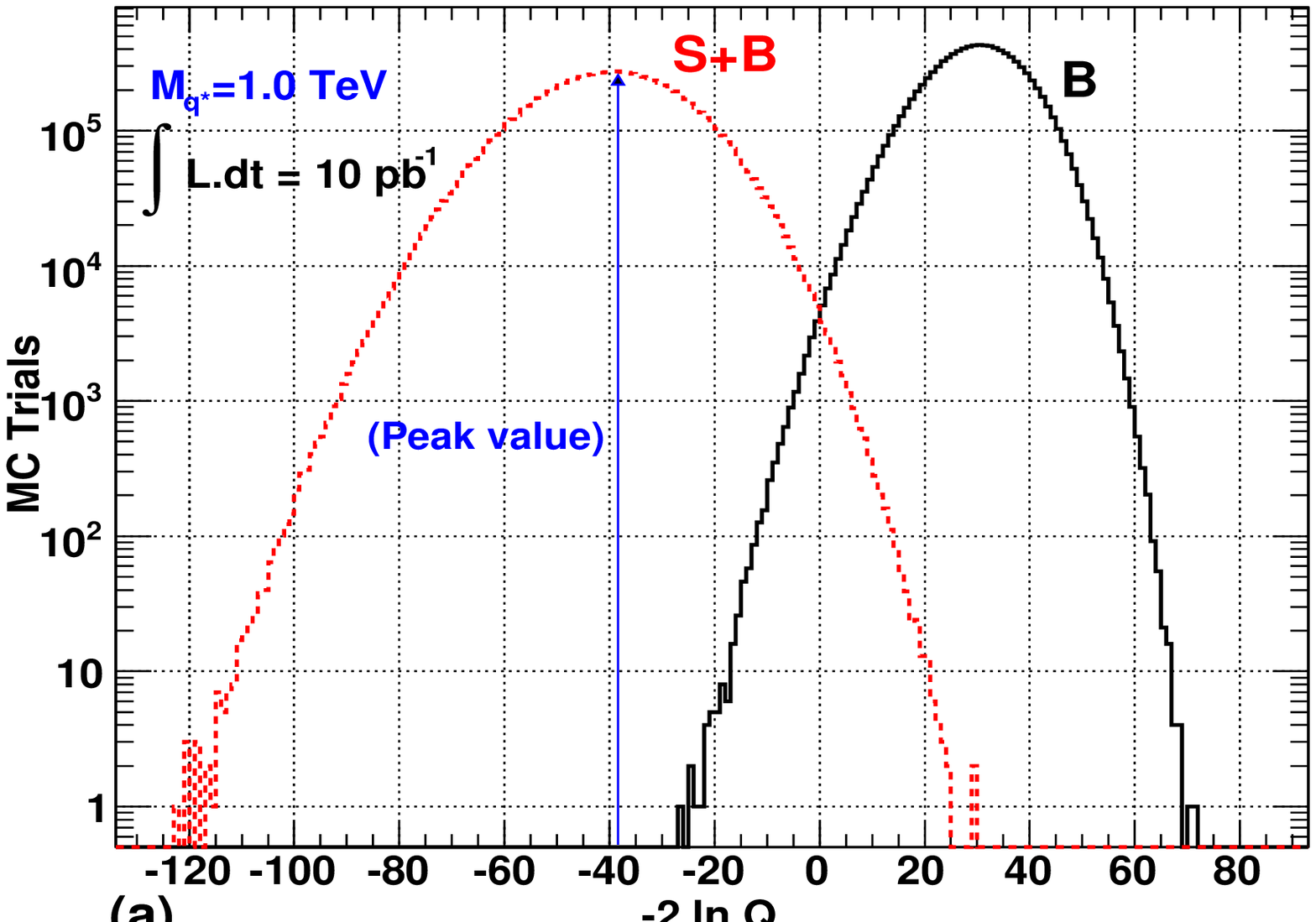}}
\vspace*{-5ex}
\scalebox{0.41}{\includegraphics{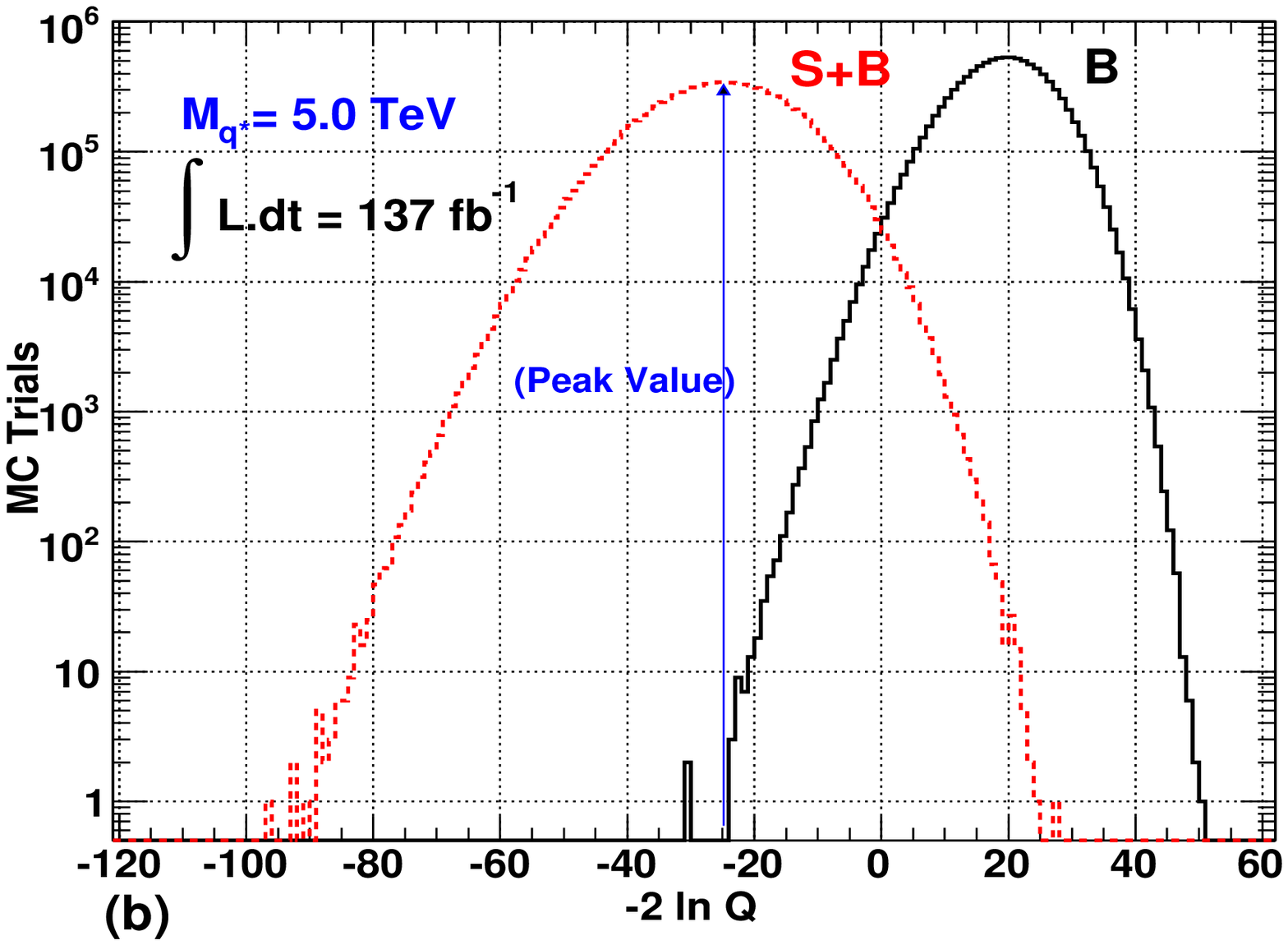}}
\caption{Log likelihood ratio distributions 
for S+B and B type hypotheses for a 5$\sigma-$significance for (a) 1.0 TeV (b) 5.0 TeV $q^*$ states.}
\label{fig:LLR_2}
\end{figure}

In Fig.~\ref{fig:LLR_2}, we show the LLR distribution for B and S+B
type hypotheses for two different mass points. The luminosities 
have been chosen so as to yield a 
5$\sigma$ significance for the S+B hypothesis. We have
used $10^{7}$ MC trials for these distributions. In this figure,
$1-CL_{B}$ is the fraction of MC trials of background type hypothesis
which falls to the left side of peak value of $LLR_{S+B}$.

\begin{widetext}

 In Fig.~\ref{fig:LLR_1} we show how the LLR discrimination between two types of hypothesis behave as a function of mass window around the $M_{q^*}$ value. As visible from the figure, beyond $\pm3 \Gamma(q*)$ the two hypthesis looks similar and hence it does not contribute to the significance level. We found similar results for all signal points and for the final selection we have used $\pm3 \Gamma(q*)$ as the mass window around $\gamma-jet$ invariant mass.

\begin{figure*}[!h]
\vspace*{-7ex}
\scalebox{0.41}{\includegraphics{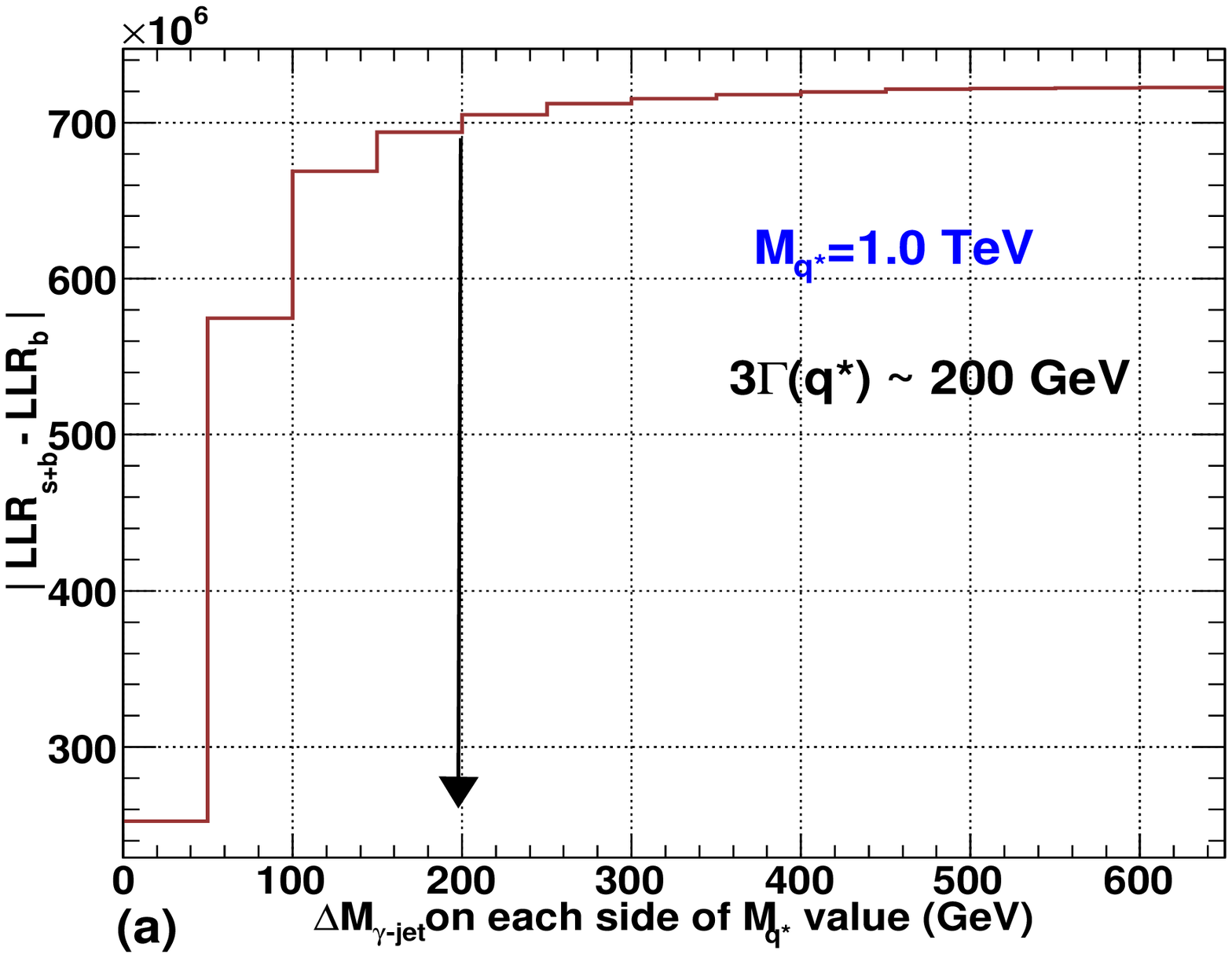}}
\scalebox{0.41}{\includegraphics{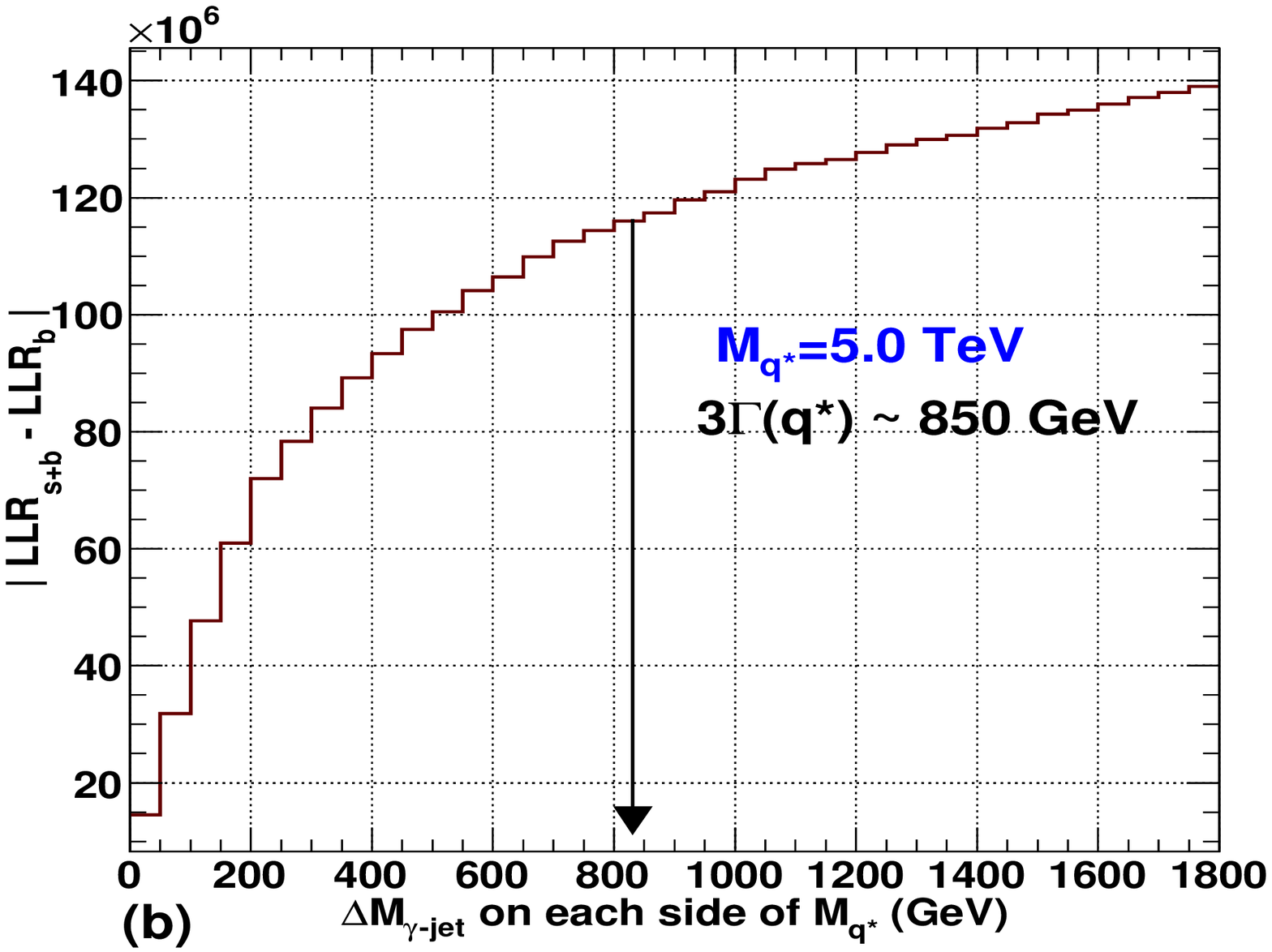}}
\caption{Effective LLR contribution as a function of $\Delta M_{\gamma-jet}$ on each side of $q^*$ state of mass (a) 1.0 TeV and (b) 5.0 TeV.}
\label{fig:LLR_1}
\end{figure*}
\end{widetext}

\section{Background subtracted Invariant mass }
    \label{sec:fitting}

\begin{figure}[!h]
\vspace*{-4ex}
\scalebox{0.41}
{\includegraphics{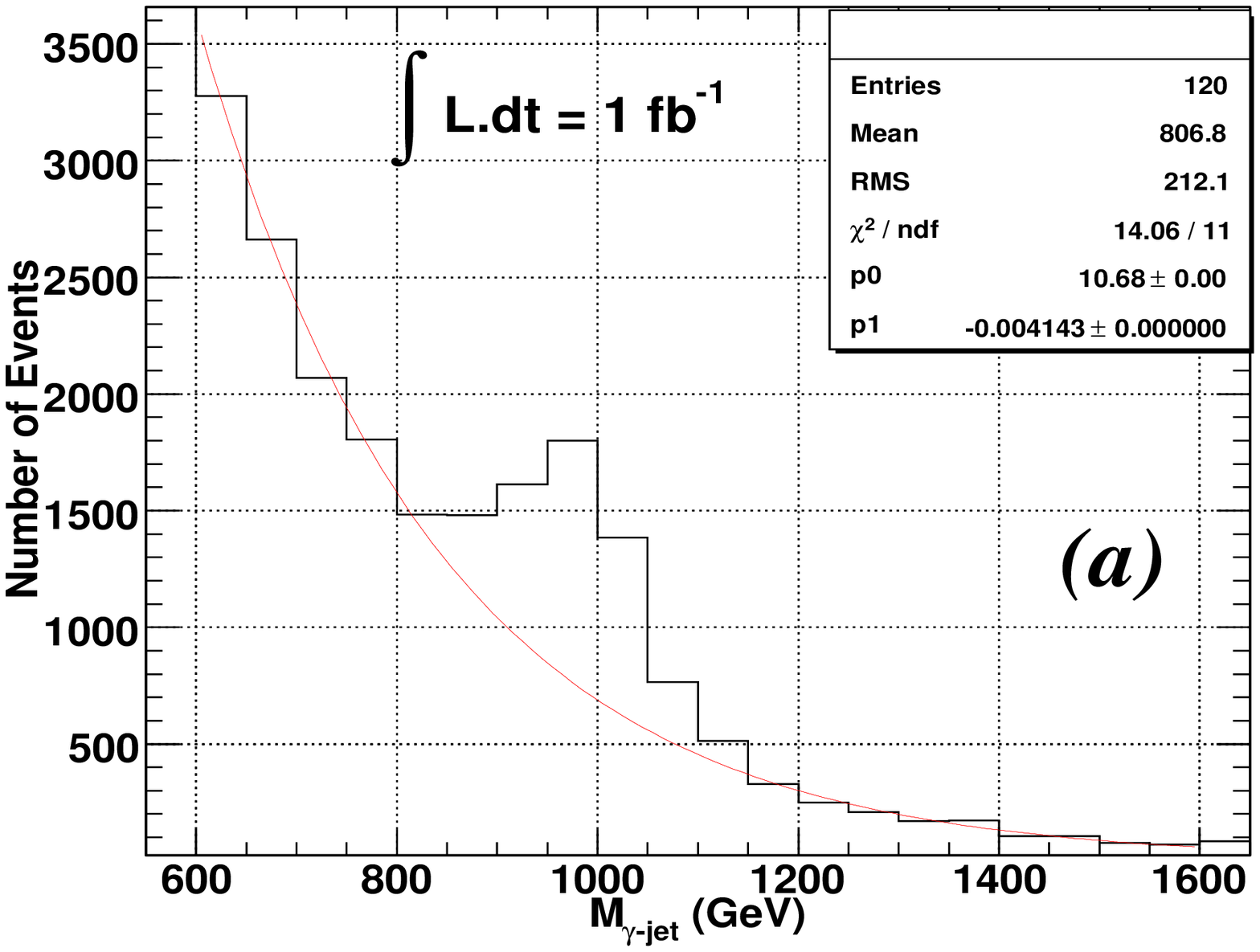}}

\vspace*{-4ex}
\scalebox{0.41}{\includegraphics{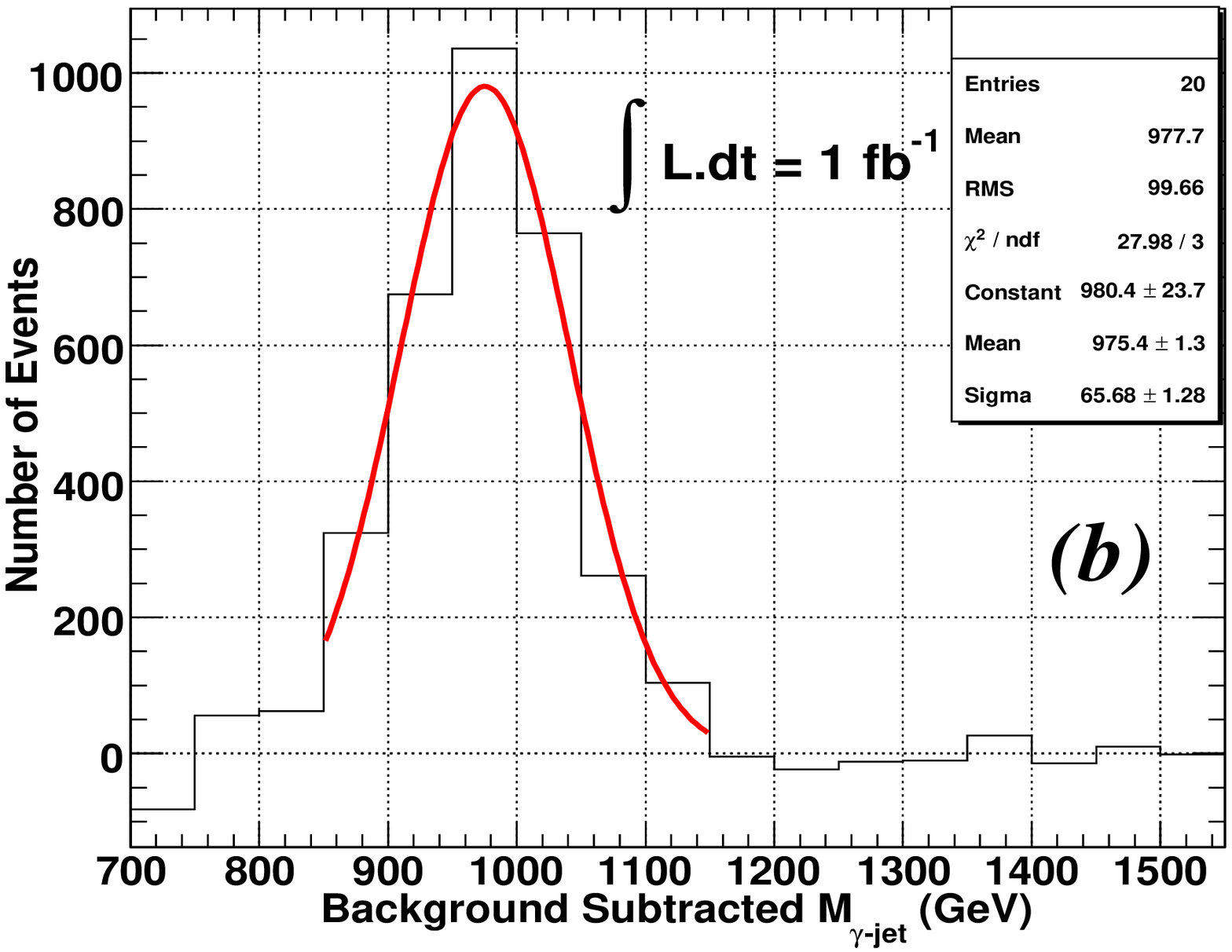}}

\caption{(a)Background fit on the (S+B) distribution with an exponential
function for 1.0 TeV  $q^*$ for an integrated luminosity of $1 \fb^{-1}$.
(b) The corresponding background subtracted invariant mass distribution.}
\label{fig:mass_subtracted_3}
\end{figure}

\begin{figure}[!h]
\scalebox{0.41}
{\includegraphics{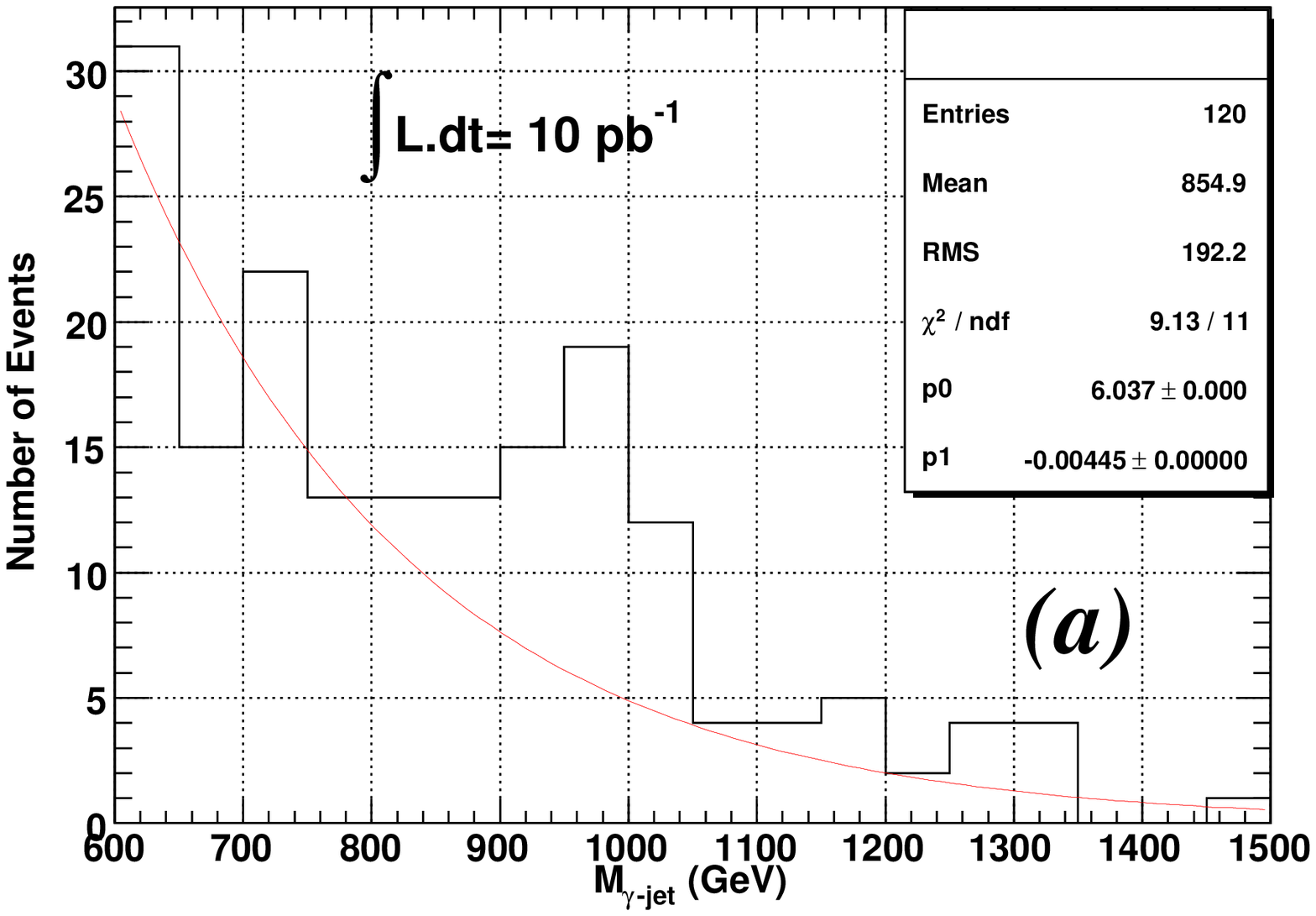}}

\scalebox{0.41}{\includegraphics{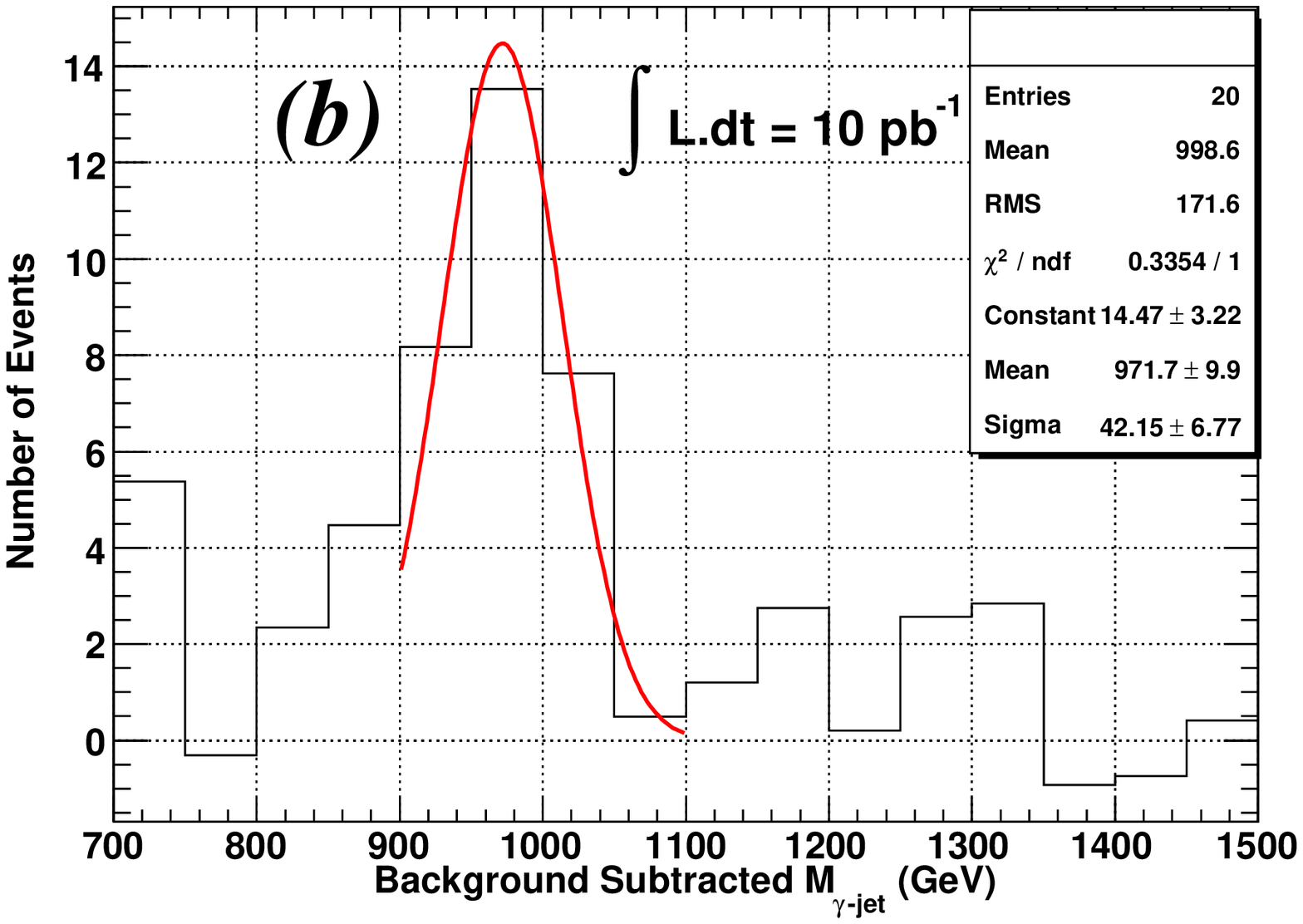}}

\caption{As in Fig.\ref{fig:mass_subtracted_3},
but for an integrated luminosity of $10 \pb^{-1}$ instead.}
\label{fig:mass_subtracted_1}
\end{figure}

In this section, we describe a procedure to estimate the number of
events under the mass peak in the case of a discovery (i.e. if the
data supports the S+B hypothesis). Assuming an excess centred
aproximately around $M_{\gamma-{\rm jet}} = M_0$, the first step
constitutes fitting the data over a $M_{\gamma-{\rm jet}}$ range
centred around $M_0$ but much wider than the region of the excess, the
aim being to fit the background as well as the sidebands.  While in a
real experiment one would attempt to fit the sidebands from data
alone, here we use a large MC sample to determine the shape of the
sidebands and find that an exponential describes them well (see
Fig.~\ref{fig:mass_subtracted_3} for $M_0 = 1 \tev$).  To generate
realistic distributions, we consider (s + b) in each bin to be an
independent Poisson distributed (and integer valued) variable with a
mean equalling the theoretically expected number of events. A random
fluctation was then used to generate the ``experimentally observed''
events in the bin concerned.  For a good background fit on the (S+B)
distribution, an identified excess has clearly to be left out.  To
this end, we leave out the range $\sim [M_0 - 3 \, \Gamma_0, M_0 + 3
\, \Gamma_0]$ consonant with the binning algorithm where $\Gamma_0 =
\Gamma(M_{q^*} = M_0)$.  For a $\chi^{2}$ minimization of the fit, the
MINUIT~\cite{minuit} package was used within the {\it ROOT}
framework~\cite{root}.  The fit in Fig.~\ref{fig:mass_subtracted_3}(a)
was done for an integrated luminosity of $1 \fb^{-1}$ although a $5
\sigma$ signal significance for $M_{q^*} = 1 \tev$ is attainable with
only $10 \pb^{-1}$ of data.  
Fig~\ref{fig:mass_subtracted_3}(b)
shows the background subtracted mass distribution for $M_{q^*}=$1
TeV. Here we have used a single Gaussian to fit the mass spectrum.
     
While an integrated luminosity of $1 \fb^{-1}$ for new physics 
mass measurements would normally 
be considered meagre when compared to the LHC design parameters, 
it is interesting to consider the physics possibilities with far lower 
luminosities. To this end, we present, in 
Fig~\ref{fig:mass_subtracted_1}, analogous distributions for 
only $10 \pb^{-1}$. While the fit for the background is, understandably,
not as good as in the earlier case, once the validity of an exponential 
fit is accepted, the background subtracted mass fit is still very 
convincing. 
In Fig~\ref{fig:mass_subtracted_1}(b) the number of signal events under
the Gaussian fit and within the 800--1200 GeV mass range was found to be
$30.5\pm 5.5(stat.)$. The uncertainty due to error on fitting
parameters are found to be at most 4.9 events.

We note in passing that in an actual detector at the LHC the mass peak will
have a tail on the lower mass side due to partial containment of
showers and fitting this may need a Gaussian modified with a Landau or some
other asymmetric distribution thereby broadening the mass peak somewhat.

The invariant mass distribution has two components, the natural Lorentzian part  
for an unstable particle with a large width and a Gaussian (or a double-Gaussian) distribution
due to resolution effects. The combined distribution is a convolution of the two above.
Although the combined distribution is thus not a simple one, a single Gaussian fits the mass peak reasonably well 
and hence we choose to fit the peak with a simple Gaussian.

\section{Results}

\begin{figure}[!t]
\scalebox{0.472}{\includegraphics{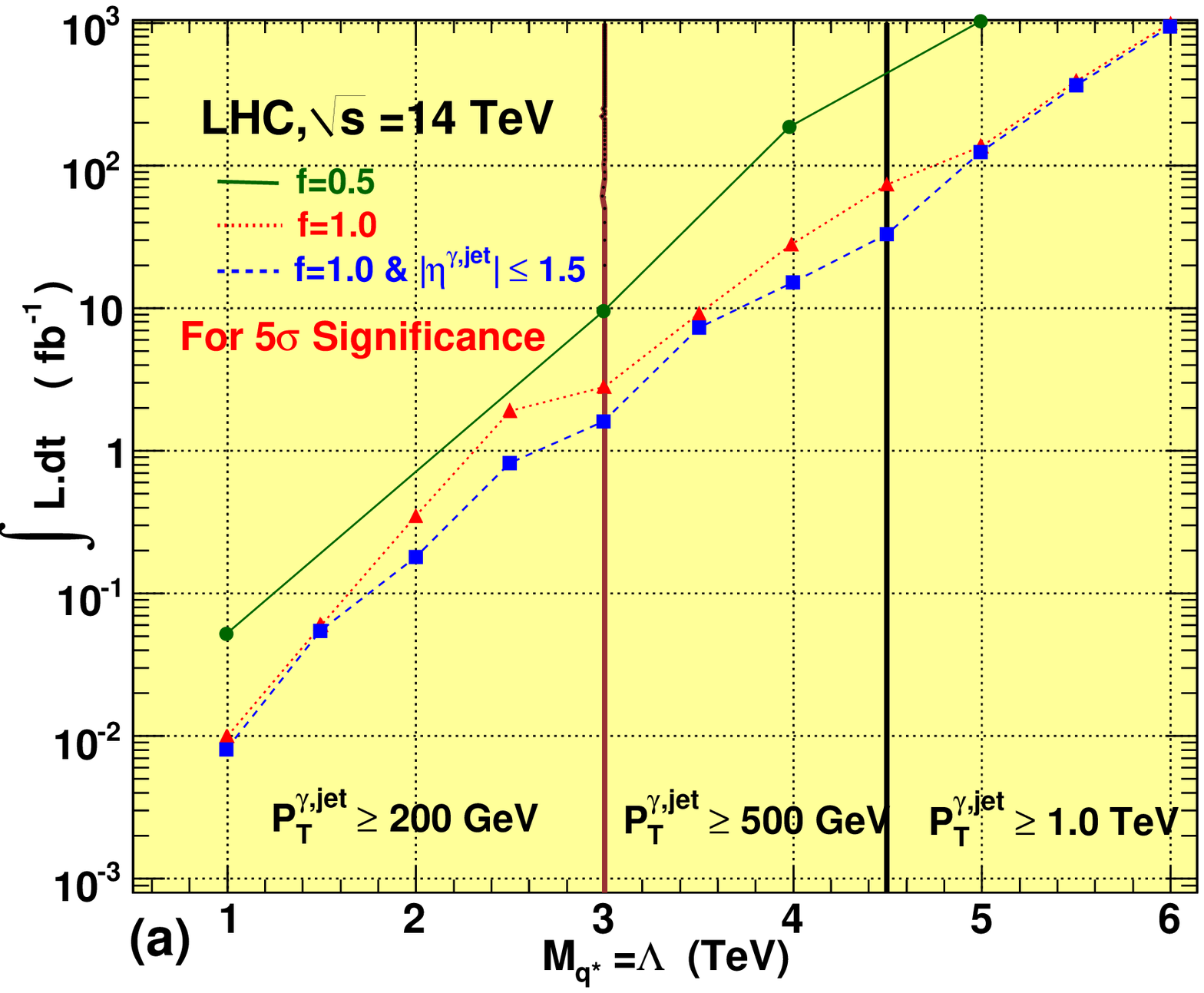}}
\scalebox{0.472}{\includegraphics{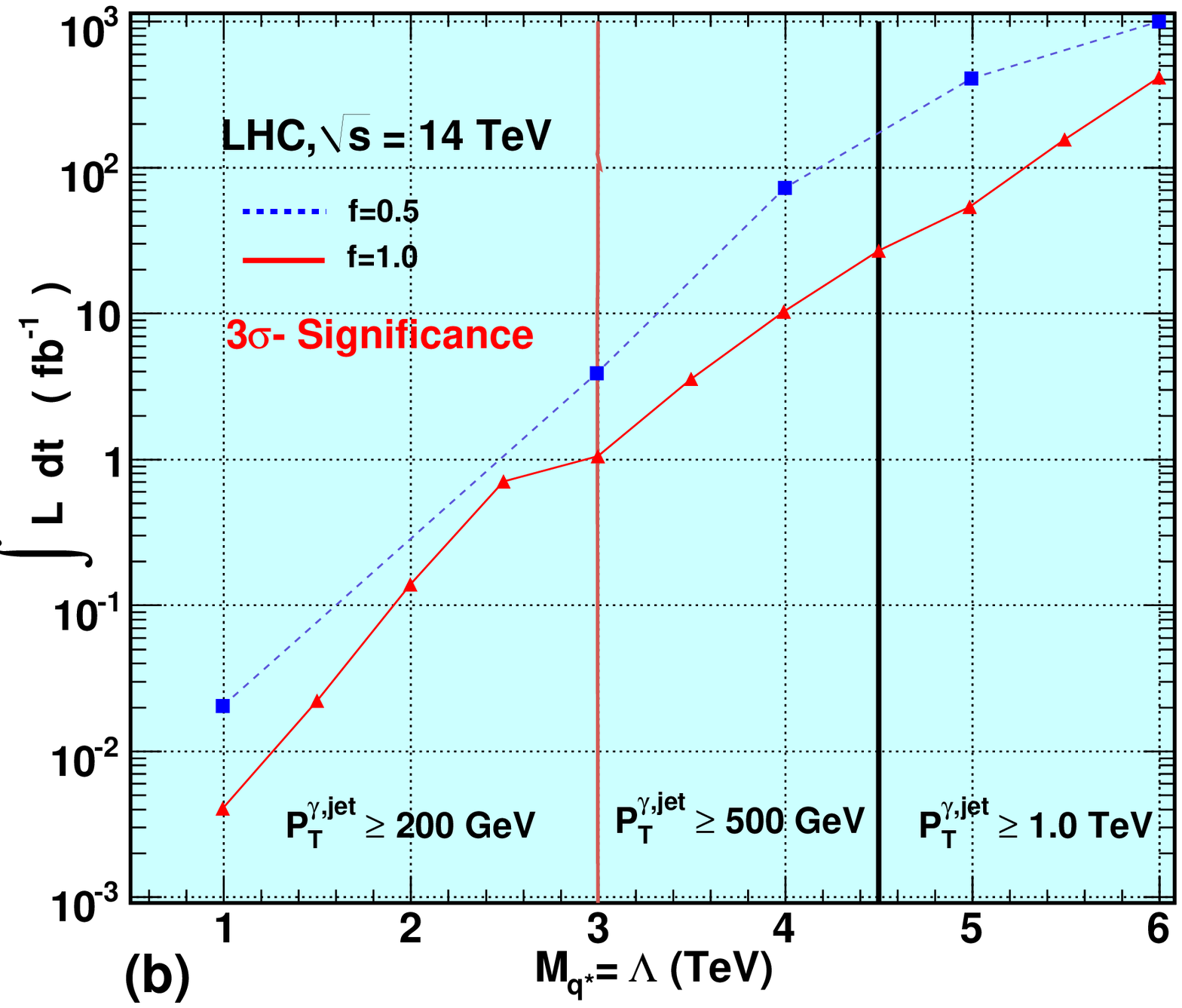}}
\caption{Required integrated luminosity as a function of $M_{q^*}$ 
for (a) 5$\sigma$ and (b) 3$\sigma$ significance for two different 
coupling strengths (systematic uncertainties are not included).}
\label{fig:result}
\end{figure}

Fig.~\ref{fig:result}(a) and (b) respectively show the 
luminosity needed to achieve 5$\sigma$ (3$\sigma$) significance for the
signal as a function of the excited quark mass. 
We find that the result obtained using ($-2 \, \ln Q$) are consistent with
those from $s/\sqrt{b}$ test statistic. In estimating the required luminosity, 
we have exploited only the mass peak region of the signal over the SM
background. We have used a mass window of $\sim \pm 3 \Gamma(q*)$
around $M_{q^*}$. In a previous section we have shown that beyond
$\pm 3\Gamma(q*)$ the discriminating statistic, namely 
LLR, looks similar for S+B and B.

We have checked the stability of the results by varying the bin width
of the invariant mass distribution from 50 GeV to 20 GeV for both
$M_{q^*} = 1.0 \tev$
and 2.0 TeV and find that the luminosity required for 5$\sigma$
significance changes by 20$\%$ and 1.1 $\%$ respectively. For 
$M_{q^*} = 5.0 \tev$, on the other hand, 
we varied the bin width from 50 GeV to 100 GeV and found that the
required luminosity changes by 2.1 $\%$. Similarly we increased the
number of Monte Carlo trials by a factor of 10 and found that the
required luminosity changes by $\sim$ 20 $\%$, 0.8 $\%$ and 2.1 $\%$
respectively for the 1.0, 2.0 and 5.0 TeV mass points.

 \begin{widetext}

In Fig.~\ref{fig:result}(a), corresponding to the $f_i=1.0$
case, we also demonstrate the effect of restricting 
the photon and
the jet to the central region of the calorimeter
on the required luminosity. 
For 5$\sigma$ significance, the latter reduces by $\sim 30 \%$ 
upto a mass of 4.5 TeV. For $M_{q^*} \gsim 5.0 \tev$, though, 
the signal
events are mostly produced in the central region and hence the
requirement $|\eta^{\gamma,jet}| \leq$ 1.5 does not affect the final
result significantly. It is also shown that within the present model
of compositenes for $M_{q^*} \gsim 5.5 \tev$, a 5$\sigma$ significance 
can not be achieved at a centre of mass energy of 14 TeV. While this might 
seem to run counter to previous work, note that, unlike in the earlier 
efforts, we have used unitarized amplitudes and hence our cross sections are 
naturally smaller than theirs.


In Table VI, we show the results for cross section and efficiency for
S+B and B after all the kinematical and isolation selection cuts 
have been imposed for
various $M_{q^*}$ values. Tabel VII shows similar information for
central events alone ($|\eta^{\gamma,jet}| \leq $1.5).


\begin{table*} \caption{Cross sections for various $M_{q^*}$ values
 after imposing all kinematical and isolation cuts. }  
\begin{ruledtabular} \begin{tabular}{|ccccccccc|} S.N. & $M_{q^*}\footnotemark[1]$ & $\Delta M_{\gamma-j}$ & $\sigma(S+B)$  & $\sigma(B)$  & $\sigma(S^*)\footnotemark[2]$ &   Efficiency(S+B)  & Efficiency(B) & Efficiency(S*) \\
     &   (TeV)   & $\pm 3\Gamma(q*)$ (GeV)  & (pb)  &  (pb) & (pb) & (\%)  & (\%)  & (\%) \\                                            
\hline

1    &  1.0 &   800-1200  & 9.26             & 4.92            & 4.34             & 1.304             & 0.699     & 60.28 \\

2    &  1.5 &  1200-1800  & 2.034            & 1.33            & 0.694               & 0.288           & 0.190    & 46.71 \\

3    &  2.0 &  1600-2400  & $6.72 \times 10^{-1}$  & $5.10 \times 10^{-1}$  & $1.61 \times 10^{-1}$      & 0.095      & 0.072    & 37.23\\

4    &  2.5 &   2000-3000 & $2.54 \times 10^{-1}$   & $2.10 \times 10^{-1}$ & $4.41 \times 10^{-2}$      & 0.036     & 0.029    & 40.67\\

5    &  3.0 &  2450-3550  & $7.85 \times 10^{-2}$   & $6.44 \times 10^{-2}$  &  $1.40 \times 10^{-2}$     & 0.011     & 0.009    & 75.95\\
\hline

6    &  3.5 &   2900-4150 & $1.11 \times 10^{-2}$  & $6.93 \times 10^{-3}$  &  $4.17 \times 10^{-3}$     & 0.274      & 0.172    & 24.70\\

7    &  4.0 &  3300-4700  & $4.90 \times 10^{-3}$   & $3.40 \times 10^{-3}$  &  $1.50 \times 10^{-3}$     & 0.121     & 0.084    & 15.60\\

8    &  4.5 &  3700-5300  & $2.20 \times 10^{-3}$    & $1.57 \times 10^{-3}$  &  $6.37 \times 10^{-4}$     & 0.054    & 0.039    & 11.48\\
\hline

9    &  5.0 &  4150-5850  & $4.60 \times 10^{-4}$   & $2.47 \times 10^{-4}$  &  $2.12 \times 10^{-4}$     & 0.628     & 0.342    & 22.49\\

10   &  5.5 &  4500-6450  & $2.17 \times 10^{-4}$   & $1.29 \times 10^{-4}$  &  $8.81 \times 10^{-5}$     & 0.299     & 0.179    & 14.91\\

11   &  6.0 &  5000-7000  & $8.39 \times 10^{-5}$   & $5.14 \times 10^{-5}$  &  $3.24 \times 10^{-5}$     & 0.115     & 0.071    & 7.85\\

\end{tabular}
\footnotetext[1]{Here $f_1=f_3=1.0$}
\footnotetext[2]{\mbox{Pure New Physics contribution evaluated by subtracting 
$B$ from $(S+B)$}}
\end{ruledtabular}
    \label{tab:cs_after_everything}
\end{table*}


\begin{table*} 
\caption{As in Table.\ref{tab:cs_after_everything}
with additional requirement of centrality ($|\eta^{\gamma,jet} |\leq 1.5$).}  
\begin{ruledtabular} \begin{tabular}{|ccccccccc|} S.N. & $M_{q^*}\footnotemark[1]$ & $\Delta M_{\gamma-j}$ & $\sigma(S+B)$  & $\sigma(B)$  & $\sigma(S^*)\footnotemark[2]$ &   Efficiency(S+B)  & Efficiency(B) & Efficiency(S*) \\
     &   (TeV)   & $\pm 3\Gamma(q*)$ (GeV)  & (pb)  &  (pb) & (pb) & (\%)  & (\%)  & (\%) \\                                            
\hline

1    &  1.0 &   800-1200  &   4.75   & 2.14     &  2.61     &    0.668    & 0.304    & 36.22 \\

2    &  1.5 &  1200-1800  &  0.87   & 0.41    & 0.45     &   0.123     & 0.059    & 30.64  \\

3    &  2.0 &  1600-2400  & $ 2.27 \times 10^{-1}$   & $1.10 \times 10^{-1}$  & $1.16  \times 10^{-1}$      &  0.032   &  0.015   & 26.82 \\

4    &  2.5 &   2000-3000 & $6.54 \times 10^{-2}$   &  $3.43 \times 10^{-2}$ &  $3.11 \times 10^{-2}$      &   0.009   & 0.004   & 28.66 \\

5    &  3.0 &  2450-3550  & $2.21 \times 10^{-2}$   & $1.27 \times 10^{-2}$  &  $9.40 \times 10^{-3}$     &  0.003   &  0.001    & 50.95 \\
\hline

6    &  3.5 &   2900-4150 & $6.67 \times 10^{-3}$  & $3.20 \times 10^{-3}$  &  $3.47  \times 10^{-3}$     &  0.165     &  0.079   & 20.55 \\

7    &  4.0 &  3300-4700  & $2.64 \times 10^{-3}$   & $ 1.30 \times 10^{-3}$  & $1.34  \times 10^{-3}$     &  0.065   &   0.032  & 13.93 \\

8    &  4.5 &  3700-5300  & $1.01 \times 10^{-3}$    & $4.59 \times 10^{-4}$  & $5.51  \times 10^{-4}$     &  0.025   &  0.011   & 9.92 \\
\hline

9    &  5.0 &  4150-5850  & $3.99 \times 10^{-4}$   & $2.00 \times 10^{-4}$  & $1.98  \times 10^{-4}$     &  0.545   &  0.277   & 20.98\\

10   &  5.5 &  4500-6450  & $1.79 \times 10^{-4}$   & $9.78  \times 10^{-5}$  & $8.15 \times 10^{-5}$     &  0.246    &  0.135   & 13.78 \\

11   &  6.0 &  5000-7000  & $6.51 \times 10^{-5}$   & $3.44 \times 10^{-5}$  & $3.07  \times 10^{-5}$     &  0.089   &  0.047   & 7.43 \\

\end{tabular}
\footnotetext[1]{Here $f_1=f_3=1.0$}
\footnotetext[2]{\mbox{Pure New Physics contribution evaluated by subtracting 
B from (S+B)}}
\end{ruledtabular}
\end{table*}


 \end{widetext}

\section{Systematic Uncertainties}

Since we have performed a detailed analysis including a realistic
simulation of various detector effects and uncertainties for the CMS
setup, here we present an estimation of systematic uncertainties. For
this we have considered only the dominant contribution both in the
signal and the backgrounds. For both the signal and the $\gamma+jet$
background we concentrate on the dominant process, viz. $qg\rightarrow
q\gamma$. For the QCD dijet background, all the available processes in
PYTHIA were used for the estimation of the uncertainty.  We did not
account for $q\bar{q}\rightarrow \gamma+jet, $$W/Z(jj)+\gamma$ and
$gg\rightarrow\gamma+jet$ as they contribute only a small fraction of
the total background and systematic uncertainities in these can safely
be neglected.

\begin{itemize}

\item

 Choice of the parton distributions (PDF): To estimate the uncertainty
in the cross sections due to the choice of the PDF, the former were
recalculated for three additional PDFs, namely CTEQ6L,
CTEQ6M~\cite{cteq_2} and MRST2001~\cite{mrst1}.  Using the LHApdf
package~\cite{LHApdf}, the results for each were compared to those for
our default choice, namely CTEQ5L~\cite{cteq_1}. While the resultant
cross sections turned out to be higher for CTEQ6M (it should be 
noted that the CTEQ6M distributions are NLO and hence their use 
with LO calculations is fraught with danger) and MRST2001
distributions, for CTEQ6L they turned out to be lower for almost all
the signal points. As can be expected, 
the uncertainty in the cross section
increases with $M_{q^*}$, simply because one starts to sample an  
ill-explored region in the $(Q^2, x)$ plane. 
 For CTEQ6M and MRST2001, the
relative deviation varies between 2.3--13.0 \% and 2.6--14.2 \% respectively
as  $M_{q^*}$ changes from 1 TeV to 6 TeV. For CTEQ6L, the
variation was found to be within $-4.5$ to +2.25 $\%$. These numbers are 
quite consistent with those applicable for the SM $\gamma+ {\rm jet}$
process alone, for which the corresponding numbers are 5.6--11.0\%
(CTEQ6M) and 6.0-12.0\%(MRS2001). Similarly, for the dijet
background an uncertatinty of 9--16\% (CTEQ6M) and 
8.7--16.5\% (MRST2001) was estimated.

 We have not only used
different PDFs but have also evaluated uncertainty due to a given 
proton PDF by varying the errors on the parameters of the PDF fit
itself. For this, we chose CTEQ6L (with NLO $\alpha_{s}$ and LO fit)
and its 40 subset PDFs. The uncerainty was found to be $\sim \pm$
1$\%$ for a 1 TeV $q^*$ state and $-8.29\%$ to $+10.93\%$ for a 5 TeV
one. For QCD di-jet and $\gamma+Jet $ background these numbers were
found to be $-9.81\%$ to $+13.74\%$ and $-8.04\%$ to $+10.54\%$
respectively.  


 \item

Scale Variation: To estimate the dependence of the signal and the
background cross-sections on the choice of the factorization scale $Q$
(default value in our analysis being $\sqrt{\hat{s}} $), they were
recalculated for two other values of the latter viz $Q^{2}= P_{T}^{2}$
and $Q^{2}= -\hat t$. Both these choices for the scale would have 
resulted in a higher cross-section compared to $Q^{2}=\hat s$. 
The deviation increases with $M_{q^*}$ 
and ranges between 2.1--11.3\% for $Q^2 = - \hat t$  and 10.6--25.0\% 
for $Q^2 = P_{T}^{2}$ case. For the QCD di-jet background the maximum
deviation was found to be $\sim$39 $\%$ while for $\gamma+jet$ it was
$\sim$26 $\%$. Thus, the overall significance of the signal remains 
largely unaltered.

\item 

Higher-order effects: For the background, these have been
studied in detail both theoretically and experimentally. For example, 
$\gamma+jet$ production in the SM has been studied in depth using the NLO
parton level Monte Carlo program
JETPHOX~\cite{phox_program1,phox_program2}. Recently, a comparison of
these predictions have been done with the Tevatron
data\cite{tevatron_nlo}. Unfortunately, the 
$P_{T}^{\gamma}$-dependent shape of the 
triple differential cross
section($d^{3}\sigma/dp^{\gamma}_{T}dy^{\gamma}dy^{jet}$) for
different pseudorapidity ranges is not explained satisfactorily by
the NLO calculation. The reason is not hard to fathom. 
A comparison with data necessitates the imposition of isolation cuts.
On the other hand, the NLO calculations depend crucially
on the choice of isolation cuts and infrared safety
needs to be taken care of. This has been discussed
in Ref.\cite{phox_program3}. 
Modulo such subtleties, an effective and easy way to incorporate higher
order effects is to include $K$-factors. For $\gamma+jet$ production, 
the $K$-factor lies in the range  1.0--1.66 depending on the details 
of jet fragmentation (primarily, to a $\gamma/\pi^{0})$\cite{k_factor}. 
While the $K$-factor for our case is not known,  in the large $M_{q^*}$ limit
it is not expected to 
be too different from the SM case.
Close to threshold, the $K$-factor is normally expected to be even bigger.
However, given the attendant theoretical complications,
we adopt a conservative approach and ignore all 
$K$-factors in this analysis.


\item Jet energy resolution: To incorporate finite detector resolutions, the
photon and jet four momenta were smeared with a energy resolution
as given in section VI. For the photon $P_{T}$ range considered
in this analysis, we expect the constant term ($C$) to be the
dominant source of error and it contributes 
about $0.55 \%$. To estimate the effect of the
jet energy resolution on this analysis, we redid this analysis 
smearing the four momenta of
the jet with an energy resolution of 100$\%$ for the barrel region and
150 $\%$ for the endcaps and the forward regions. The effect was
studied for two different mass state, viz. 1.0 TeV and 5.0 TeV. It was
found that such a large worsening of the
jet energy resolution would increase the 
luminosity required for a 5$\sigma$ significance by about $30 \%$ ($1\%$)
for 1.0 TeV (5.0 TeV) mass states respectively. However, if we increase
the number of MC trials by a factor of 10 (to stablize the peak value
of $LLR_{S+B}$) then these numbers were found to be well within 2$\%$.


\item Uncertainty due to pre-selection: The systematic uncertainty due
to preselection in the $P_{T}$ range of this study is found to be less
than  $1\%$.


\item Luminosity error: For the CMS
experiment, this error is expected to be $\sim 10 \% $ for an integrated
luminosity of $1 \fb^{-1}$ ~\cite{syslumi_1} and $\sim 3 \%$ for an
integrated luminosity of $30 \fb^{-1}$~\cite{syslumi_2}.


\item To estimate the combined effect of the aforementioned
uncertatinties we lowered the signal
cross-section by 5$\%$ for the 5.0 TeV mass point and found that the
luminosity for 5$\sigma$ significance increases by $\sim10.9\%$. 
On the other hand, if we increase the background cross-section
by $5\%$ for the same mass point then the required luminosity
increases by $\sim 5.1 \%$.

\end{itemize}

\section{Conclusions}

To summarise, we have investigated the potential of using a direct
photon (in association with a single hard jet) final state at the LHC
to probe possible quark excitations.  Such states arise naturally in a
variety of scenarios, ranging from Kaluza-Klein excitations in
extra-dimensional models to theories wherein the quarks themselves are
composed of more fundamental objects (preons). And, as far as the
concerns of the present analysis go, even other fundamental quarks (as
often appear in theories with extended symmetries, gauged or global)
could lead to similar signals and, hence, be discoverable.

In any such model, the excited states may couple to their SM
counterparts only through a generalised (chromo-)magnetic transition
term in an effective Lagrangian. Consequently, the presence of such
states would alter the direct photon cross section, whether through an
on-shell production (and subsequent decay) of the $q^*$ or through an
off-shell exchange (both $s$- and $t$-channel).  The extent of the
deviation depends on both the mass $M_{q^*}$ and the
compositeness/excitation scale $\Lambda$.  The deviation concentrates
in the large $p_T$ regime, especially for larger $M_{q^*}$ and can be
quite substantial.

Two points need to be noted here. The first and straightforward one
relates to the width of the excited state. With $\Gamma(q^*)$ being
typically quite large, a narrow-width approximation does not hold and
the full matrix element needs to be incorporated. The second issue is
more subtle and is connected to the non-renormalizable nature of the
effective Lagrangian. Since a naive use of a (chromo-)magnetic dipole
moment vertex leads to a cross section constant or even growing with
the center of mass energy, the amplitude needs to be unitarized. This,
understandably, leads to a suppression of the cross sections, a fact
often ignored in experimental analyses, but included here.

Using the photon and jet reconstruction algorithms as used for the CMS
dectector at the LHC, we have performed a realistic estimation of the
deviation caused by the excited quark exchange contribution to the
direct photon plus a single hard jet rate.  We have accounted for all
major backgrounds to evaluate the limits in the $\Lambda-M_{q^*}$
parameter space. With the imposition of moderate restrictions on the
rapidities (as dictated by the detector acceptances), but stringent
cuts on the transverse momenta, the background can be beaten down
severely without any damaging loss of signal. A most crucial
ingredient is the application of reasonably stringent isolation
criteria, as it helps control the orders of magnitude larger
backgrounds from QCD dijet production with one jet faking a photon.
The consequent exclusion limits that may be reached are very strong.
While it may seem that these are still not as strong as some quoted in
the literature, it should be realised that most of the latter have
worked with a non-unitarised cross section and hence the two cannot be
compared directly.



\vspace*{2ex}
 \begin{acknowledgments}
BCC would like to thank S. Mrenna for discussions on $W\gamma$
production. SB and DC acknowledge
support from the Department of Science and Technology(DST), India
under project number SR/S2/RFHEP-05/2006. BCC acknowledges support
from the DST, India under project number SP/S2/K-25/96-V. SSC would
like to express gratitude to the Council of Scientific and Industrial
Research, India for financial assistance and to Prof. R.K. Shivpuri
and Prof. Raghuvir Singh for support and encouragement.
 \end{acknowledgments}

\def\Journal#1#2#3#4{{#1} {#2} (#4) #3}
\def\ANNP{\em Ann. Phys. (N.Y.)}
\def\ARNS{\em Ann.~Rev.~Nucl.~Sci.}
\def\EPJC{{\em Eur.~Phys.~J.}{\bf C}}
\def\IJMPE{{\em Int. J. Mod. Phys.} E}
\def\IJMPA{{\em Int. J. Mod. Phys.} A}
\def\JETPLC{{\em Sov. Phys. JETP Lett.} C}
\def\JHEP{{\em J.~High.~E.~Phys.}}
\def\MPLA{{\em Mod.~Phys.~Lett.} A}
\def\NIMA{{\em Nucl.~Instr.~and~Meth.} A}
\def\NIMB{{\em Nucl.~Instr.~and~ Meth.} B}
\def\NCA{{\em Nuovo Cimento} A}
\def\NPA{{\em Nucl. Phys.} A}
\def\NPB{{\em Nucl.~Phys.} B}
\def\NJP{{\em New~J.~Phys.}}
\def\PHYS{{\em Physica}}
\def\PLA{{\em Phys. Lett.} A}
\def\PLB{{\em Phys. Lett.} {\bf B}}
\def\PLD{{\em Phys. Lett.} D}
\def\PL{{\em Phys. Lett.}}
\def\PRL{\em Phys. Rev. Lett.}
\def\PREV{\em Phys. Rev.}
\def\PREP{\em Phys. Rep.}
\def\PRA{{\em Phys. Rev.} A}
\def\PRD{{\em Phys. Rev.} D}
\def\PRC{{\em Phys. Rev.} C}
\def\PRB{{\em Phys. Rev.} B}
\def\PRO{{\em Prog. Theor. Phys.}}
\def\RMP{{\em Rev. Mod. Phys.}}
\def\ZPC{{\em Z. Phys.} C}
\def\ZPA{{\em Z. Phys.} A}

\def\etal{{\em et al.}}
\def\idem{{\em idem}}

\end{document}